\documentclass{elsart}

\usepackage[square,comma]{natbib}
\usepackage{graphicx}
\usepackage{pxfonts}
\usepackage{lineno}

\usepackage{amssymb}

\journal{}

\begin{document}

\thispagestyle{empty}
\begin{Large}
\textbf{DEUTSCHES ELEKTRONEN-SYNCHROTRON}

\textbf{\large{Ein Forschungszentrum der Helmholtz-Gemeinschaft}\\}
\end{Large}

DESY 12-159

September 2012

\begin{eqnarray}
\nonumber &&\cr \nonumber && \cr \nonumber &&\cr
\end{eqnarray}
\begin{eqnarray}
\nonumber
\end{eqnarray}
\begin{center}
\begin{Large}
\textbf{Optimization of a dedicated bio-imaging beamline at the
European X-ray FEL}
\end{Large}
\begin{eqnarray}
\nonumber &&\cr
\end{eqnarray}

\begin{large}
Gianluca Geloni,
\end{large}
\textsl{\\European XFEL GmbH, Hamburg}
\begin{large}

Vitali Kocharyan and Evgeni Saldin
\end{large}
\textsl{\\Deutsches Elektronen-Synchrotron DESY, Hamburg}
\begin{eqnarray}
\nonumber
\end{eqnarray}
\begin{eqnarray}
\nonumber
\end{eqnarray}
ISSN 0418-9833
\begin{eqnarray}
\nonumber
\end{eqnarray}
\begin{large}
\textbf{NOTKESTRASSE 85 - 22607 HAMBURG}
\end{large}
\end{center}
\clearpage
\newpage

\begin{frontmatter}



\title{Optimization of a dedicated bio-imaging beamline at the European X-ray FEL}


\author[XFEL]{Gianluca Geloni\thanksref{corr},}
\thanks[corr]{Corresponding Author. E-mail address: gianluca.geloni@xfel.eu}
\author[DESY]{Vitali Kocharyan}
\author[DESY]{and Evgeni Saldin}

\address[XFEL]{European XFEL GmbH, Hamburg, Germany}
\address[DESY]{Deutsches Elektronen-Synchrotron (DESY), Hamburg,
Germany}

\begin{abstract}
We recently proposed a basic concept for design and layout of the
undulator source for a dedicated bio-imaging beamline at the
European XFEL. The goal of the optimized scheme proposed here is to
enable experimental simplification and performance improvement. The
core of the scheme is composed by soft and hard X-ray self-seeding
setups. Based on the use of an improved design for both
monochromators it is possible to increase the design electron energy
up to 17.5 GeV in photon energy range between 2 keV and 13 keV,
which is the most preferable for life science experiments. An
advantage of operating at such high electron energy is the increase
of the X-ray output peak power. Another advantage is that 17.5 GeV
is the preferred operation energy for SASE1 and SASE2 beamline
users. Since it will be necessary to run all the XFEL lines at the
same electron energy, this choice will reduce the interference with
other undulator lines and increase the total amount of scheduled
beam time. In this work we also propose a study of the performance
of the self-seeding scheme accounting for spatiotemporal coupling
caused by the use of a single crystal monochromator. Our analysis
indicates that this distortion is easily suppressed by the right
choice of diamond crystal planes and that the proposed undulator
source yields about the same performance as in the case for a X-ray
seed pulse with no coupling. Simulations show that the FEL power
reaches 2 TW in the 3 keV - 5 keV photon energy range, which is the
most preferable for single biomolecule imaging.

\end{abstract}

%
%
%
\end{frontmatter}




\section{\label{sec:intro} Introduction}

The availability of free undulator tunnels at the European XFEL
facility offers a unique opportunity to build a beamline optimized
for coherent diffraction imaging of complex molecules, like proteins
and other biologically interesting structures. Crucial parameters
for such bio-imaging beamline are photon energy range, peak power,
and pulse duration \cite{HAJD}-\cite{SEIB}.

The highest diffraction signals are achieved at the longest
wavelength that supports a given resolution, which should be better
0.3 nm. With photon energy of about 3 keV one can reach a resolution
better than 0.3 nm with a detector designed to collect diffracted
light in all forward directions, that is at angles $2\theta <
\pi/2$. Higher photon energies up to about 13 keV give access to
absorption edges of specific elements used for phasing by anomalous
diffraction. The most useful edges to access are the K-edge of Fe
(7.2 keV) and Se (12.6 keV), \cite{BERG}. Access to the sulfur
K-edge (2.5 keV) is required too. Finally, the users of the
bio-imaging beamline also wish to investigate large biological
structures in the soft X-ray photon energy range down to the water
window (0.3 keV - 0.5 keV), \cite{BERG}.

Overall, one aims at the production of pulses containing enough
photons to produce measurable diffraction patterns, and yet short
enough to avoid radiation damage in a single pulse. This is, in
essence, the principle of imaging by "diffraction before
destruction" \cite{NEUT}. These capabilities can be obtained by
reducing the pulse duration to $5$ fs or less, and simultaneously
increasing the peak power to the TW power level or higher, at photon
energies between 3 keV and 5 keV, which are optimal for imaging of
macromolecular structures \cite{BERG}.

The requirements for a dedicated bio-imaging beamline are the
following. The X-ray beam should be delivered in ultrashort pulses
with TW peak power and within a very wide photon energy range
between 0.3 keV and 13 keV. The pulse duration should be adjustable
from 10 fs in hard X-ray regime to 2 fs - 5 fs in photon energy
range between 3 keV and 5 keV. At the European XFEL it will be
necessary to run all undulator beamlines at the same electron energy
and bunch charge. However, bio-imaging experiments should be
performed without interference with other main SASE1, SASE2
beamlines. This assumes the use of nominal electron energy and
electron beam distribution.

A key component of the bio-imaging beamline is the undulator source.
A basic concept for layout and design of the undulator system for a
dedicated bio-imaging beamline at the European XFEL was proposed in
\cite{OURCC}. All the requirements in terms of photon beam
characteristics can be satisfied by the use a very efficient
combination of self-seeding, fresh bunch, and undulator tapering
techniques \cite{TAP1}-\cite{BZVI}, \cite{HUAN}-\cite{WU}. A
combination of self-seeding and undulator tapering techniques would
allow to meet the design TW output power. The bio-imaging beamline
would be equipped with two different self-seeding setups, one
provide monochromatization in the soft X-ray range, and one to
provide monochromatization in the hard X-ray range.  The most
preferable solution in the photon energy range for single
biomolecule imaging consists in using a fresh bunch technique in
combination with self-seeding and undulator tapering techniques. In
\cite{OURCC} it was shown how the installation of an additional
(fresh bunch) magnetic chicane behind the soft X-ray self-seeding
setup enables an output power in the TW level for the photon energy
range between 3 keV and 5 keV. Additionally, the pulse duration can
be tuned between 2 fs and 10 fs with the help of this chicane, still
operating with the nominal electron bunch distribution \cite{S2ER}.

The overall setup proposed in \cite{OURCC} is composed of four
undulators separated by three magnetic chicanes. The undulator parts
consist of 4,3,4 and 29 cells. Each magnetic chicane compact enough
to fit one 5 m-long undulator segment and the FODO lattice will not
be perturbed. The undulator system will be realized in a similar
fashion as other European XFEL undulators. In order to make use of
standard components we favor the use of SASE3 type of undulator
segments, which are optimized for the generation of soft X-rays. The
present layout of the European XFEL enables to accommodate such new
beamline. The previously proposed undulator source provides access
to a photon energy range between 3 keV and 5 keV only at the reduced
electron beam energy of 10.5 GeV. Although the 10.5 GeV is one of
the nominal electron energy, it may not be the preferable mode of
operation for SASE1, SASE2 beamline users. Note that the SASE3
undulator type would enable operation down to 0.7 keV at an electron
energy of 17.5 GeV. However, the delay of photons induced in the
grating monochromator (3 ps) and, consequently, the delay of the
electrons required in the magnetic chicane of the soft X-ray
self-seeding setup sets a limit to the electron energy.

This paper constitutes an update to the scheme proposed in
\cite{OURCC}. The present design assumes the use of the same 40
cells undulator system, with an improved design of both self-seeding
setups. To avoid any interference with other beamlines, we propose
to extend the photon energy range of the self-seeding setup with a
single crystal monochromator down to 3 keV \cite{OURCC2}.  As a
result, the design electron energy can be increased up to 17.5 GeV
in the photon energy range most preferable for bio-imaging. This is
achieved exploiting 0.1 mm diamond crystals in symmetric Bragg
geometry. Based on the use C(111), C(220), and C(400) reflections
($\sigma$-polarization) it will be possible to cover the photon
energy range between 3 keV and 13 keV. In particular, we exploit
C(111) reflection ($\sigma$-polarization) in photon energy range
between 3 keV and 5 keV. Combination of self-seeding and fresh bunch
techniques, as in the case of the original design, has the advantage
that the pulse duration can be tuned between 2 fs and 10 fs.

The users of the bio-imaging beamline also wish to investigate their
samples around sulfur K-edge, i.e. in the photon energy range
between 2 keV and 3 keV. A solution suitable for this spectral range
constitutes a major challenge for self-seeding designers. In fact,
on the one hand crystals with right lattice parameters are difficult
to be obtained. On the other hand, grating monochromator throughput
is usually too low due to high absorption. As for the original
design we propose a method around this obstacle, which is based in
essence on a fresh bunch technique, and exploits a self-seeding
setup based on grating monochromator in the photon energy range
between 0.7 keV and 1 keV. It should be noted that due to extension
of the single crystal monochromator setup down to 3 keV, the maximal
photon energy of operation for the grating monochromator is reduced
from 1.7 keV in the original design down to 1 keV in the current
design.

Also, here we adopt an improved design of grating monochromator,
which was recently proposed for the soft X-ray self-seeding setup at
the LCLS \cite{FENG3}, substituting a previously proposed one
\cite{FENG, FENG2}. In this novel design the optical delay is
reduced down to below 1 ps. As a result, a self-seeding setup with
such grating monochromator allows for reduced constraints on the
magnetic chicane, and can operate at the European XFEL down to 0.7
keV at the highest nominal electron energy of 17.5 GeV. Such high
electron energy enables to increase the X-ray output peak power in
the most preferable photon energy range for bio-imaging experiments
up to 2 TW.

\section{Setup description}

\begin{figure}[tb]
\includegraphics[width=1.0\textwidth]{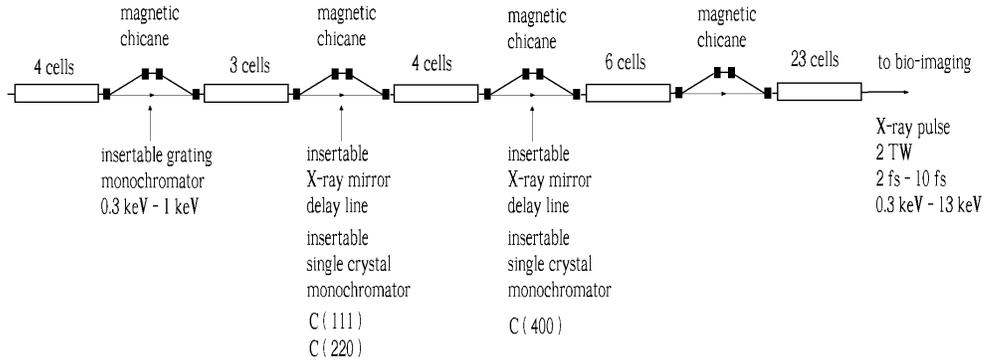}
\caption{Design of the undulator system for the bio-imaging
beamline. The method exploits a combination of self-seeding, fresh
bunch, and undulator tapering technique. Each magnetic chicane
accomplishes three tasks by itself. It creates an offset for
monochromator or X-ray mirror delay line installation, it removes
the electron microbunching produced in the upstream undulator, and
it acts as a magnetic delay line. } \label{bio3f1}
\end{figure}

\begin{figure}[tb]
\includegraphics[width=1.0\textwidth]{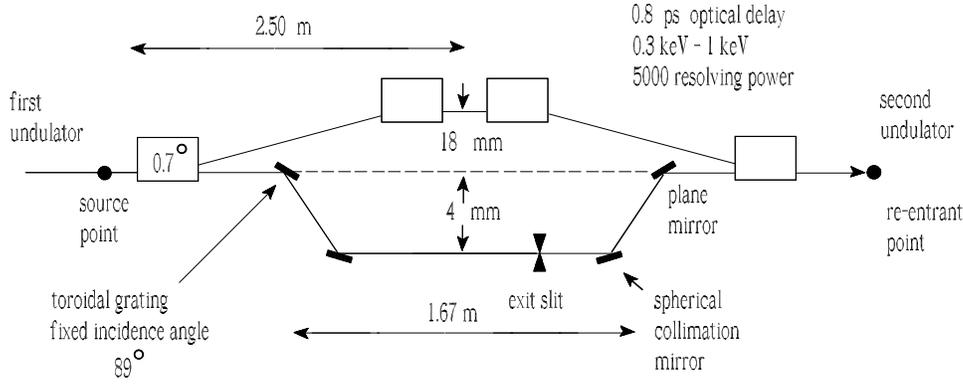}
\caption{Compact grating monochromator originally proposed at SLAC
\cite{FENG3} for soft X-ray self-seeding setup. The chicane fits in
one European XFEL undulator undulator section (5 m). }
\label{bio3f7}
\end{figure}
Self-seeding is a promising approach to significantly narrow the
SASE bandwidth and to produce nearly transform-limited X-ray pulses
\cite{SELF}-\cite{WUFEL2}. In its simplest configuration, a
self-seeding setup in the hard X-ray regime consists of two
undulators separated by photon monochromator and electron bypass
beamline, typically a 4-dipole chicane. The two undulators are
resonant at the same radiation wavelength. The SASE radiation
generated by the first undulator passes through the narrow-band
monochromator, thus generating a transform-limited pulse, which is
then used as a coherent seed in the second undulator. Chromatic
dispersion effects in the bypass chicane smear out the microbunching
in the electron bunch produced by the SASE lasing in the first
undulator. Electrons and monochromatized photon beam are recombined
at the entrance of the second undulator, and the radiation is
amplified by the electron bunch in the second undulator, until
saturation is reached. The required seed power at the beginning of
the second undulator must dominate over the shot noise power within
the gain bandpass, which is order of a few kW.

Despite the unprecedented increase in peak power of the X-ray pulses
for SASE X-ray FELs (see e.g. \cite{LCLS2}), some applications,
including single biomolecule imaging, require still higher photon
flux. The most promising way to extract more FEL power than that at
saturation is by tapering the magnetic field of the undulator
\cite{TAP1}-\cite{LAST}. Also, a significant increase in power is
achievable by starting the FEL process from a monochromatic seed
rather than from noise \cite{OURY3}-\cite{WUFEL2}. Tapering consists
in a slow reduction of the field strength of the undulator in order
to preserve the resonance wavelength, while the kinetic energy of
the electrons decreases due to the FEL process. The undulator taper
could be simply implemented at discrete steps from one undulator
segment to the next. The magnetic field tapering is provided by
changing the undulator gap.

The setup suggested in this article constitutes an optimization of
the original proposal in \cite{OURCC} and is composed of five
undulator parts separated by four magnetic chicanes as shown in Fig.
\ref{bio3f1}. These undulators consist of $4$, $3$, $4$, $6$ and
$23$ undulator cells, respectively. Each magnetic chicane is compact
enough to fit one undulator segment. The installation of chicanes
does not perturb the undulator focusing system. The implementation
of the self-seeding scheme for soft X-ray would exploit the first
magnetic chicane. The second and third magnetic chicanes create an
offset for the installation of a single crystal monochromator or an
X-ray mirror delay line, and act as a magnetic delay line. Both
self-seeding setups should be compact enough to fit  one undulator
module.

For soft X-ray self-seeding, the monochromator usually consists of a
grating \cite{SELF}. Recently, a very compact soft X-ray
self-seeding scheme has appeared, based on a grating monochromator
\cite{FENG3}. The proposed monochromator is composed of a toroidal
grating followed by three mirrors, and is equipped with an exit slit
only. The delay of the photons is about $1$ ps. The monochromator is
continuously tunable in the photon energy range between $0.3$ keV
and $1$ keV.  The resolution is about $5000$. The transmission of
the monochromator beamline is close to $10 \%$. The magnetic chicane
delays the electron electron bunch accordingly, so that the photon
beam passing through the monochromator system recombines with the
same electron bunch. The chicane provides a dispersion strength of
about $0.6$ mm in order to match the optical delay and also smears
out the SASE microbunching generated in the first $4$ cells of the
undulator. It should be noted that in \cite{OSOF} we studied the
performance of a previous scheme of a grating monochromator for a
soft X-ray self-seeding setup \cite{FENG,FENG2}. For the present
investigation we consider the new scheme in \cite{FENG3}. The layout
of the bypass and of the monochromator optics is schematically shown
in Fig. \ref{bio3f7}.

\begin{figure}[tb]
\includegraphics[width=1.0\textwidth]{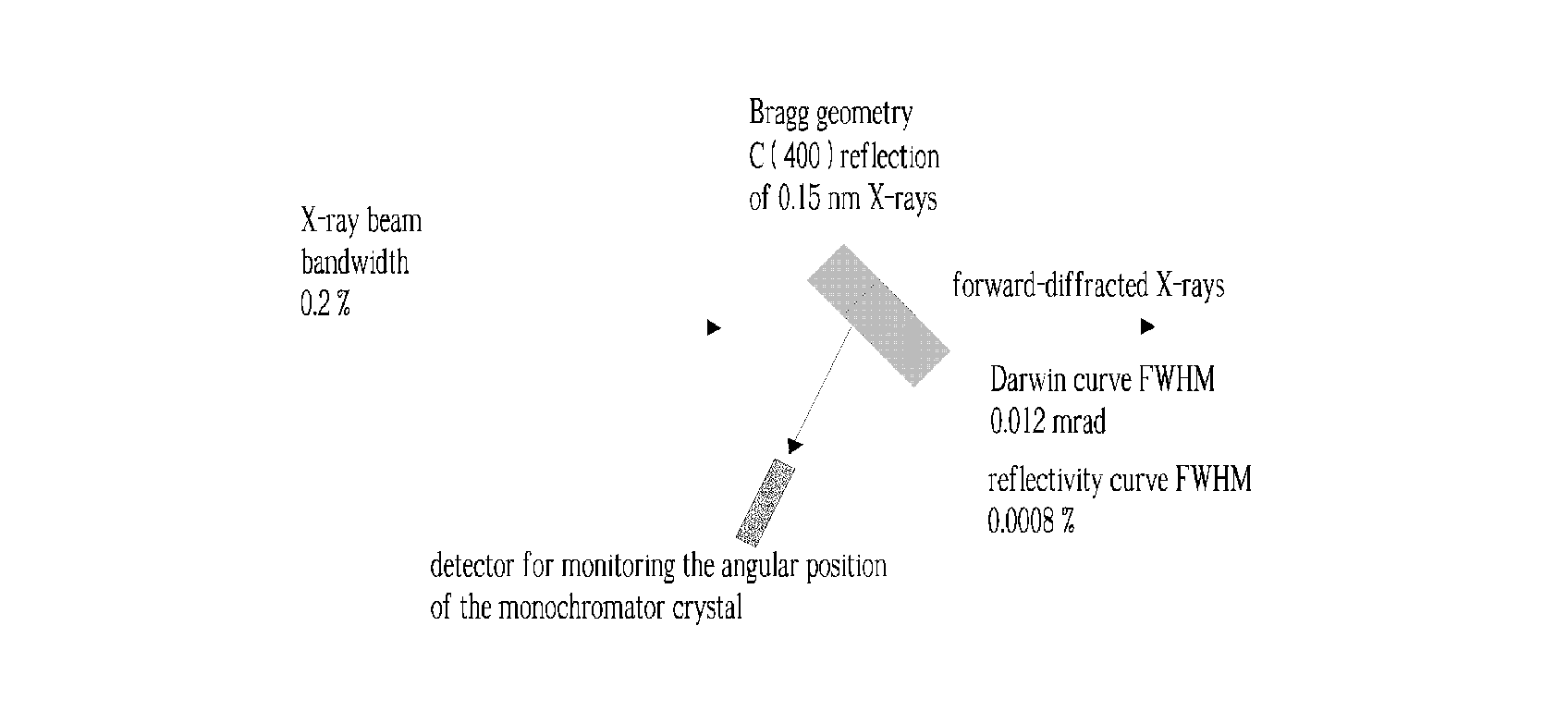}
\includegraphics[width=0.5\textwidth]{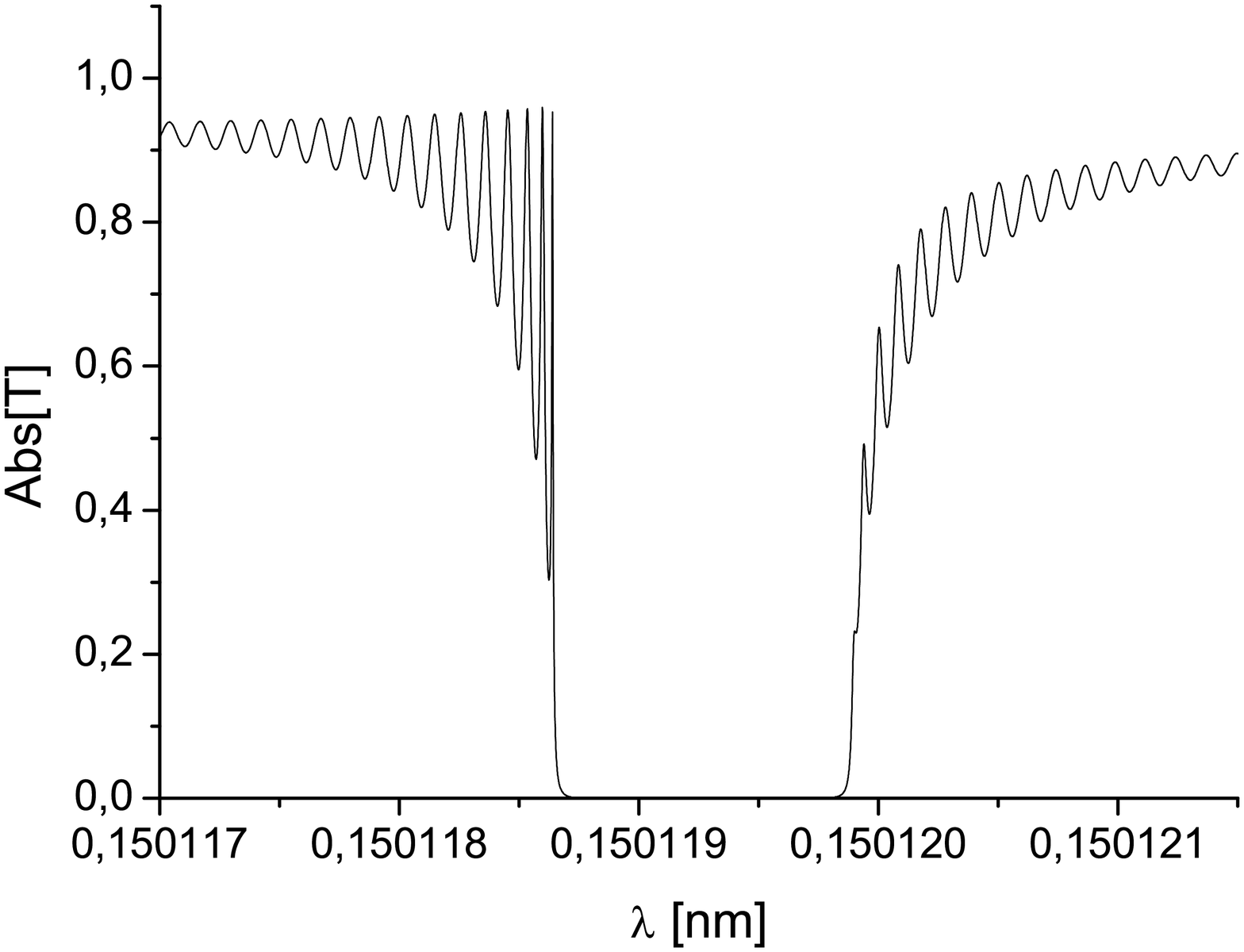}
\includegraphics[width=0.5\textwidth]{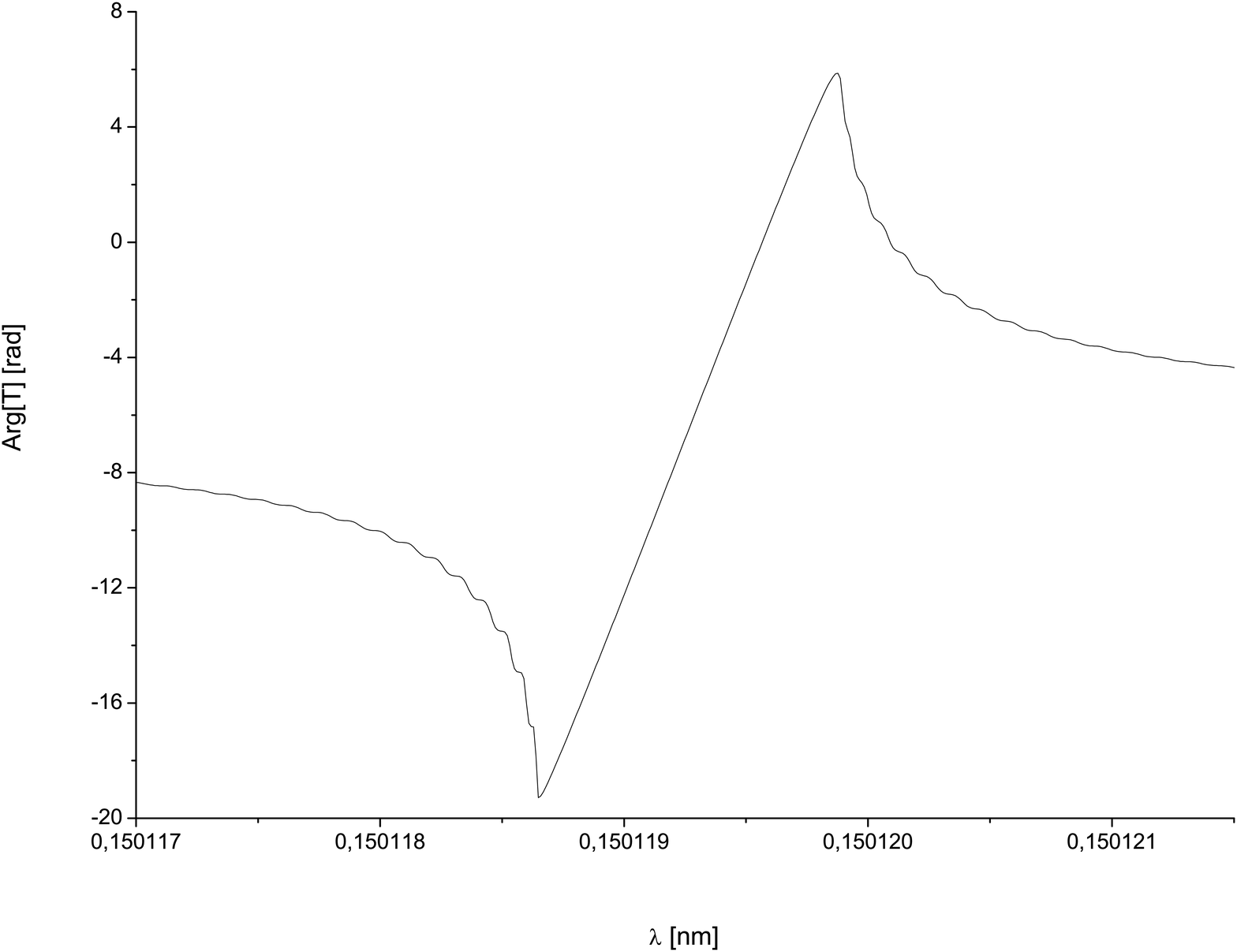}
\caption{X-ray optics for compact crystal monochromator originally
proposed in \cite{OURY5b} for a hard X-ray self-seeding setup, based
on the C(400) reflection ($\sigma$-polarization). Modulus and phase
of the transmissivity are shown in the two lower plots.}
\label{biof1}
\end{figure}

\begin{figure}[tb]
\includegraphics[width=1.0\textwidth]{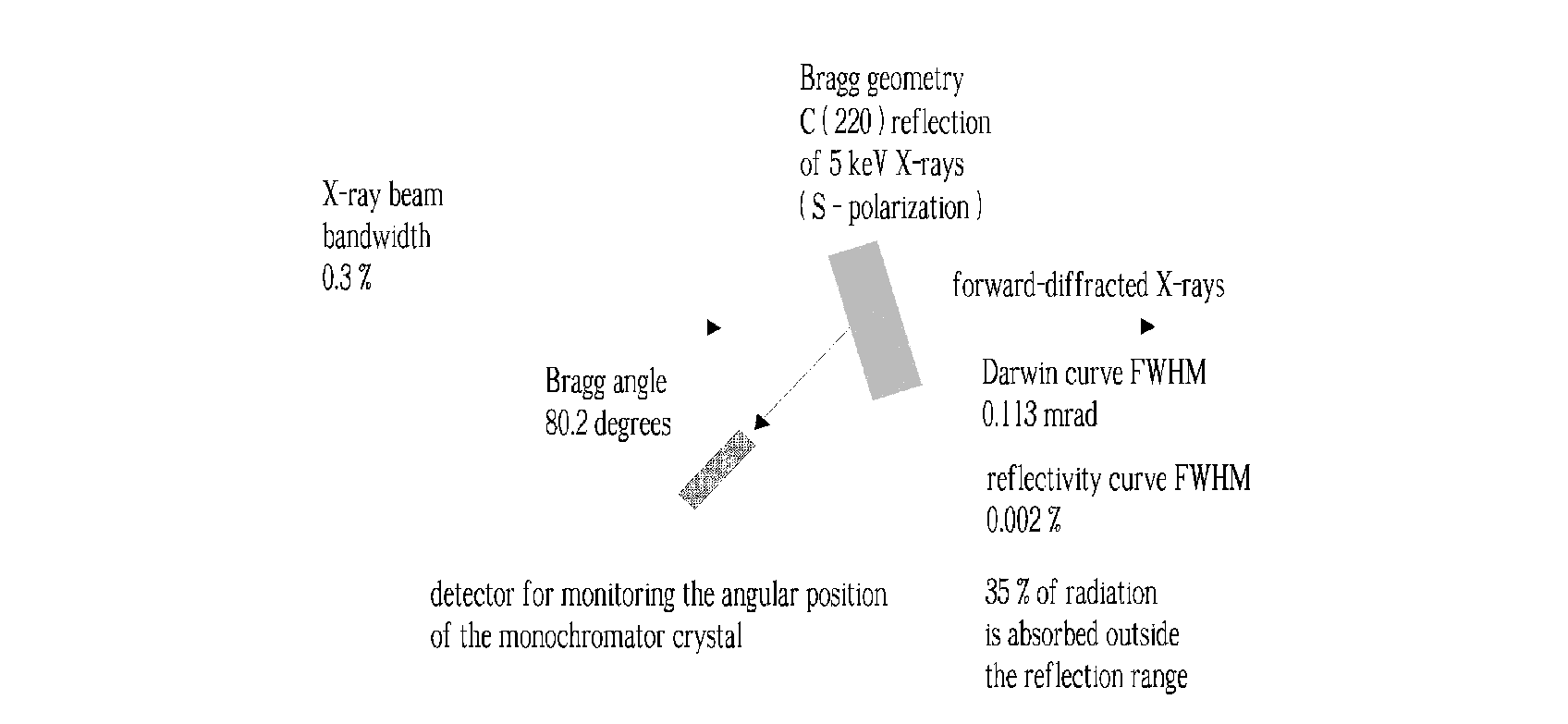}
\includegraphics[width=0.5\textwidth]{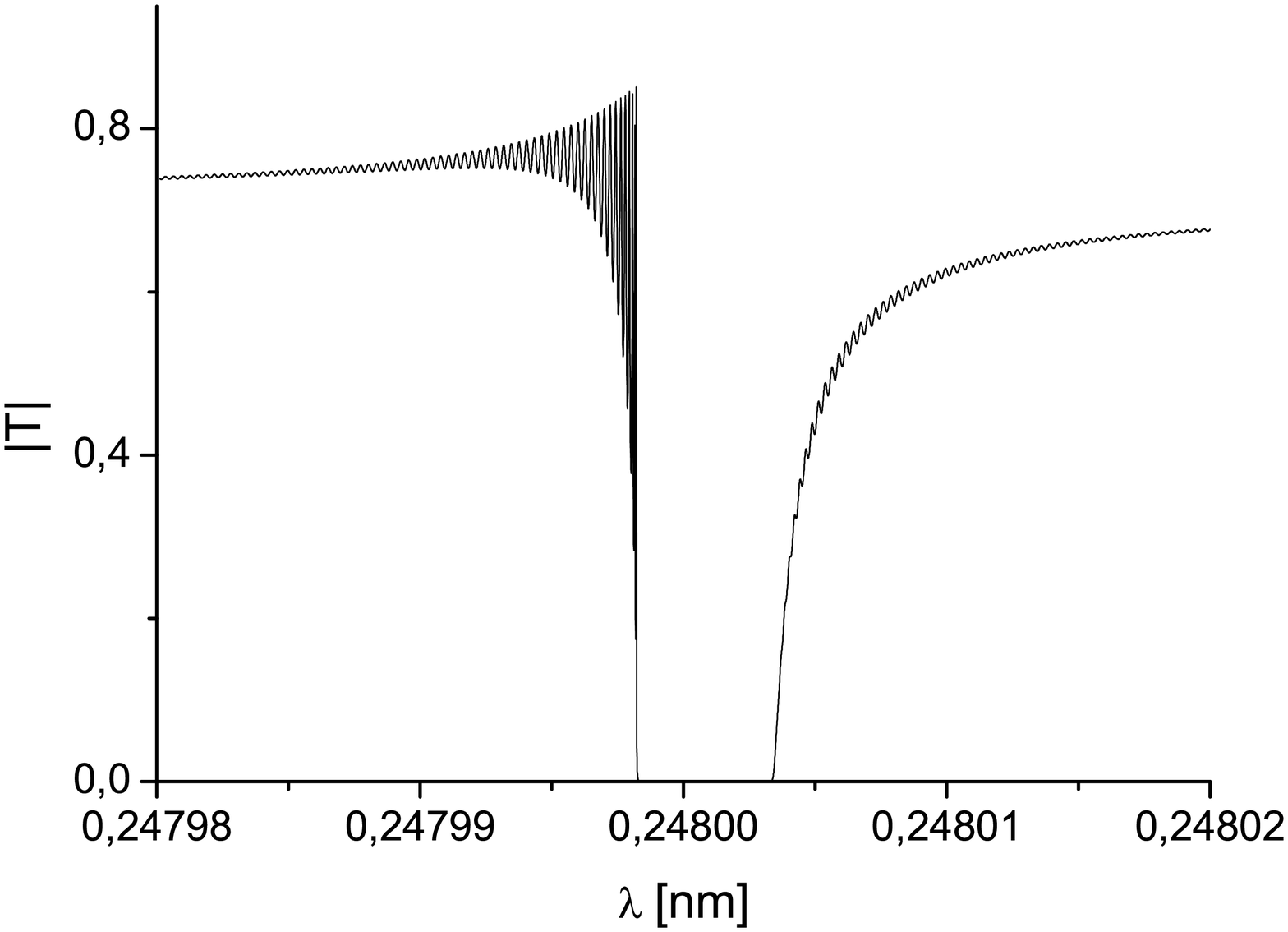}
\includegraphics[width=0.5\textwidth]{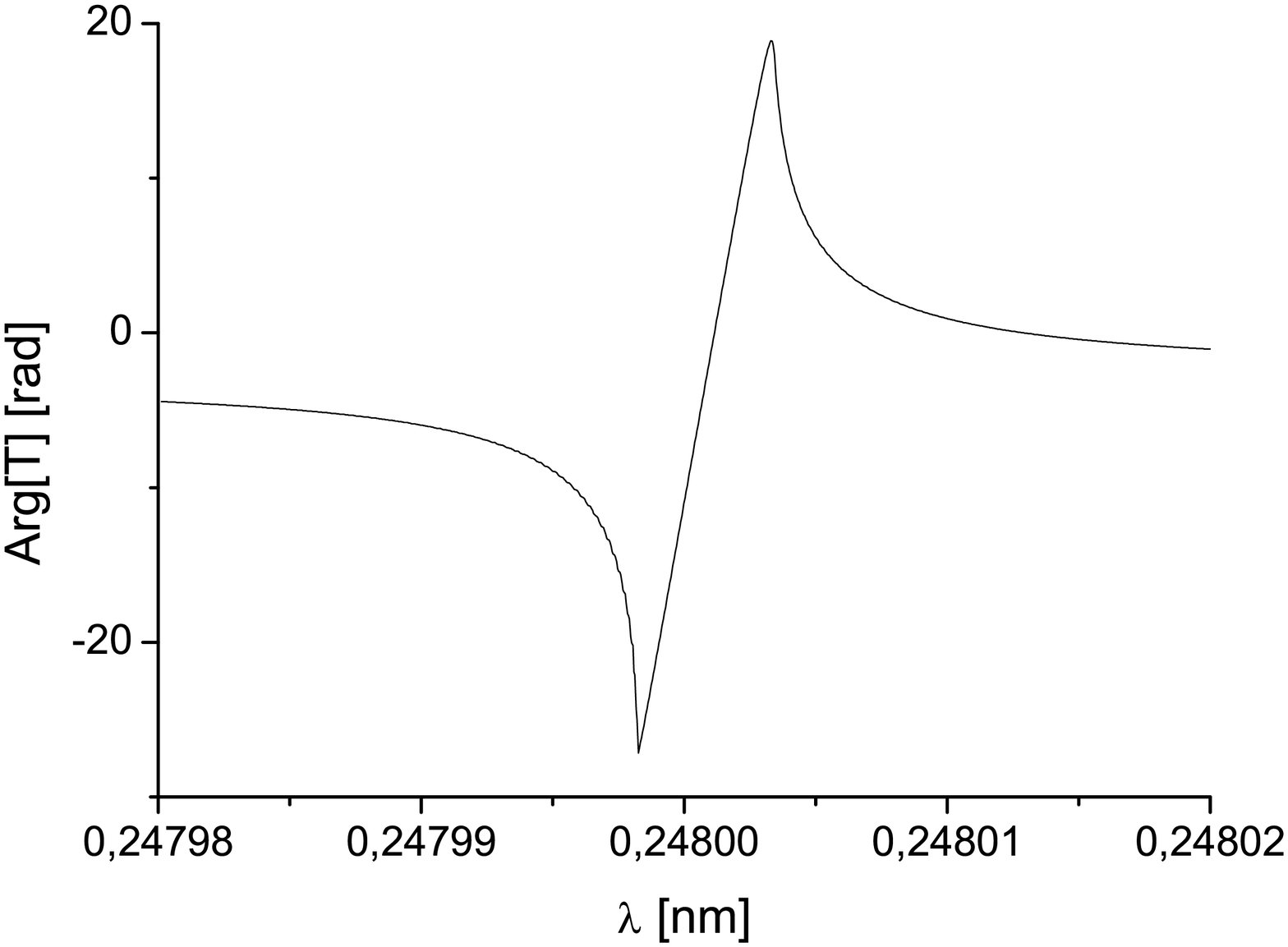}
\caption{Schematic of single crystal monochromator  for operation in
photon energy range between 5 keV and 7 keV. In this range the
C(220) reflection will be exploited. Modulus and phase of the
transmissivity are shown in the two lower plots.} \label{bioff3}
\end{figure}

\begin{figure}[tb]
\includegraphics[width=1.0\textwidth]{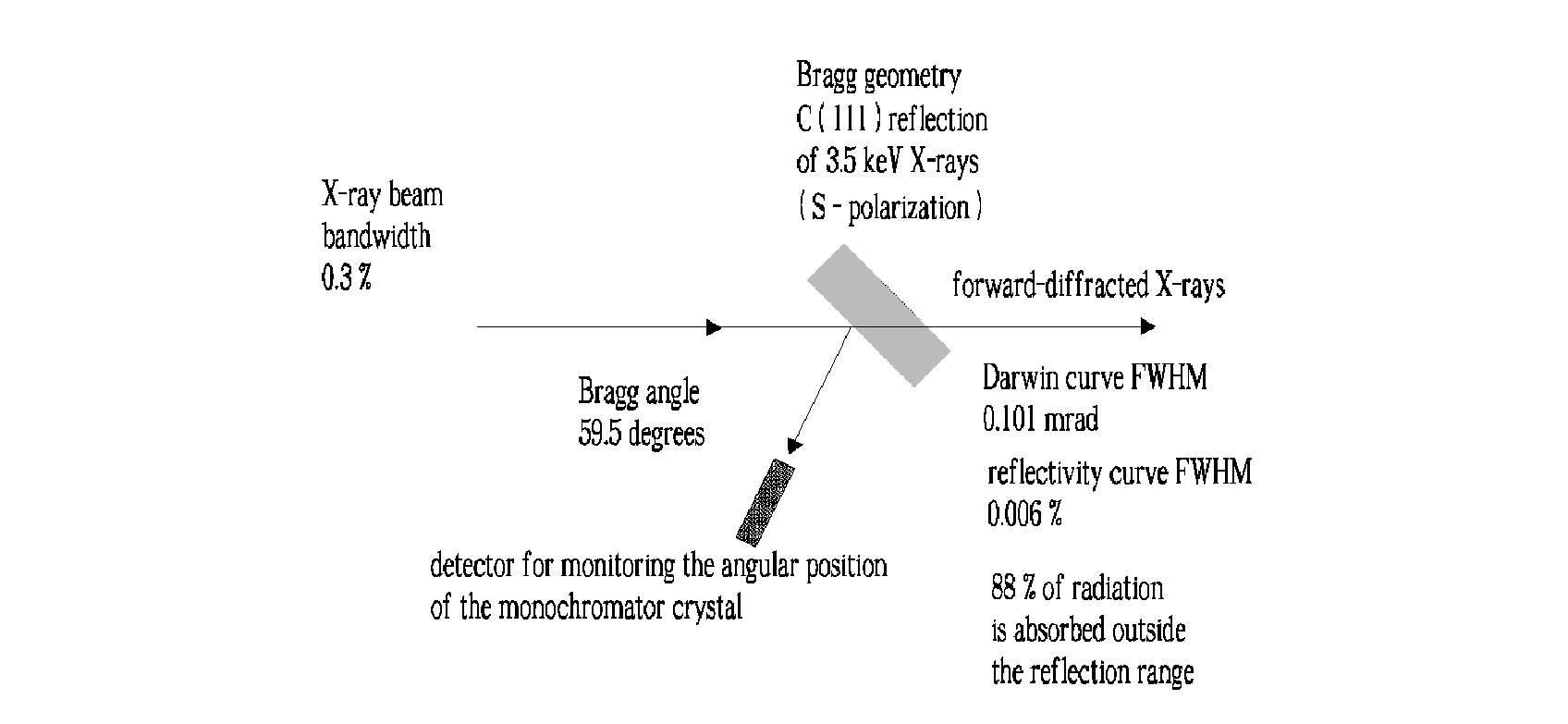}
\includegraphics[width=0.5\textwidth]{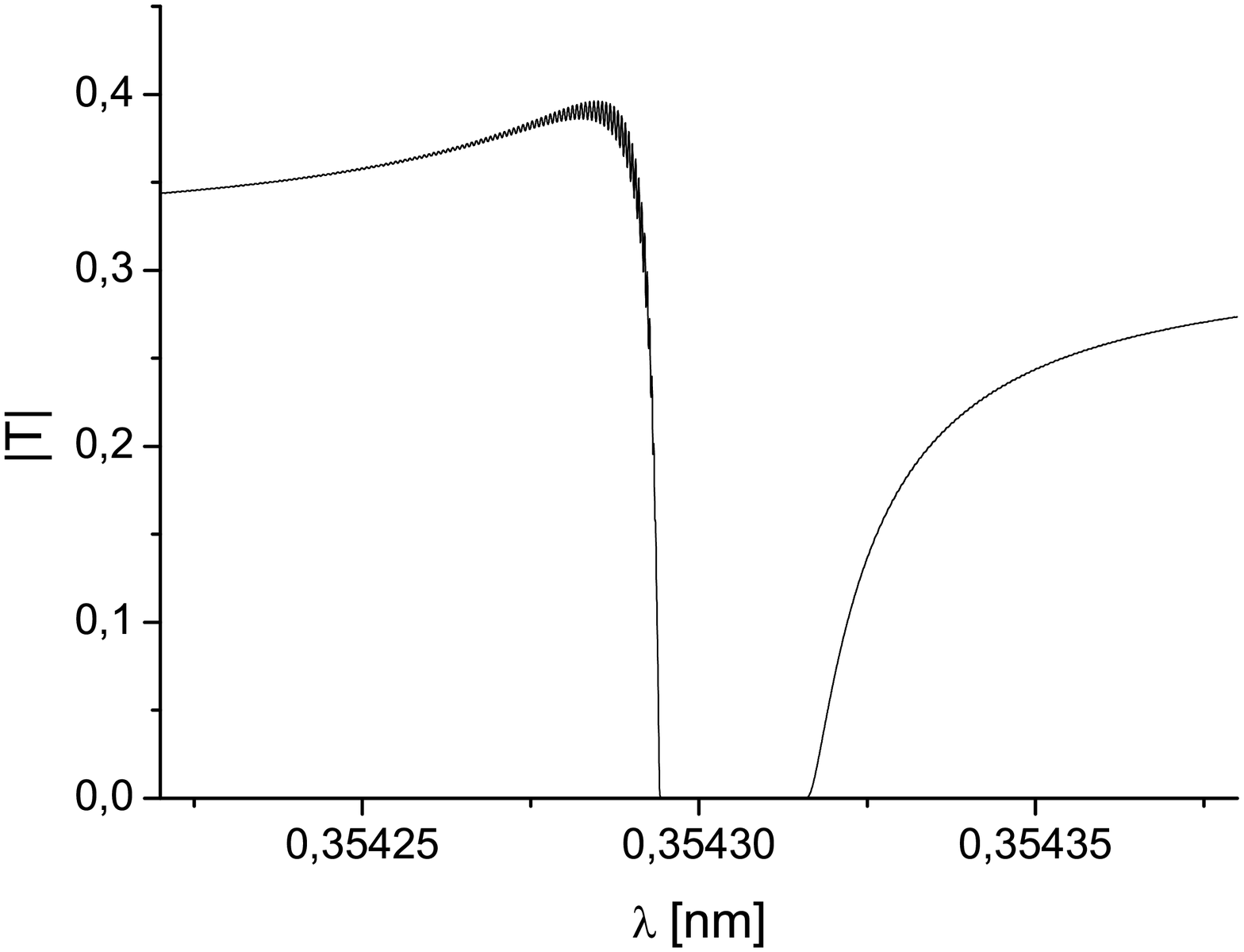}
\includegraphics[width=0.5\textwidth]{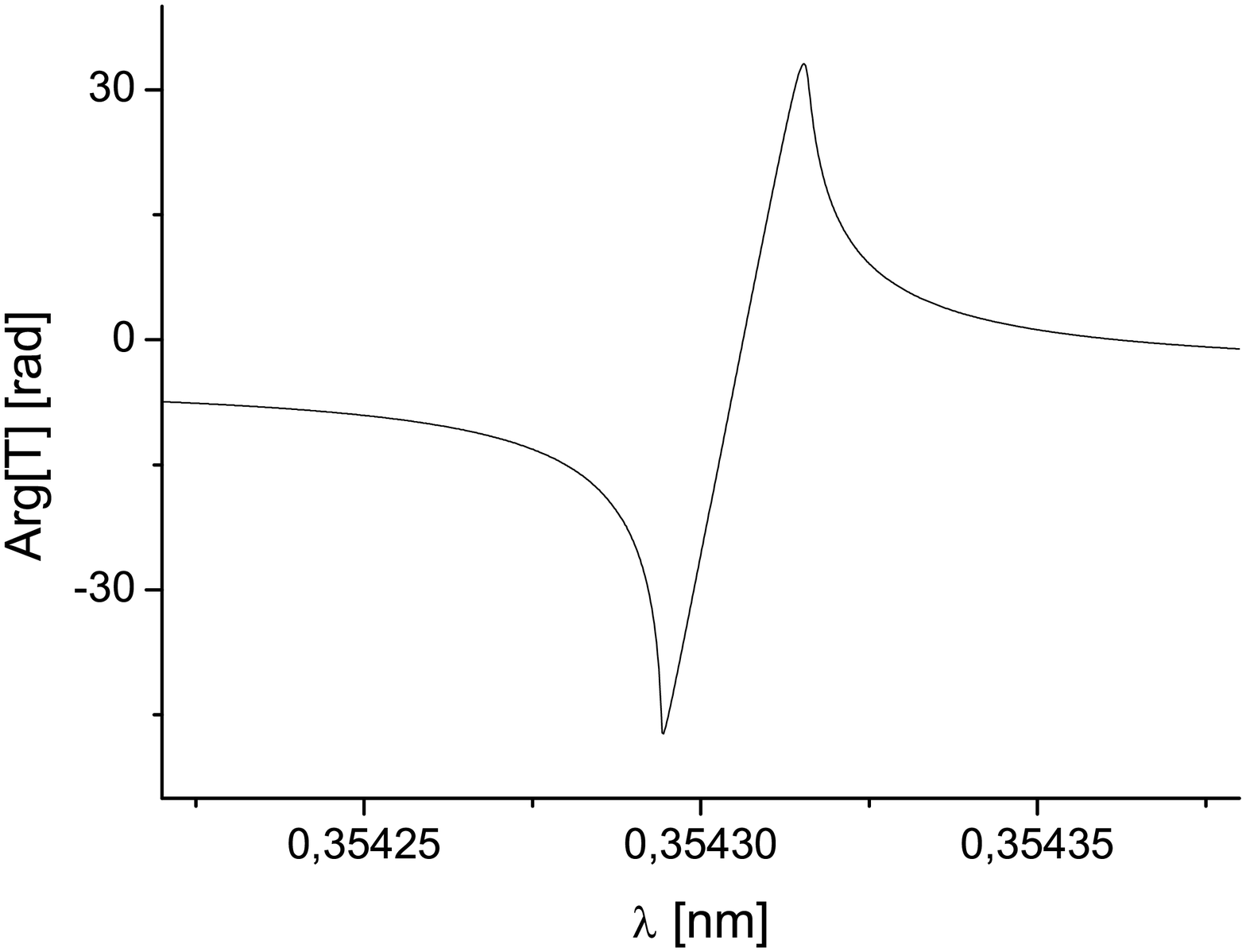}
\caption{Schematic of single crystal monochromator  for operation in
photon energy range between 3 keV and 5 keV. In this range the
C(111) reflection will be exploited. Modulus and phase of the
transmissivity are shown in the two lower plots.} \label{bioff2}
\end{figure}

\begin{figure}[tb]
\includegraphics[width=1.0\textwidth]{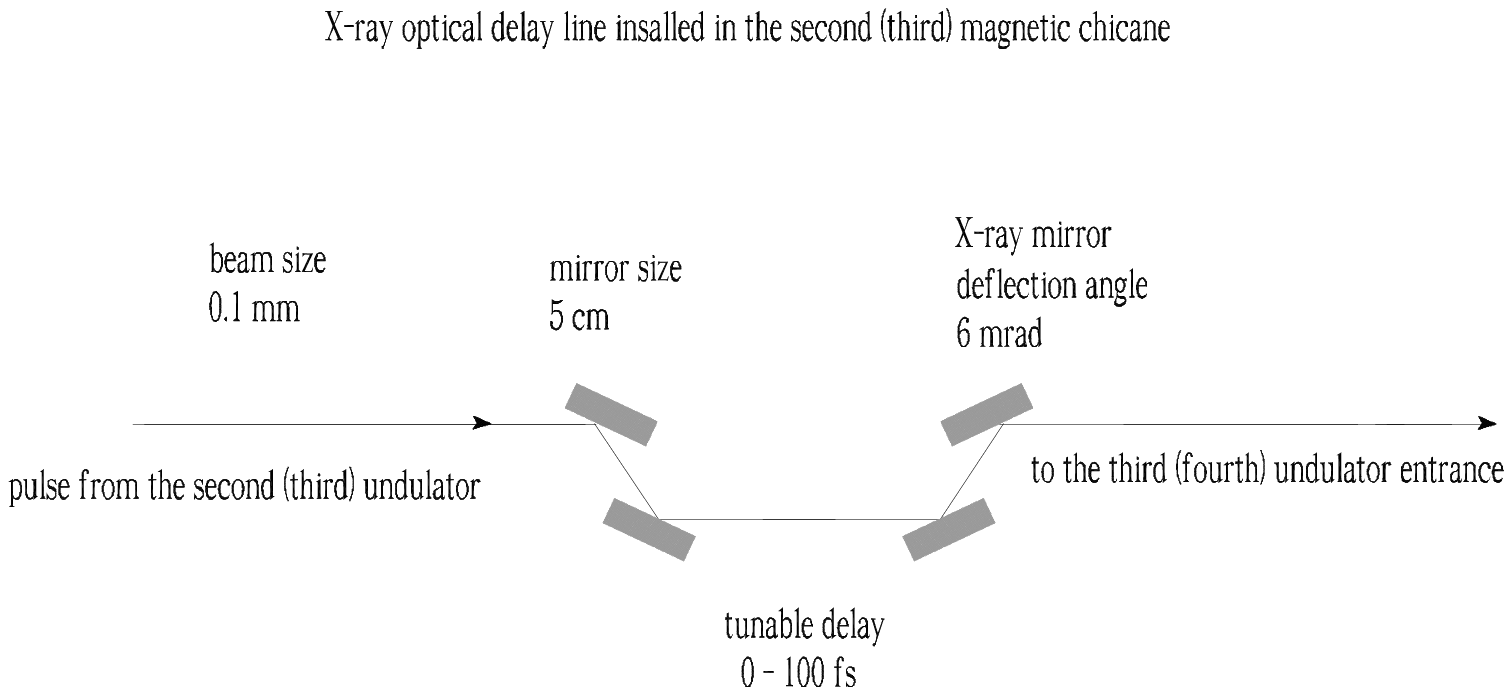}
\caption{X-ray optical system for delaying the soft X-ray pulse with
respect to the electron bunch. Two distinct X-ray optical systems
can be installed within the second and the third magnetic chicane. }
\label{biof2}
\end{figure}

For hard X-ray self-seeding, a monochromator usually consists of
crystals in the Bragg geometry. A conventional 4-crystal, fixed exit
monochromator introduces optical delay of, at least, a few
millimeters, which has to be compensated with the introduction of an
electron bypass longer than one undulator module. To avoid this
difficulty, a simpler self-seeding scheme was proposed in
\cite{OURY5b}, which uses the transmitted X-ray beam from the single
crystal to seed the same electron bunch. Here we propose to use a
diamond crystal with a thickness of $0.1$ mm. Using the symmetric
C(400) Bragg reflection, it will be possible to cover the photon
energy range from $7$ keV to $9$ keV, Fig. \ref{biof1}. The range
between $5$ keV and $7$ keV can be covered with the C(220)
reflection, Fig. \ref{bioff3}, while the range between $3$ keV and
$5$ keV can be obtained using the C(111) reflection, Fig.
\ref{bioff2}.

One of the main technical problems for self-seeding designers is to
provide bio-imaging capabilities in $2$ keV - $3$ keV photon energy
range. Here we will use the same method already exploited in
\cite{OURCC} to get around this obstacle. Our solution is based in
essence on the fresh bunch technique \cite{BZVI} and exploits the
above described conservative design of self-seeding setup based on a
grating monochromator.  The hardware requirement is minimal, and in
order to implement a fresh bunch technique it is sufficient to
install an additional magnetic chicane at a special position behind
the soft X-ray self-seeding setup. The function of this second
chicane is both to smear out the electron bunch microbunching, and
to delay the electron bunch with respect to the monochromatic soft
X-ray pulse produced in the second undulator. In this way, only half
of the electron bunch is seeded, and saturates in the third
undulator. Finally, the second half of the electron bunch, which
remains unspoiled, is seeded by the third harmonic of the
monochromatic radiation pulse generated in the third undulator,
which is also monochromatic. The final delay of the electron bunch
with respect to the seed radiation pulse can be obtained with a
third, hard X-ray self-seeding magnetic chicane, which in this mode
of operation is simply used to provide magnetic delay. The
monochromatic third harmonic radiation pulse used as seed for the
unperturbed part of the electron bunch is in the GW power level, and
the combination of self-seeding and fresh bunch technique is
extremely insensitive to non-ideal effects. The final undulator,
composed by $29$ cells, is tuned to the third harmonic frequency,
and is simply used to amplify the X-ray pulse up to the TW power
level.

In order to introduce a tunable delay of the photon beam with
respect to the electron beam,  a mirror chicane can be installed
within the second magnetic chicane, as shown in Fig. \ref{biof2}.
The function of the mirror chicane is to delay the radiation in the
range between $0.7$ keV and $1$ keV relatively to the electron
bunch. The glancing angle of the mirrors is as small as $3$ mrad. At
the undulator location, the transverse size of the photon beam is
smaller than $0.1$ mm, meaning that the mirror length would be just
about $5$ cm. The single-shot mode of operation will relax the
heat-loading issues. The mirror chicane can be built in such a way
to obtain a delay of the radiation pulse of about $23~\mu$m. This is
enough to compensate a bunch delay of about $20~\mu$m from the
magnetic chicane, and to provide any desired shift in the range
between $0~\mu$m and $3~\mu$m. Note that for the European XFEL
parameters, $1$ nm microbunching is washed out with a weak
dispersive strength corresponding to an $R_{56}$ in the order of ten
microns. The dispersive strength of the proposed magnetic chicane is
more than sufficient to this purpose. Thus, the combination of
magnetic chicane and mirror chicane removes the electron
microbunching produced in the second undulator and acts as a tunable
delay line within $0~\mu$m and $3~\mu$m with the required choice of
delay sign.

\subsection{\label{sub:verysoft} Operation into the water window}

\begin{figure}[tb]
\includegraphics[width=1.0\textwidth]{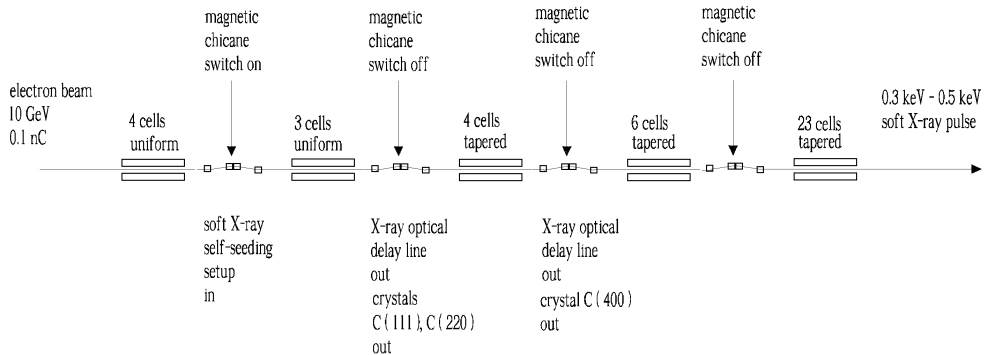}
\caption{Design of the undulator system for high power mode of
operation in the water window. The method exploits a combination of
self-seeding scheme with grating monochromator and an undulator
tapering technique. } \label{bio3f2}
\end{figure}

The five-undulator configuration in Fig. \ref{bio3f1} can be
naturally taken advantage of at different photon energies ranging
from soft to hard X-rays. Fig. \ref{bio3f2} shows the basic setup
for the high-power mode of operation in the soft X-ray wavelength
range. The second, the third and the fourth chicane are not used for
such regime, and must be switched off. After the first undulator (4
cells-long) and the grating monochromator, the output undulator
follows. The first section of the output undulator (consisting of
second and third undulator) is composed by $3$ untapered cells,
while tapering is implemented starting from the second cell of the
fourth undulator. The monochromatic seed is exponentially amplified
by passing through the first untapered section of the output
undulator. This section is long enough to allow for saturation, and
yields an output power of about $100$ GW. Such monochromatic FEL
output is finally enhanced up to $1$ TW in the second
output-undulator section, by tapering the undulator parameter over
the last cells after saturation. Under the constraints imposed by
undulator and chicane parameters it is only possible to operate at
the nominal electron beam energy of $10.5$ GeV. The setup was
optimized based on results of start-to-end simulations for a nominal
electron beam with 0.1 nC charge. Results were presented in
\cite{OSOF}, where we studied the performance of this scheme for the
SASE3 upgrade.

\subsection{Operation around the sulfur K-edge}

\begin{figure}[tb]
\includegraphics[width=1.0\textwidth]{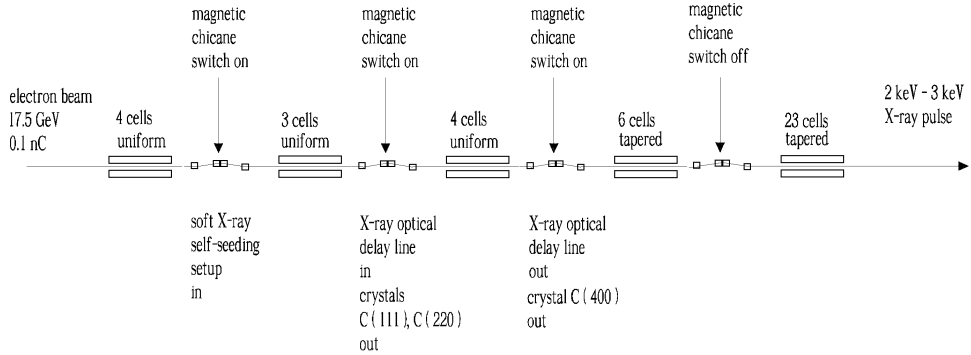}
\caption{Design of the undulator system for high power mode of
operation around the sulfur K-edge. The method exploits a
combination of self-seeding scheme with grating monochromator, fresh
bunch and undulator tapering techniques. } \label{bio3f3}
\end{figure}

\begin{figure}[tb]
\includegraphics[width=1.0\textwidth]{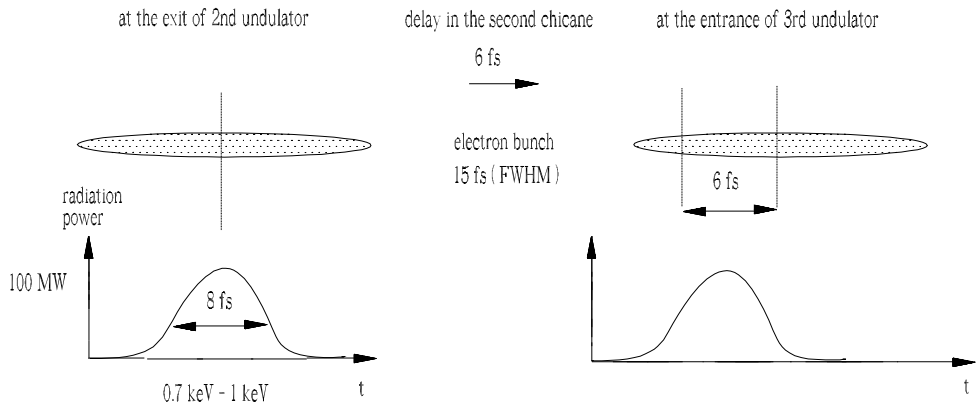}
\caption{Principle of the fresh bunch technique for the high power
mode of operation in the photon energy range between $2$ keV and $3$
keV. The second chicane smears out the electron microbunching and
delays the monochromatic soft X-ray pulse with respect to the
electron bunch of $6$ fs. In this way, half of of the electron bunch
is seeded and saturates in the third undulator.} \label{biof9}
\end{figure}
%
\begin{figure}[tb]
\includegraphics[width=1.0\textwidth]{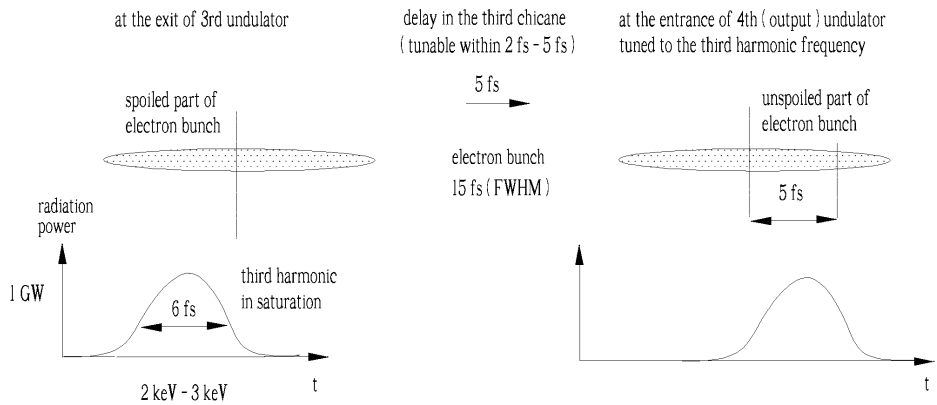}
\caption{Principle of the fresh bunch technique for the high power
mode of operation in the photon energy range between $2$ keV and $3$
keV. The third magnetic chicane smears out the electron
microbunching and delays the electron bunch with respect to the
radiation pulse. The unspoiled part of electron bunch is seeded by a
GW level monochromatic pulse at third harmonic frequency. Tunability
of the output pulse duration can be easily obtained by tuning the
magnetic delay of the third chicane.} \label{biof10}
\end{figure}
%
%
Fig. \ref{bio3f3} shows the basic setup for high power mode of
operation in the photon energy range between 2 keV and 3 keV. The
first three chicanes are used for such regime, and must be switched
on, while the last fourth chicane is off. The third chicane is used
as a magnetic delay only, and the crystal must be removed from the
light path. We propose to perform monochromatization at photon
energies ranging between $0.7$ keV and $1$ keV  with the help of a
grating monochromator, and to amplify the seed in the second
undulator up to the power level of $0.2$ GW. The second chicane
smears out the electron microbunching and delays the monochromatic
soft X-ray pulse of $2~\mu$m with respect to the electron bunch. In
this way, half of the electron bunch is seeded and saturates in the
third undulator up to $40$ GW. At saturation, the electron beam
generates considerable monochromatic radiation at the third harmonic
in the GW power level. The third magnetic chicane smears out the
electron microbunching and delays the electron bunch with respect to
the radiation of $2~\mu$m. Thus, the unspoiled part of the electron
bunch is seeded by the GW-level monochromatic pulse at the third
harmonic frequency, Fig. \ref{biof10}. The fourth, 29 cells-long
undulator is tuned to the third harmonic frequency (between $2$ keV
and $3$ keV), and is used to amplify the radiation pulse up to $1$
TW.  The additional advantage of the proposed setup for bio-imaging
is the tunability of the output pulse duration, which is obtained by
tuning the magnetic delay of the third chicane. Simulations show
that the X-ray pulse duration can be tuned from $2$ fs to $5$ fs.
The production of such pulses is of great importance when it comes
to single biomolecule imaging experiments.

The soft X-ray background can be easily eliminated by using a
spatial window positioned downstream of the fourth undulator exit
\cite{OURCC}. Since the soft X-ray radiation has an angular
divergence of about $0.02$ mrad FWHM, and the slits are positioned
more than $100$ m downstream of the third undulator, the background
has much larger spot size compared with the $2$ keV - $3$ keV
radiation spot size, which is about 0.1 mm at the exit of the fourth
undulator. Therefore, the background radiation power can be
diminished of more than two orders of magnitude without any
perturbations of the main pulse.

With the monochromator design in \cite{FENG3}, it will be possible
to operate at an electron beam energy of $17.5$ GeV. The setup was
optimized based on results from start-to-end simulations for a
nominal electron bunch with a charge of $0.1$ nC. Results are
presented in the following Sections of this article. The proposed
undulator setup uses the electron beam coming from the SASE1
undulator. We assume that SASE1 operates at the photon energy of
$12$ keV, and that the FEL process is switched off for one single
dedicated electron bunch within each macropulse train. A method to
control the FEL amplification process is based on the betatron
switcher technique described in \cite{SWIT1,SWIT2}. Due to quantum
energy fluctuations in the SASE1 undulator, and to wakefields in the
SASE1 undulator pipe, the energy spread and the energy chirp of the
electron bunch at the entrance of the bio-imaging beamline
significantly increase compared with the same parameters at the
entrance of the SASE1 undulator. The dispersion strength of the
first chicane has been taken into account from the viewpoint of the
electron beam dynamics, because it disturbs the electron beam
distribution. The other two chicanes have tenfold smaller dispersion
strength compared with the first one. The electron beam was tracked
through the first chicane using the code Elegant \cite{ELEG}. The
electron beam distortions complicate the simulation procedure.
However, simulations show that the proposed setup is not
significantly affected by perturbations of the electron phase space
distribution, and yields about the same performance as in the case
for an electron beam without the tracking through the first chicane
(see below).

\subsection{Operation in the $3$ keV - $7$ keV photon energy range}

\begin{figure}[tb]
\includegraphics[width=1.0\textwidth]{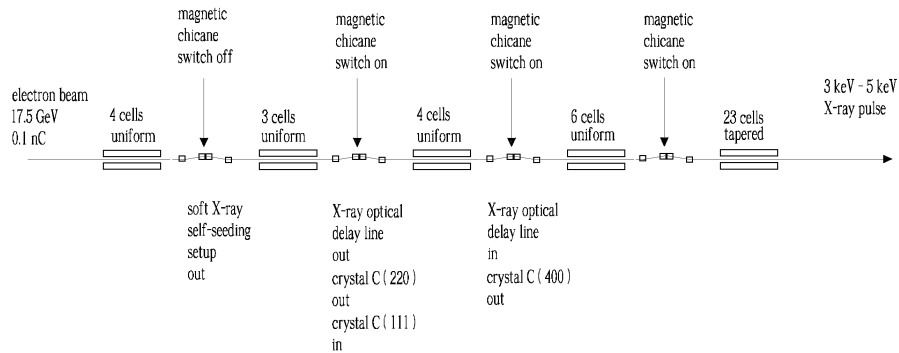}
\caption{Design of the undulator system for high power mode of
operation in the most preferable photon energy range for single
molecule imaging, between 3 keV and 5 keV. The method exploits a
combination of the self-seeding scheme with single crystal
monochromator, fresh bunch and undulator tapering techniques. }
\label{bio3f4}
\end{figure}

\begin{figure}[tb]
\includegraphics[width=1.0\textwidth]{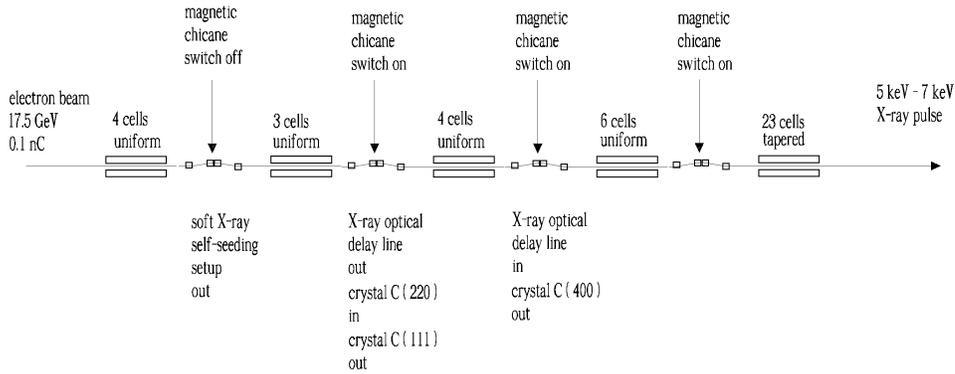}
\caption{Design of the undulator system for high power mode of
operation in the photon energy range between 5 keV and 7 keV. The
method exploits a combination of self-seeding scheme with single
crystal monochromator, fresh bunch and undulator tapering
techniques. } \label{bio3f5}
\end{figure}

\begin{figure}[tb]
\includegraphics[width=1.0\textwidth]{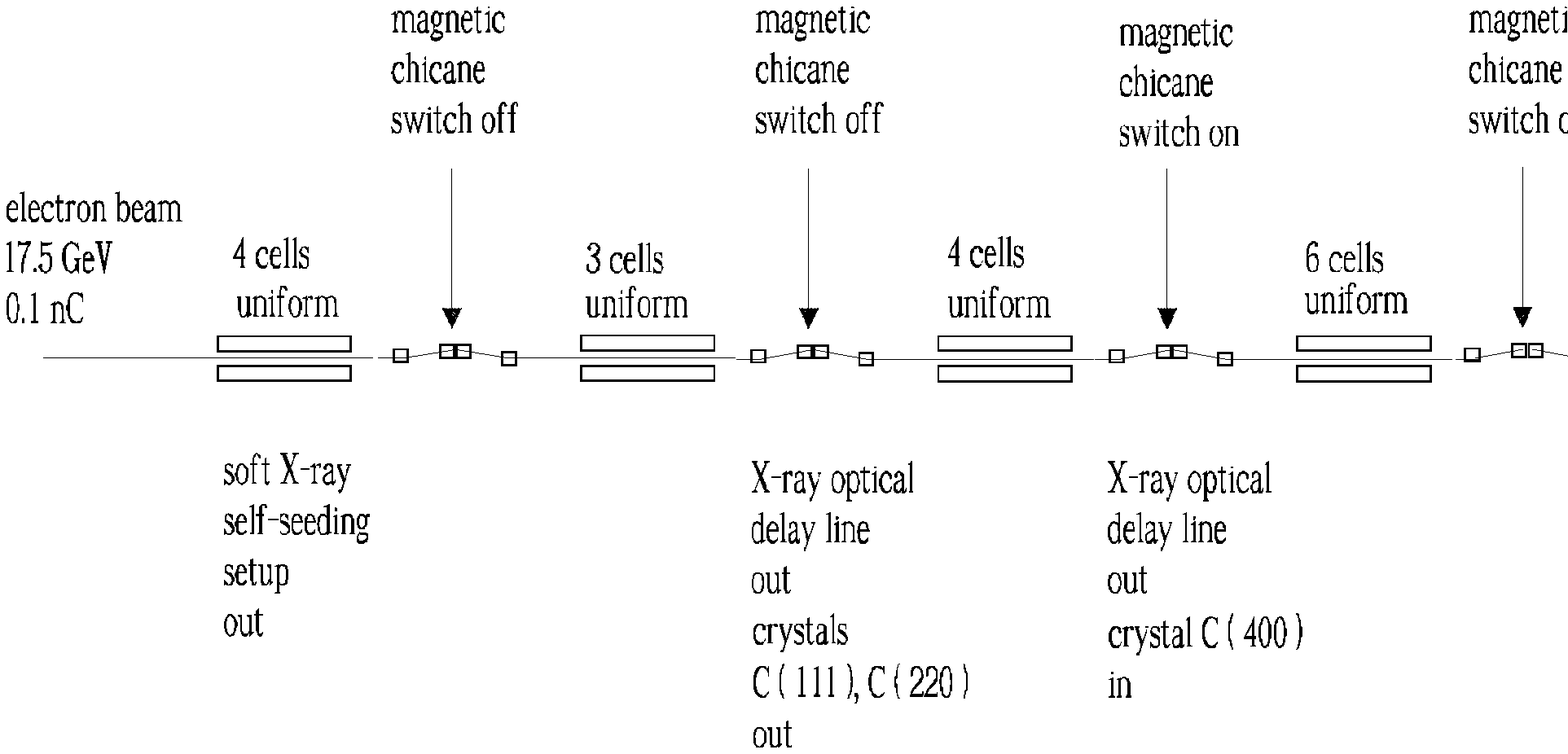}
\caption{Design of the undulator system for high power mode of
operation in the photon energy range in the photon energy range
between 7 keV and 9 keV. The method exploits a combination of
self-seeding scheme with single crystal monochromator and an
undulator tapering technique. } \label{bio3f6}
\end{figure}
Starting with the energy range of $3$ keV it is possible to use a
single crystal monochromator instead of a grating monochromator at
an electron energy of $17.5$ GeV. Different crystal reflections and
different positions of the monochromator down the undulator enable
self-seeding for different spectral ranges.

For the range between 3 keV and 5 keV, Fig. \ref{bio3f4}, the first
chicane is not used and is switched off. After the first 7 cells the
electron and the photon beams are separated with the help of the
second magnetic chicane, and the C(111) reflection is used to
monochromatize the radiation. The seed is amplified in the next 4
cells. After that, the electron and the photon beam are separated
again by the third chicane, and an X-ray optical delay line allows
for the introduction of a tunable delay of the photon beam with
respect to the electron beam. The following 6 cells use only a part
of the electron beam as a lasing medium. A magnetic chicane follows,
which shifts the unspoiled part of the electron bunch on top of the
of the photon beam. In this way, a fresh bunch technique can be
implemented. Since the delays are tunable, the photon pulse length
can also be tuned. Finally, radiation is amplified into the last
$23$ tapered cells to provide pulses with about 2 TW power. The
photon energy range between 5 keV and 7 keV can be achieved
similarly, Fig. \ref{bio3f5}. The only difference is that now the
C(220) reflection is used, instead of the C(111).

It may be worth to point out the difference between the operation in
the $3$ keV - 7 keV range and the previously discussed range between
2 keV and 3 keV. In the 3 keV - 7 keV range we use seeding in
combination of a fresh bunch technique, but we do not exploit
harmonic generation. Moreover, the fresh bunch technique is only
used for tuning the duration of the radiation pulse.

\subsection{Operation in the $7$ keV - $9$ keV photon energy range}

The energy range between 7 keV and 9 keV can be achieved by
deactivating the first and the second magnetic chicane, thus letting
the SASE process building up the radiation pulse to be
monochromatized for 11 cells. After that, the third chicane is used
for the monochromator setup, which makes use of the C(400)
reflection. The last chicane is switched off, and the the output
undulator is long enough to reach $1$ TW power. The duration of the
output pulses is of about $10$ fs. If tunability of the pulse
duration is requested in this energy range, this is most easily
achieved by providing additional delay with the fourth magnetic
chicane installed behind the hard X-ray self-seeding setup.

\subsection{Operation around the selenium K-edge}

Finally, for the energy range between 9 keV and 13 keV, a
combination of self-seeding, fresh bunch tachnique and harmonic
generation is used. The undulator line is configured as for the
range between 3 keV and 5 keV, Fig. \ref{bio3f4}, the only
difference being that the final undulator segments are tuned at the
third harmonic of the fundamental thus enabling the 9 keV - 13 keV
energy range. As before, the first chicane is not used and is
switched off. After the first 7 cells the electron and the photon
beams are separated with the help of the second magnetic chicane,
and the C(111) reflection is used to monochromatize the radiation.
The seed is amplified in the next 4 cells. After that, the electron
and the photon beam are separated again by the third chicane, and an
X-ray optical delay line allows for the introduction of a tunable
delay of the photon beam with respect to the electron beam. The
second chicane smears out the electron microbunching and delays the
monochromatic soft X-ray pulse with respect to the electron bunch of
$6$ fs. In this way, half of of the electron bunch is seeded and
saturates in the following 6 cells. A magnetic chicane follows,
which shifts the unspoiled part of the electron bunch on top of the
of the photon beam. In this way, a fresh bunch technique can be
implemented. Since the delays are tunable, the photon pulse length
can also be tuned. Since third harmonic bunching is considerable,
the last $23$ tapered cells are tuned at the third harmonic of the
fundamental providing pulses with about 0.5 TW power.

\subsection{\label{subloc} Possible location of the bio-imaging
line}

\begin{figure}[tb]
\includegraphics[width=1.0\textwidth]{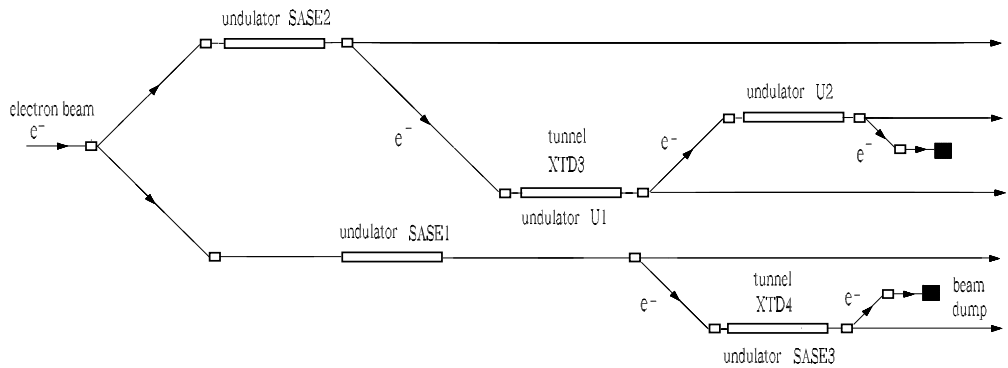}
\caption{Original design of the European XFEL facility.}
\label{biof12}
\end{figure}

\begin{figure}[tb]
\includegraphics[width=1.0\textwidth]{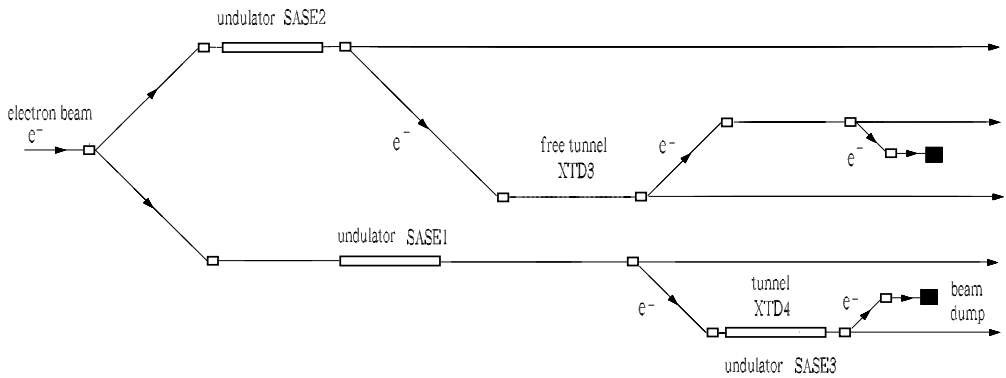}
\caption{Current design of the European XFEL facility.}
\label{biof13}
\end{figure}
%


\begin{figure}[tb]
\includegraphics[width=1.0\textwidth]{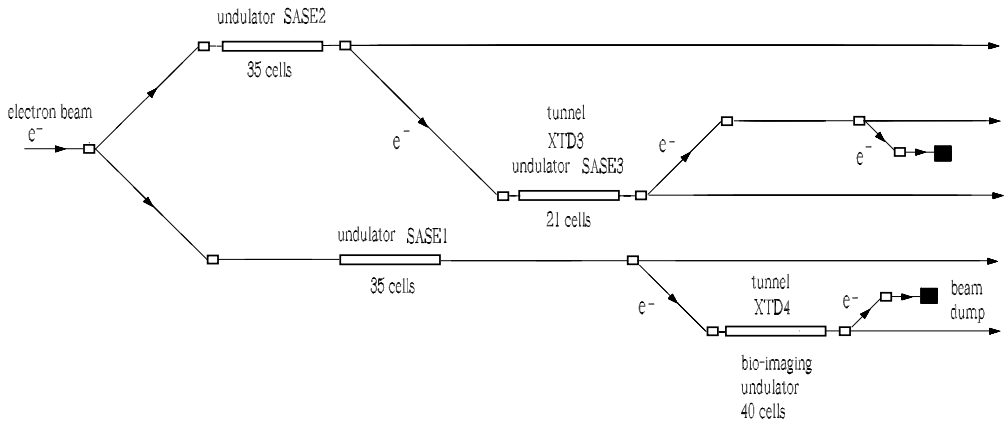}
\caption{Schematic of the proposed extension of the European XFEL
facility.} \label{biof15}
\end{figure}

The original design of the European XFEL \cite{tdr-2006} was
optimized to produce XFEL radiation at $0.1$ nm, simultaneously at
two undulator lines, SASE1 and SASE2. Additionally, the design
included one FEL line in the soft X-ray range, SASE3, and two
undulator lines for spontaneous synchrotron radiation, U1 and U2,
Fig. \ref{biof12}. The soft X-ray SASE3 beamline used the spent
electron beam from SASE1, and the U1 and U2 beamlines used the spent
beam from SASE2. In fact, although the electron beam performance is
degraded by the FEL process, the beam can still be used in
afterburner mode in the SASE3 undulator, which will be equipped with
a $126$ m-long undulator system, for a total of $21$ cells.

After a first design report, the layout of the European XFEL
changed. In the last years after the achievement of the LCLS, and
the subsequent growth of interest in XFEL radiation by the
scientific community, it became clear that the experiments with XFEL
radiation, rather than with spontaneous synchrotron radiation, had
to be prioritized. In the new design, the two beamlines behind SASE2
are now free for future XFEL undulators installations, Fig.
\ref{biof13}.

Recently it was also realized that the amplification process in the
XFEL undulators can be effectively controlled by betatron FEL
switchers \cite{SWIT1,SWIT2}. The SASE3 undulator was then optimized
for generating soft X-rays. However, due to the possibility of
switching the FEL process in SASE1, it is possible to produce high
power SASE3 radiation in a very wide photon energy range between
$0.3$ keV and $13$ keV.  The SASE3 beamline is now expected to
provide excellent performance, and to take advantage of its location
in the XTD4 tunnel, which is close to the experimental hall and has
sufficient free space behind the undulator for future expansion
($140$ m). After this section, the electron beam will be separated
from the photon beam and will be bent down to an electron beam dump,
Fig. \ref{biof12}.  In the photon energy range between $3$ keV and
$13$ keV, the SASE3 beamline is now expected to provide even better
conditions for users than SASE1 and SASE2.

In this article we propose to build the bio-imaging beamline in the
XTD4 tunnel. The SASE3 undulator, which is composed by 21 cells, can
be installed from the very beginning in the free XTD3 tunnel, which
is shorter than the XTD4 tunnel but sufficiently long for such
installation, Fig. \ref{biof15}. The undulator can be placed within
the straight beam path that is defined by the last upstream and
first downstream dipole of the electron deflection system. The
limiting length given by these constraints is referred to as
"available length". For the XTD3 and the XTD4 tunnels this available
length respectively amounts to 255 m and 460 m, see \cite{DECKP}.
However, the electron beam optics requires sufficient space in front
and the above-mentioned dipoles. The length that fits these electron
optics restrictions is referred to as "potential length" and can be
actually used for installations. For the XTD3 and the XTD4 tunnels,
this potential length respectively amounts to 215 m and 400 m
\cite{DECKP}. It should be mentioned that the potential length of
the XTD4 tunnel is practically the same as the main SASE1 and SASE2
tunnels, and nicely fits with the undulator system for a dedicated
bio-imaging beamline. It offers thus a great potential for future
upgrades of this new beamline.

The bio-imaging beamline would support experiments carried out over
a rather wide photon energy range. It is therefore proposed that the
photon beam transport of the new beamline includes two lines. Line A
uses $0.5$m-long mirrors operating at a grazing angle of $2$ mrad.
This line is dedicated to the transport of X-ray radiation in the
photon energy range from $3$ keV up to $13$ keV. This would be
complementary to the Line B that is now optimized in the soft X-ray
range between $0.3$ keV and $3$ keV. The distance from the
40-cells-long undulator exit to the first mirror system will be only
of about $100$m\footnote{This is in contrast with SASE1 and SASE2
beamlines, where an opening angle of $0.003$ mrad at $3$ keV FEL
radiation leads to unacceptable mirror length of 2 m due to long
distance of about 500 m between the source and mirror system. For
these beamlines there is no possibility to use identical
configuration of mirrors within the photon energy range from 3 keV
to 13 keV. }.

\section{\label{spatio} Spatiotemporal transformation caused by the use of a single crystal
monochromator}

The development of  self-seeding schemes in the hard X-ray
wavelength range necessarily involves crystal monochromators. Any
device like a crystal monochromator introduces spatiotemporal
deformations of the seeded X-ray pulse, which can be problematic for
seeding. The spatiotemporal coupling in the electric field relevant
to self-seeding schemes with crystal monochromators has been
analyzed in the frame of classical dynamical theory of X-ray
diffraction \cite{SHVID}. This analysis shows that a crystal in
Bragg reflection geometry transforms the incident electric field
$E(x,t)$ in the $\{x,t\}$ domain into $E(x- a t, t)$, that is the
field of a pulse with a less well-known distortion, first studied in
\cite{GABO}. The physical meaning of this distortion is that the
beam spot size is independent of time, but the beam central position
changes as the pulse evolves in time. Here we  show in a simple
manner that, based on the use only Bragg law, we may arrive directly
to explanation of spatiotemporal coupling phenomena in the dynamical
theory of X-ray diffraction \cite{OURTILT}.

Let us consider an electromagnetic plane wave in the X-ray frequency
range incident on an infinite, perfect crystal. Within the
kinematical approximation, according to the Bragg law, constructive
interference of waves scattered from the crystal occurs if the angle
of incidence, $\theta_i$ and the wavelength, $\lambda$, are related
by the well-known relation

\begin{eqnarray}
\lambda = 2 d \sin \theta_i ~. \label{bragg}
\end{eqnarray}
assuming reflection into  the first order. This equation shows that
for a given wavelength of the X-ray beam, diffraction is possible
only at certain angles determined by the interplanar spacings $d$.
It is important to remember the following geometrical relationships:

1. The angle between the incident X-ray beam and normal to the
reflection plane is equal to that between the normal and the
diffracted X-ray beam. In other words, Bragg reflection is a mirror
reflection, and the incident angle is equal to the diffracted angle
($\theta_i = \theta _D$).

2. The angle between the diffracted X-ray beam and the transmitted
one is always $2 \theta_i$. In other words, incident beam and
forward diffracted (i.e. transmitted) beam have the same direction.

We now turn our attention beyond the kinematical approximation to
the dynamical theory of X-ray diffraction by a crystal. An optical
element inserted into the X-ray beam is supposed to modify some
properties of the beam as its width, its divergence, or its spectral
bandwidth. It is useful to describe the modification of the beam by
means of a transfer function. The transmisivity curve - the
transmittance - in Bragg geometry can be expressed in the frame of
dynamical theory as

\begin{eqnarray}
T(\theta_i,\omega) = T(\Delta \omega +  \omega_B \Delta \theta \cot
\theta_B  ) ~, \label{reflectance}
\end{eqnarray}
where $\Delta \omega = (\omega - \omega_B)$ and $\Delta \theta =
(\theta_i - \theta_B)$ are the deviations of frequency and incident
angle of the incoming beam from the Bragg frequency and Bragg angle,
respectively. The frequency $\omega_B$ and the angle $\theta_B$ are
given by the Bragg law: $\omega_B \sin \theta_B = \pi c/d$. Here we
followed the usual procedure of expanding $\omega$ in a Taylor
series about $\omega_B$, so that

\begin{eqnarray}
\omega = \omega_B + (d \omega/d \theta)_B (\theta - \theta_B) + ...
~. \label{taylor}
\end{eqnarray}

Consider now a perfectly collimated, white beam incident on the
crystal. In kinematical approximation $T$ is a Dirac
$\delta$-function, which is simply represented by the differential
form of Bragg law:

\begin{eqnarray}
d \lambda/d \theta_i = \lambda \cot \theta_i  ~. \label{difform}
\end{eqnarray}
In contrast to this, in dynamical theory the reflectivity width is
finite. This means that there is a reflected beam even when incident
angle and wavelength of the incoming beam are not related exactly by
Bragg equation. It is interesting to note that the geometrical
relationships 1. and 2. are still valid in the framework of
dynamical theory. In particular, reflection in dynamical theory is
always a mirror reflection. We underline here that if we have a
perfectly collimated, white incident beam, we also have a perfectly
collimated reflected and transmitted beam. Its bandwidth is related
with the width of the reflectivity curve. We will regard the beam as
perfectly collimated when the angular spread of the beam is much
smaller than the angular width of the transfer function $T$. It
should be realized that the crystal does not introduce an angular
dispersion similar to a grating or a prism. However, a more detailed
analysis based on the expression for the reflectivity, Eq.
(\ref{reflectance}), shows that a less well-known spatiotemporal
coupling exists. The fact that the reflectivity is invariant under
angle and frequency transformations obeying

\begin{eqnarray}
\Delta \omega +  \omega_B \Delta \theta \cot \theta_B  =
\mathrm{const}~ \label{transform}
\end{eqnarray}
is evident, and corresponds to the coupling in the Fourier domain
$\{k_x, \omega\}$. The origin of this relation is kinematical, in
the sense that it follows from Bragg diffraction. One might be
surprised that the field transformation derived in \cite{SHVID} for
an XFEL pulse after a crystal in the $\{x,t\}$ domain is given by

\begin{eqnarray}
E_{\mathrm{out}} (x,t) =  FT[T(\Delta \omega,
k_x)E_{\mathrm{in}}(\Delta \omega, k_x)] = E(x- c t\cot \theta_B ,
t)~, \label{Eoutxt}
\end{eqnarray}
where $FT$ indicates a Fourier transform from the $\{k_x,\omega\}$
to the $\{x,t\}$ domain, and $k_x = \omega_B \Delta \theta/c$. In
general, one would indeed expect the transformation to be symmetric
in both the $\{k_x,\omega\}$ and in the $\{x,t\}$ domain due to the
symmetry of the transfer function\footnote{There is a breaking of
the symmetry in the diffracted beam in the $\{k_x,\omega\}$ domain.
While the symmetry is present at the level of the transfer function,
it is not present anymore when one considers the incident beam. We
point out that symmetry breaking in \cite{SHVID} is a result of the
approximation of temporal profile of the incident wave to a Dirac
$\delta$-function.}. However, it is reasonable to expect the
influence of a nonsymmetric input beam distribution. In the
self-seeding case, the incoming XFEL beam is well collimated,
meaning that its angular spread is  a few times smaller than angular
width of the transfer function. Only the bandwidth of the incoming
beam is much wider than the bandwidth of the transfer function. In
this limit, we can approximate the transfer function in the
expression for the inverse temporal Fourier transform as a Dirac
$\delta$-function. This gives

\begin{eqnarray}
&& E_\mathrm{out}(x,t) = FT[R(\Delta \omega, k_x) E_\mathrm{in}
(\Delta \omega, k_x)]\cr && = \xi(t) \cdot \frac{1}{2\pi}  \int d
k_x \exp(-i k_x c t \cot \theta_B) \exp(i k_x x)
E_\mathrm{in}(0,k_x)\cr && = \xi(t) b(x - c t\cot \theta_B) ~,
\label{eoutdel}
\end{eqnarray}
where we applied the Shift Theorem twice, and where

\begin{eqnarray}
\xi(t) = \frac{1}{2\pi} \int d Y \exp( i Y t) R(Y) \label{tempFT}
\end{eqnarray}
is the inverse temporal Fourier transform of the reflectivity curve.

The spatial shift given by Eq. (\ref{Eoutxt}) is proportional to
$\cot(\theta_B)$, and is maximal in the range for small $\theta_B$.
If we want to use long delay $c t  \sim 15 \mu$m, a rms transverse
size of radiation pulse of about $15 \mu$m limits the maximum
acceptable value of $\cot(\theta_b)$ to $1-1.5$. Thus, the
spatiotemporal coupling is an issue, and efforts are necessarily
required to avoid distortion. It is worth mentioning that this
distortion is easily suppressed by the right choice of crystal
planes.

\section{FEL studies}

\begin{figure}[tb]
\includegraphics[width=0.5\textwidth]{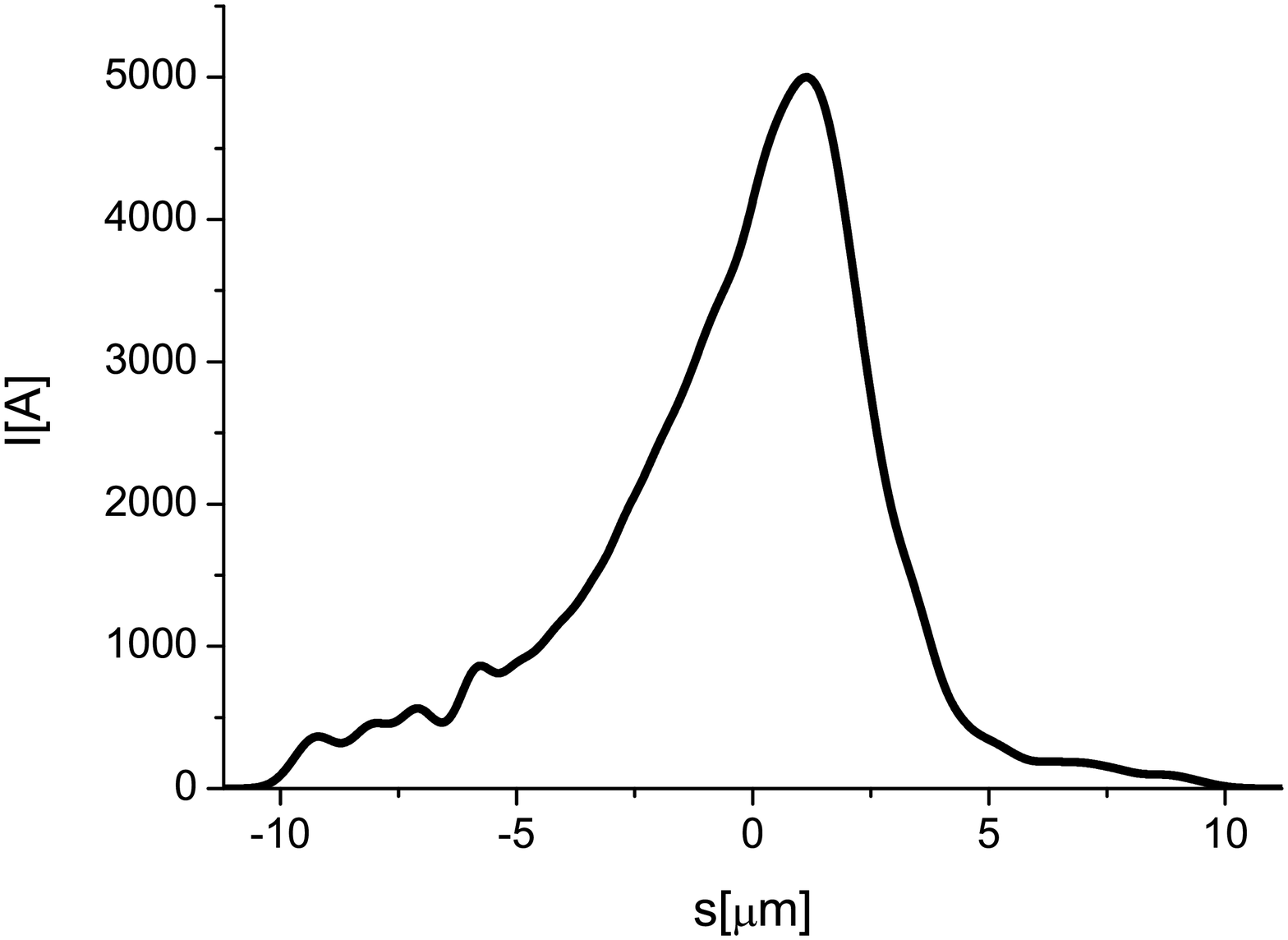}
\includegraphics[width=0.5\textwidth]{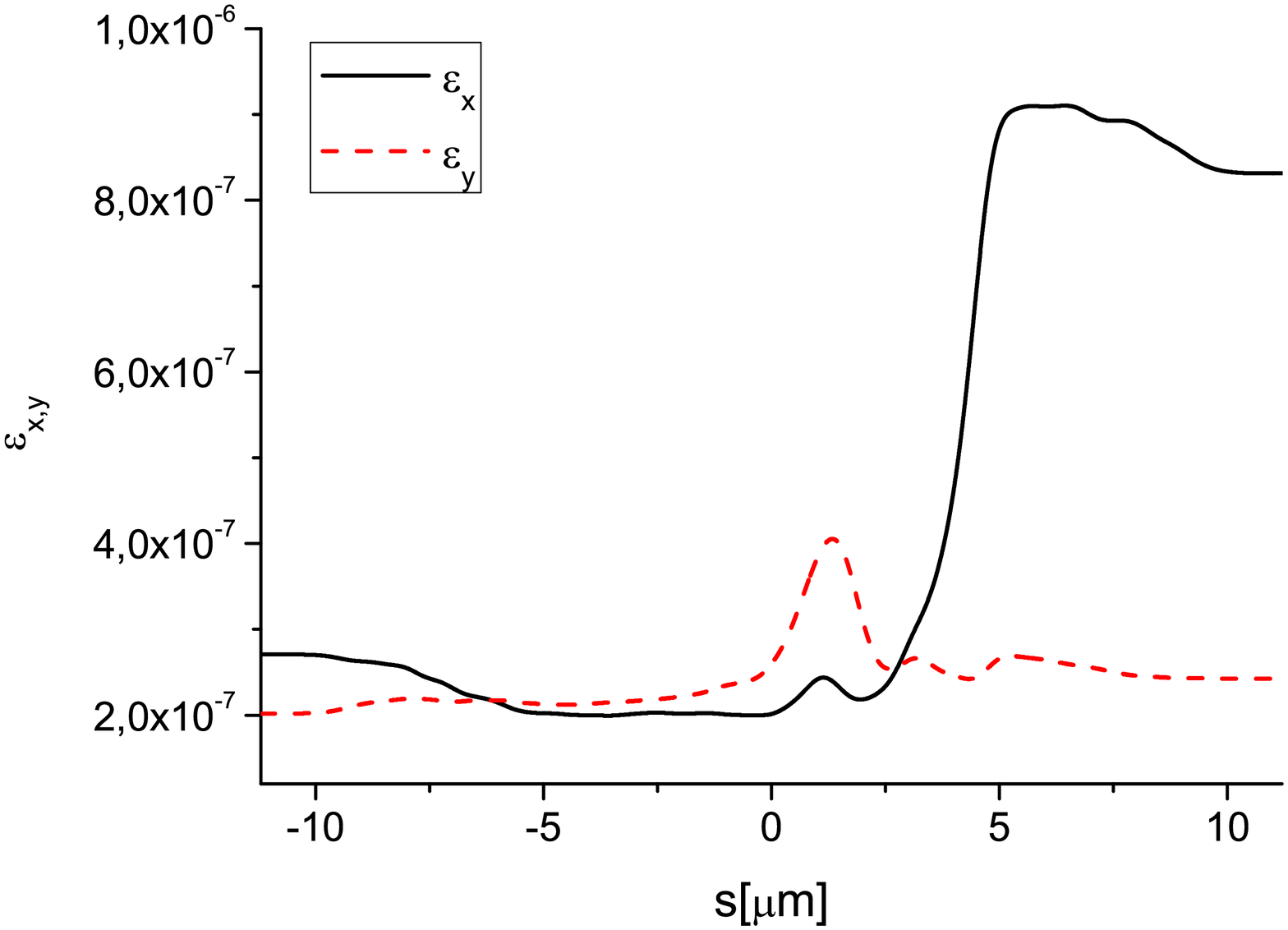}
\includegraphics[width=0.5\textwidth]{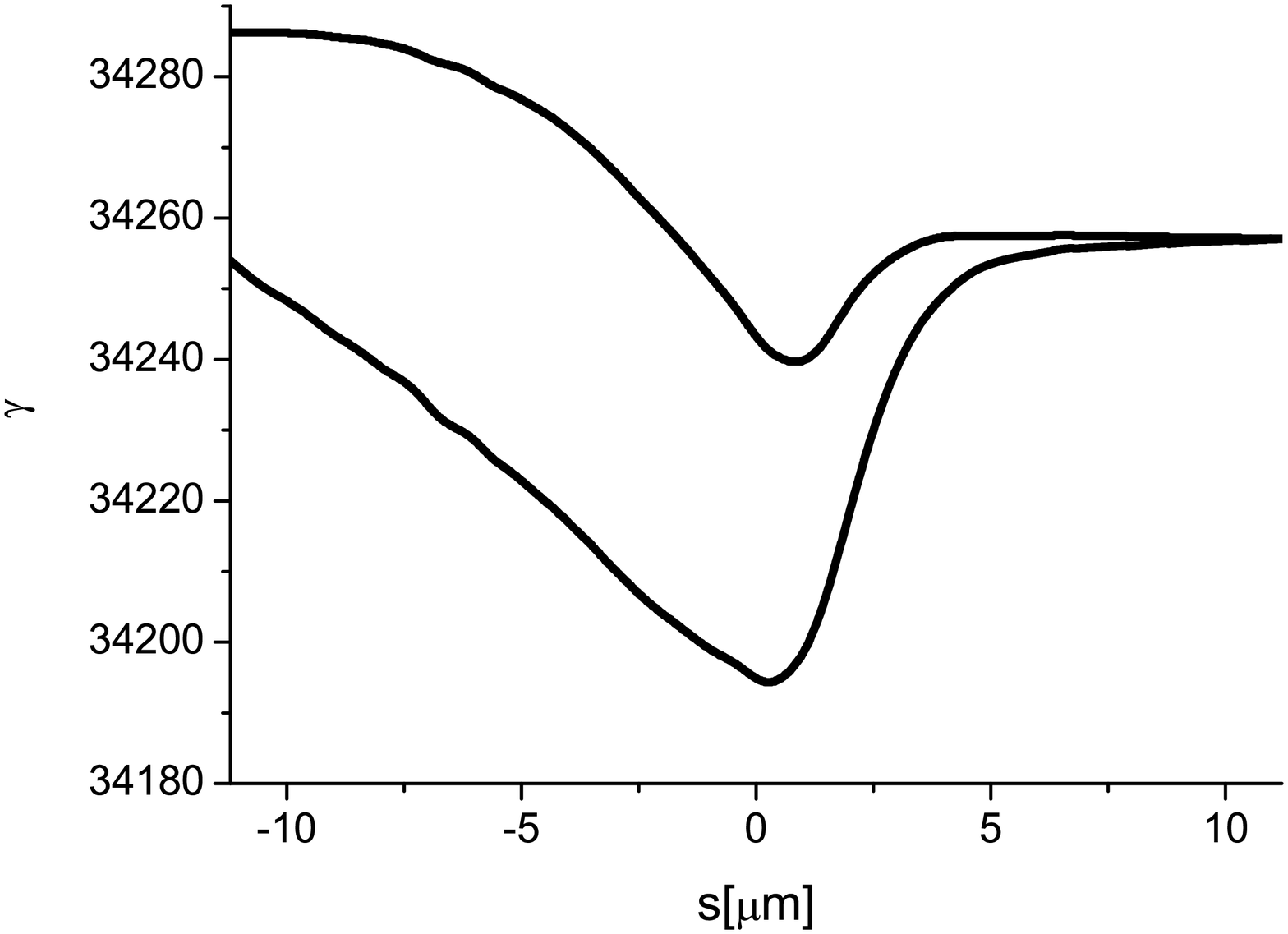}
\includegraphics[width=0.5\textwidth]{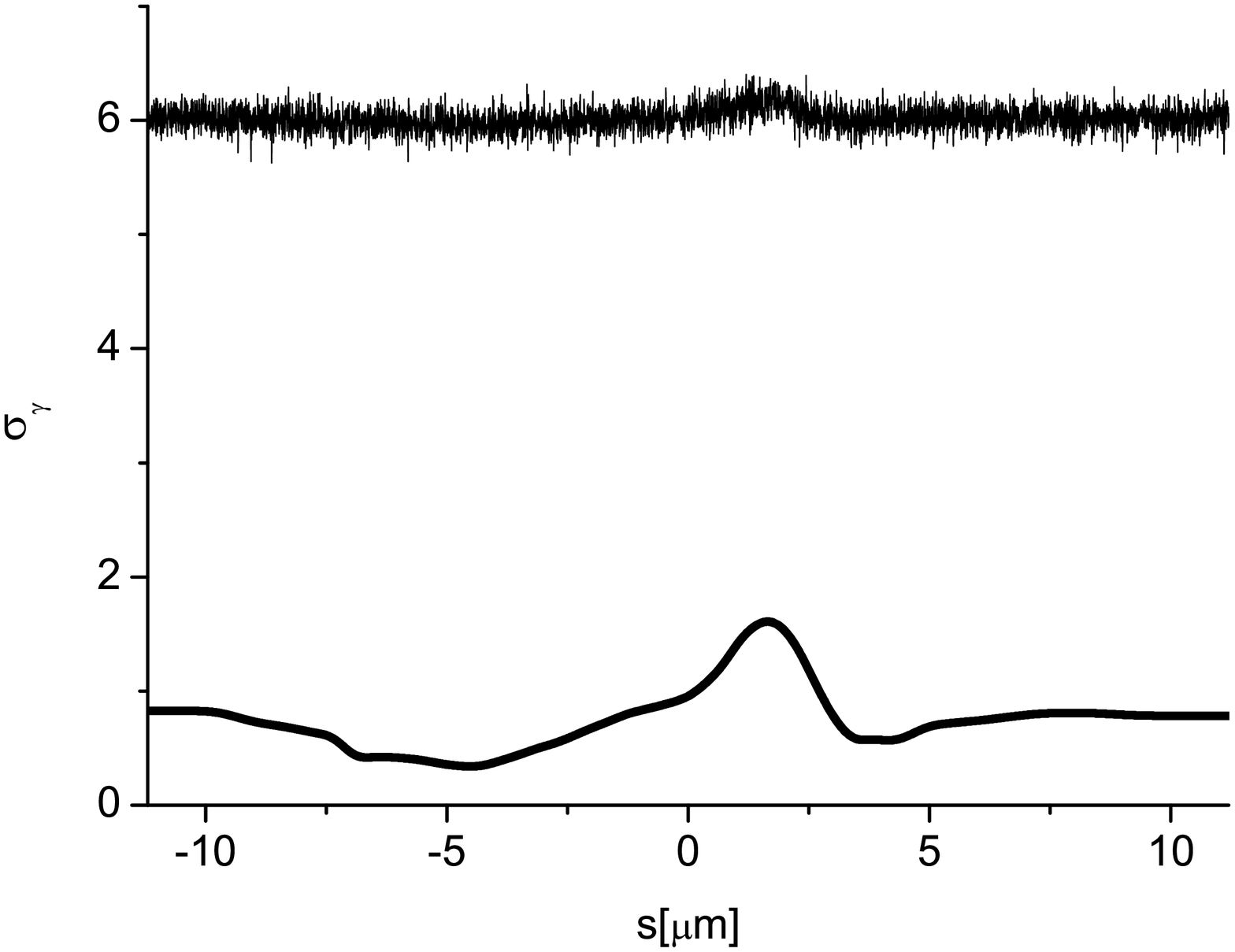}
\begin{center}
\includegraphics[width=0.5\textwidth]{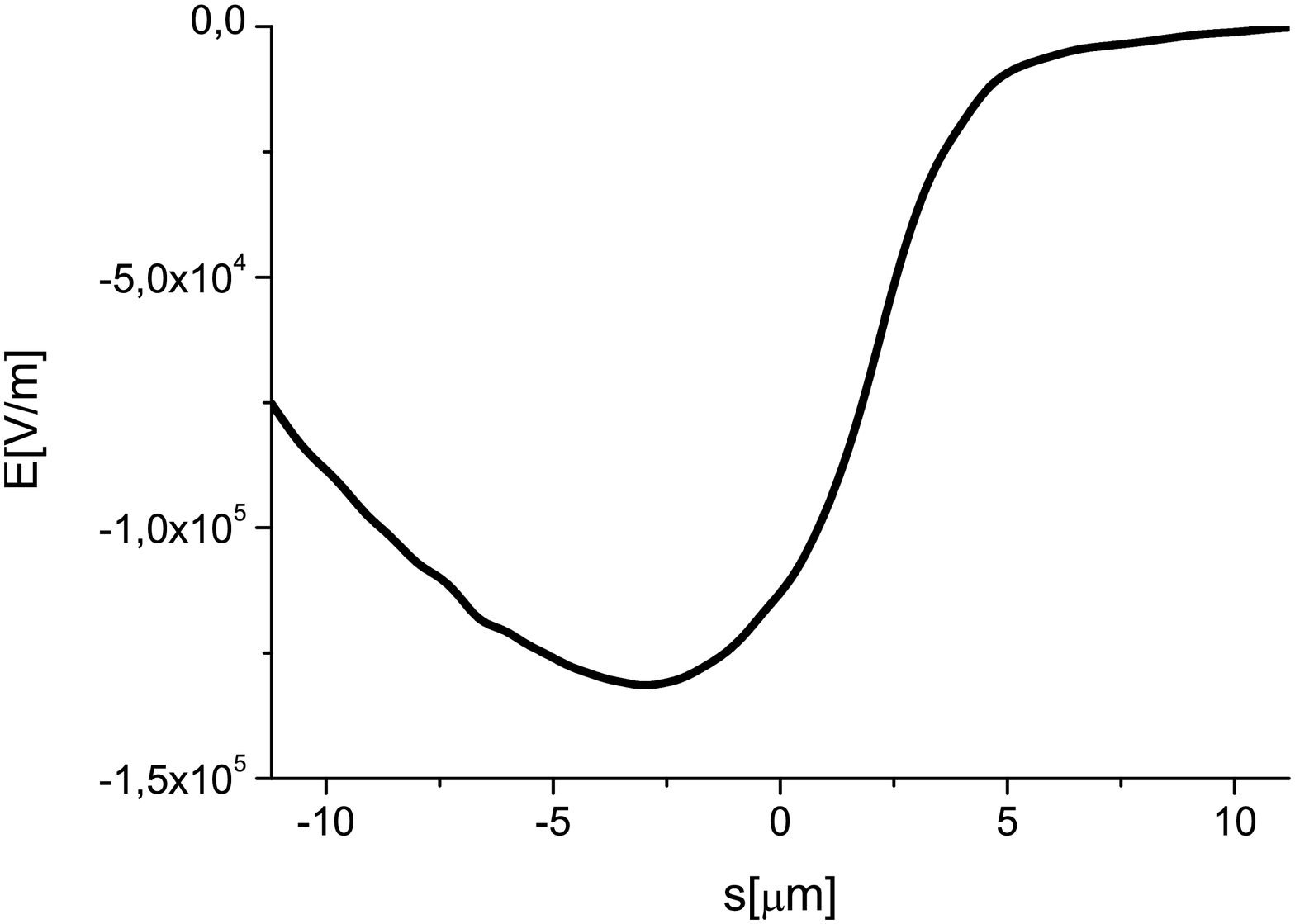}
\end{center}
\caption{Results from electron beam start-to-end simulations at the
entrance of the undulator system of the bio-imaging beamline
\cite{S2ER} for the hard X-ray case for the 17.5 GeV mode of
operation. (First Row, Left) Current profile. (First Row, Right)
Normalized emittance as a function of the position inside the
electron beam. (Second Row, Left) Energy profile along the beam,
lower curve. The effects of resistive wakefields along SASE1 are
illustrated by the comparison with the upper curve, referring to the
entrance of SASE1 (Second Row, Right) Electron beam energy spread
profile, upper curve. The effects of quantum diffusion along SASE1
are illustrated by the comparison with the lower curve, referring to
the entrance of SASE1. (Bottom row) Resistive wakefields in the
SASE3 undulator \cite{S2ER}.} \label{biof2f3}
\end{figure}

With reference to Fig. \ref{bio3f1} we performed feasibility studies
pertaining different energy ranges considered in this article. These
studies were performed with the help of the FEL code GENESIS 1.3
\cite{GENE} running on a parallel machine. Simulations are based on
a statistical analysis consisting of $100$ runs.

The main undulator parameters are reported in Table \ref{tt1}.
Operation is foreseen at two different energies: $10.5$ GeV and
$17.5$ GeV. The lower energy is used in the very soft X-ray regime,
between $0.3$ keV and $0.5$ keV. For this case, we refer to
\cite{OSOF} for a summary of the electron beam characteristics at
the entrance of the setup. The case for $17.5$ GeV instead is
summarized in Fig. \ref{biof2f3}, where we plot the results of
start-to-end simulations \cite{S2ER}.

\begin{table}
\caption{Undulator parameters}

\begin{small}\begin{tabular}{ l c c}
\hline & ~ Units &  ~ \\ \hline
Undulator period      & mm                  & 68     \\
Periods per cell      & -                   & 73   \\
Total number of cells & -                   & 40    \\
Intersection length   & m                   & 1.1   \\
Photon energy         & keV                 & 0.3-13 \\
\hline
\end{tabular}\end{small}
\label{tt1}
\end{table}

\subsection{Soft X-ray photon energy range}

Production of soft X-rays with photon energies below $1$ keV is
enabled by configuring the setup as described in Fig. \ref{bio3f2}.
A feasibility study for this case has been already carried out in
\cite{OSOF}, to which we refer the reader for further details and
simulation results.

\subsection{Photon energy range between 2 keV and 3 keV}

We now turn to analyze the case described in Fig. \ref{bio3f3},
which pertains the energy range between $2$ keV and $3$ keV. The
electron beam energy here is $17.5$ GeV.

\begin{figure}[tb]
\includegraphics[width=0.5\textwidth]{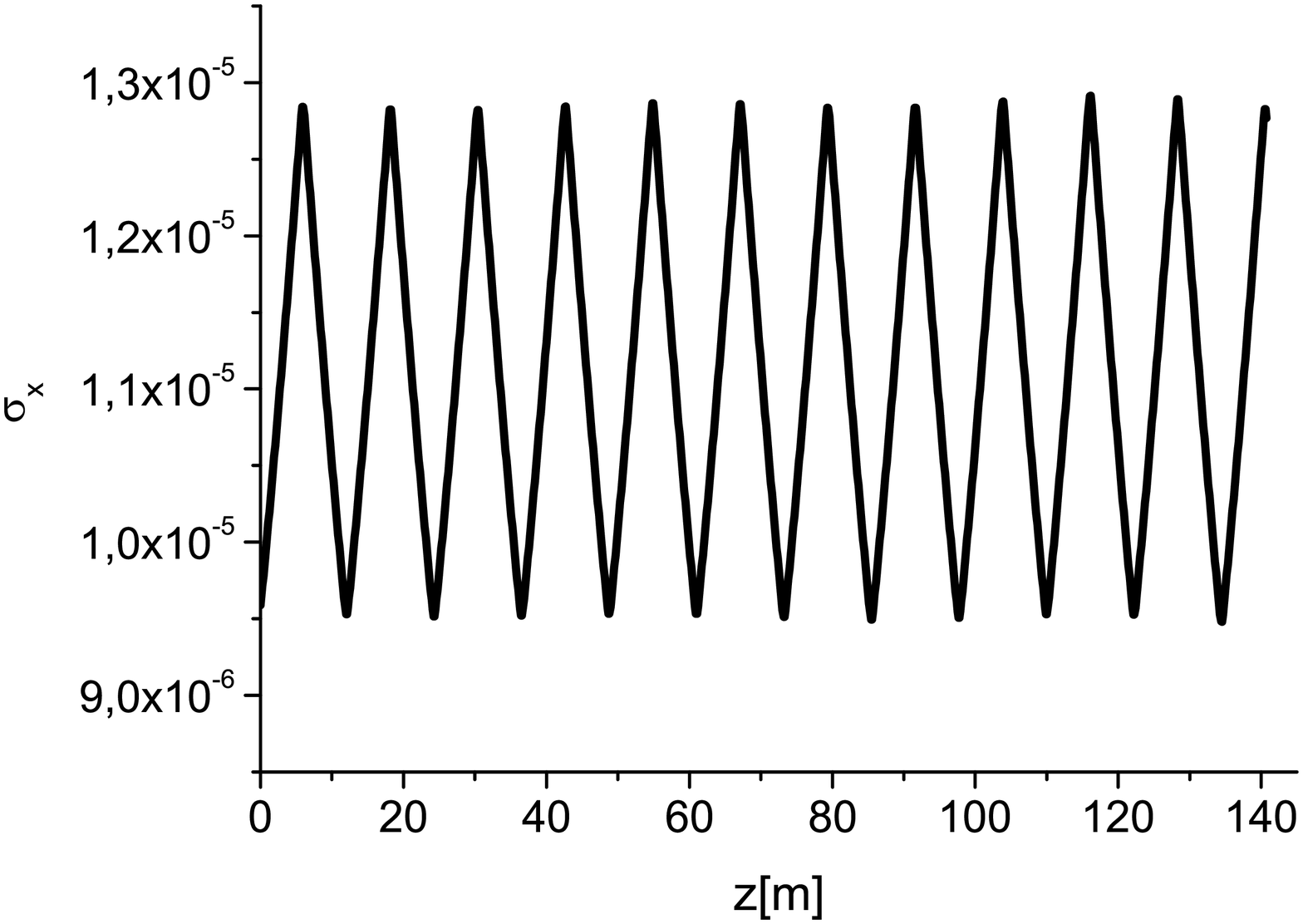}
\includegraphics[width=0.5\textwidth]{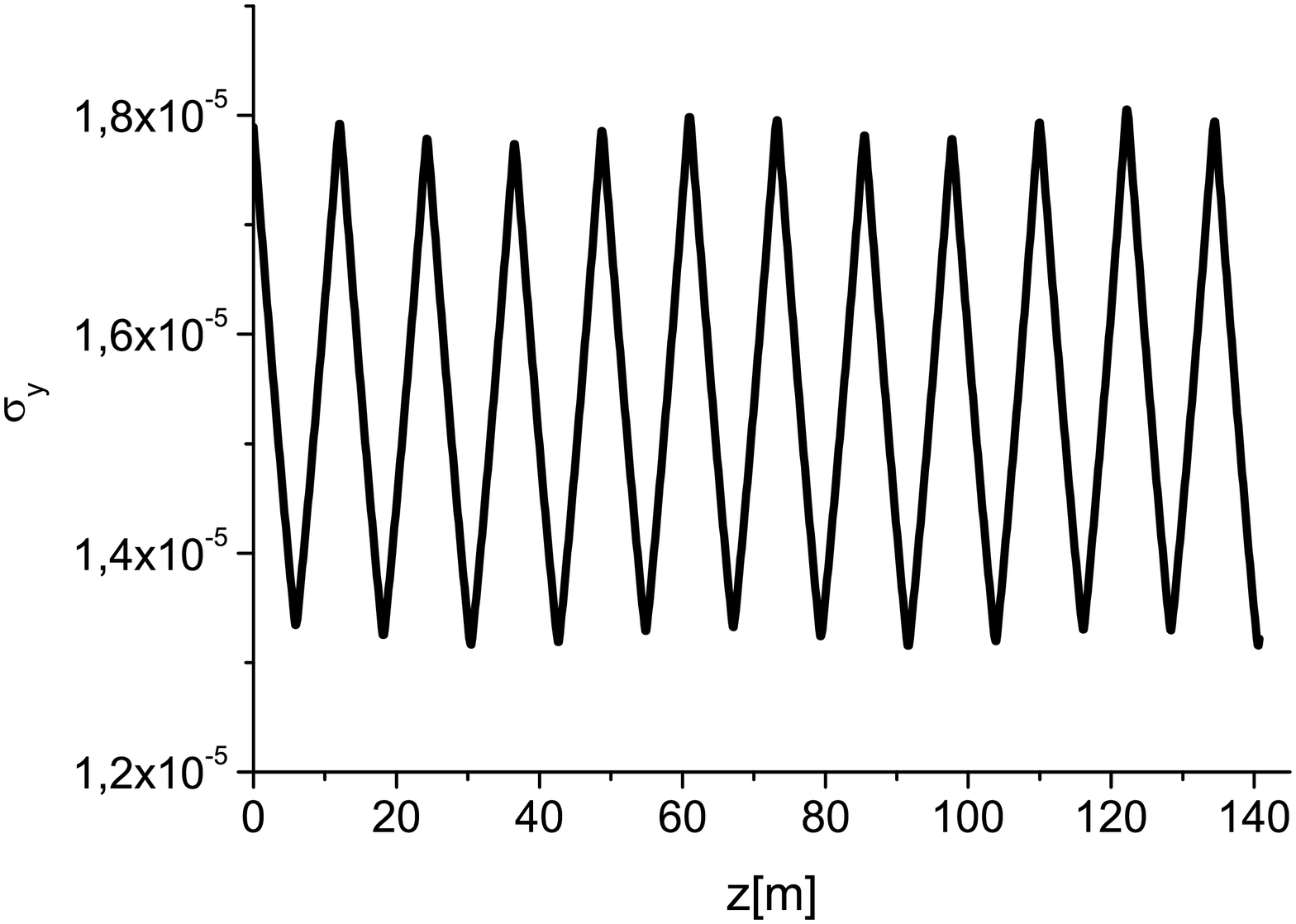}
\caption{Evolution of the horizontal (left plot) and vertical (right
plot) dimensions of the electron bunch as a function of the distance
inside the undulator. The plots refer to the longitudinal position
inside the bunch corresponding to the maximum current value.}
\label{sigma}
\end{figure}
The expected beam parameters at the entrance of the bio-imaging
beamline undulator, and the resistive wake inside the undulator are
shown in Fig. \ref{biof2f3}. The evolution of the transverse
electron bunch dimensions are plotted in Fig. \ref{sigma}. Since the
electron energy is fixed to $17.5$ GeV, both expected beam
parameters and evolution of the transverse beam dimensions before
the tapered part are valid for the different energy ranges treated
in the following Sections.

\begin{figure}[tb]
\includegraphics[width=0.5\textwidth]{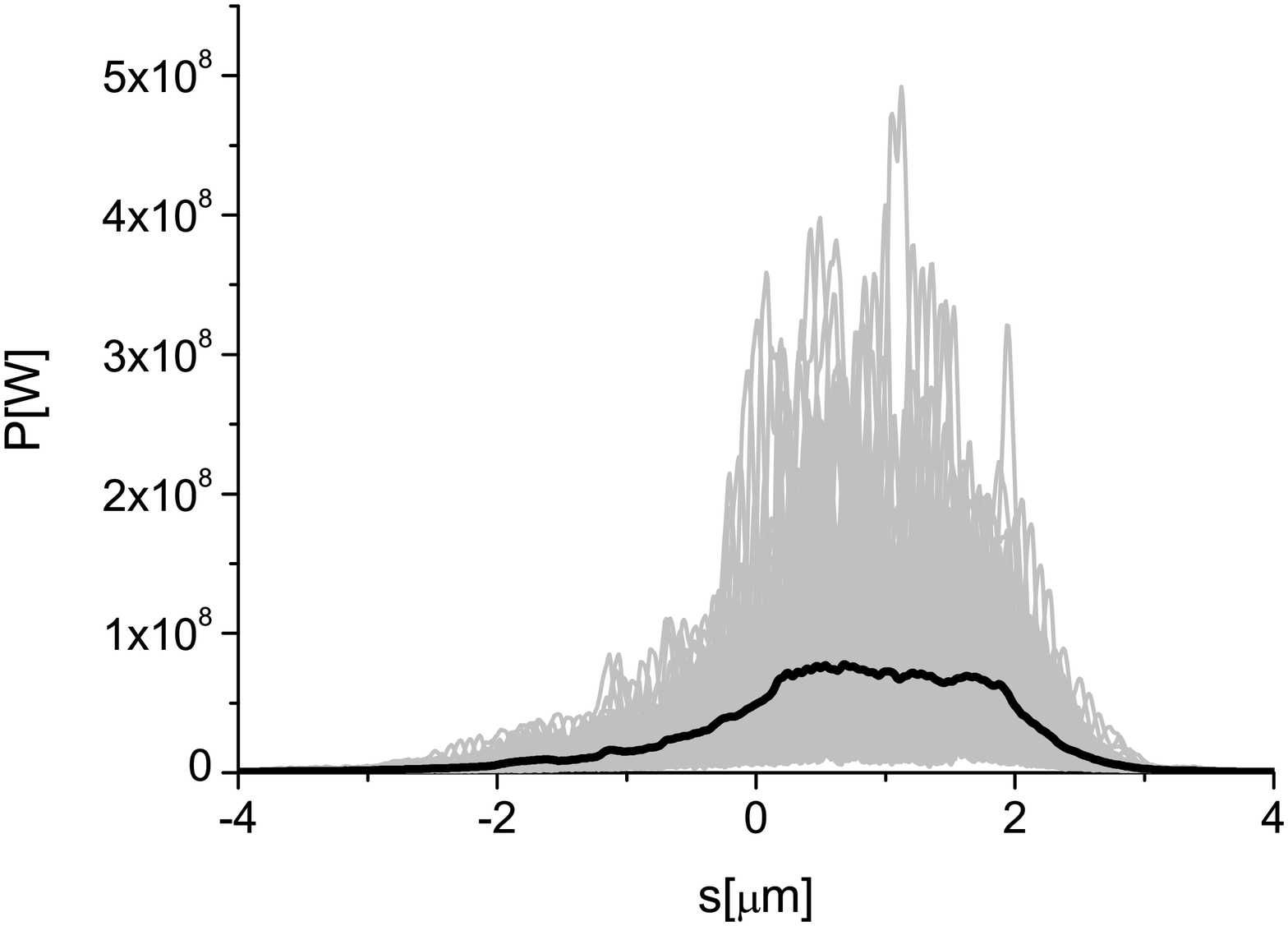}
\includegraphics[width=0.5\textwidth]{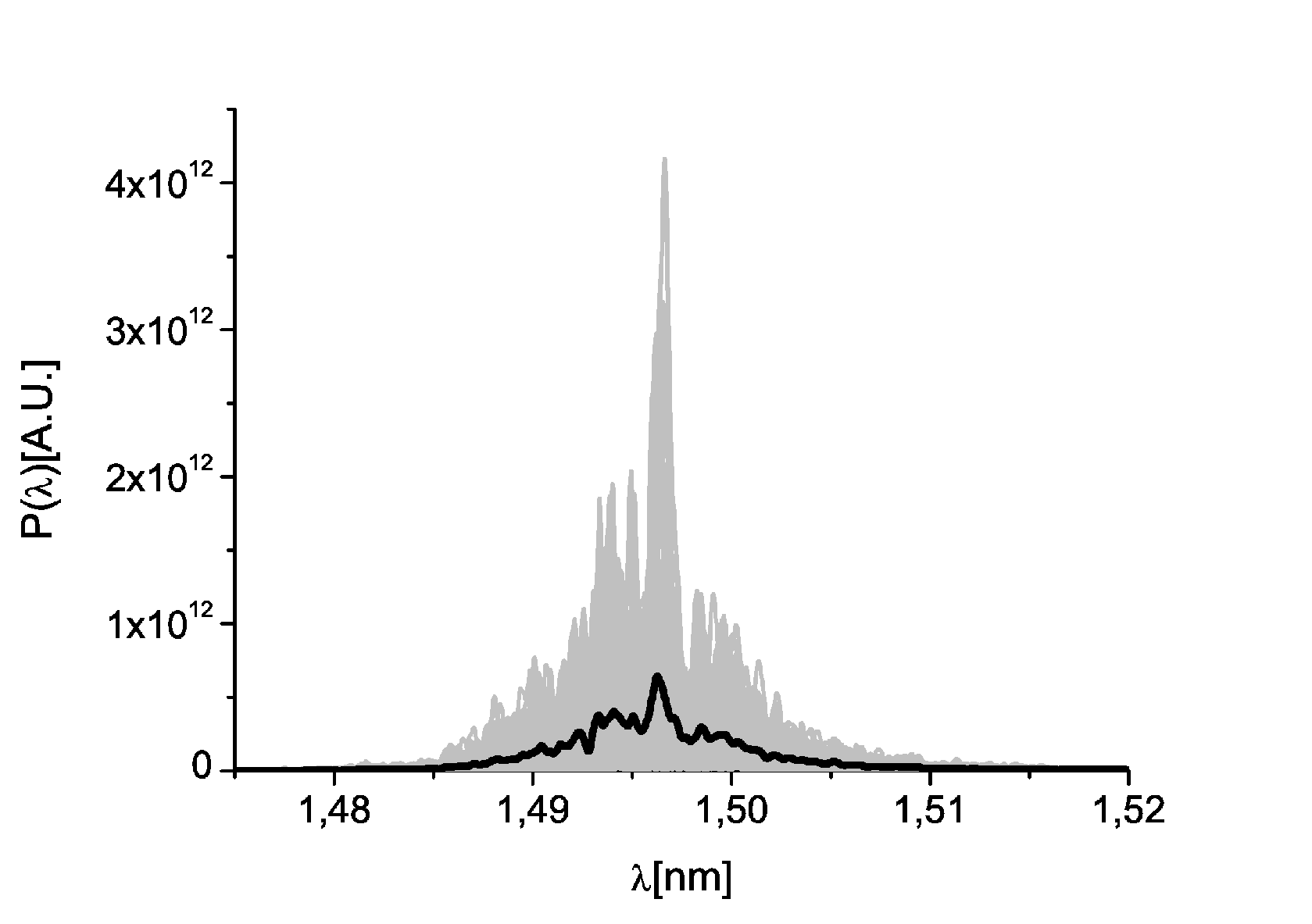}
\caption{Power and spectrum before the first magnetic chicane. Grey
lines refer to single shot realizations, the black line refers to
the average over a hundred realizations.} \label{biof17}
\end{figure}
We begin our investigation by simulating the SASE power and spectrum
after the first $4$ undulator cells, that is before the first
magnetic chicane. Results are shown in Fig. \ref{biof17}.

\begin{figure}[tb]
\includegraphics[width=0.5\textwidth]{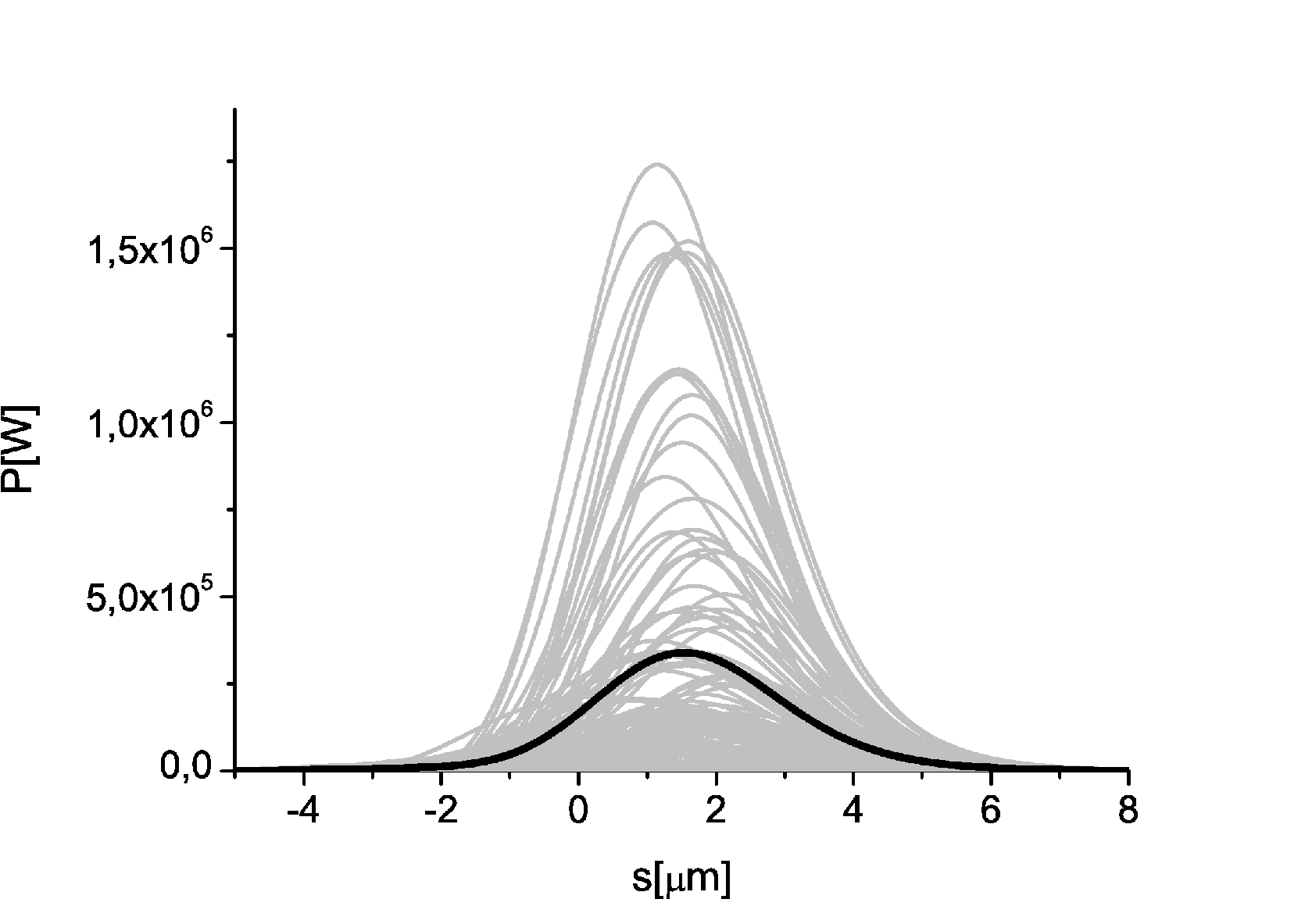}
\includegraphics[width=0.5\textwidth]{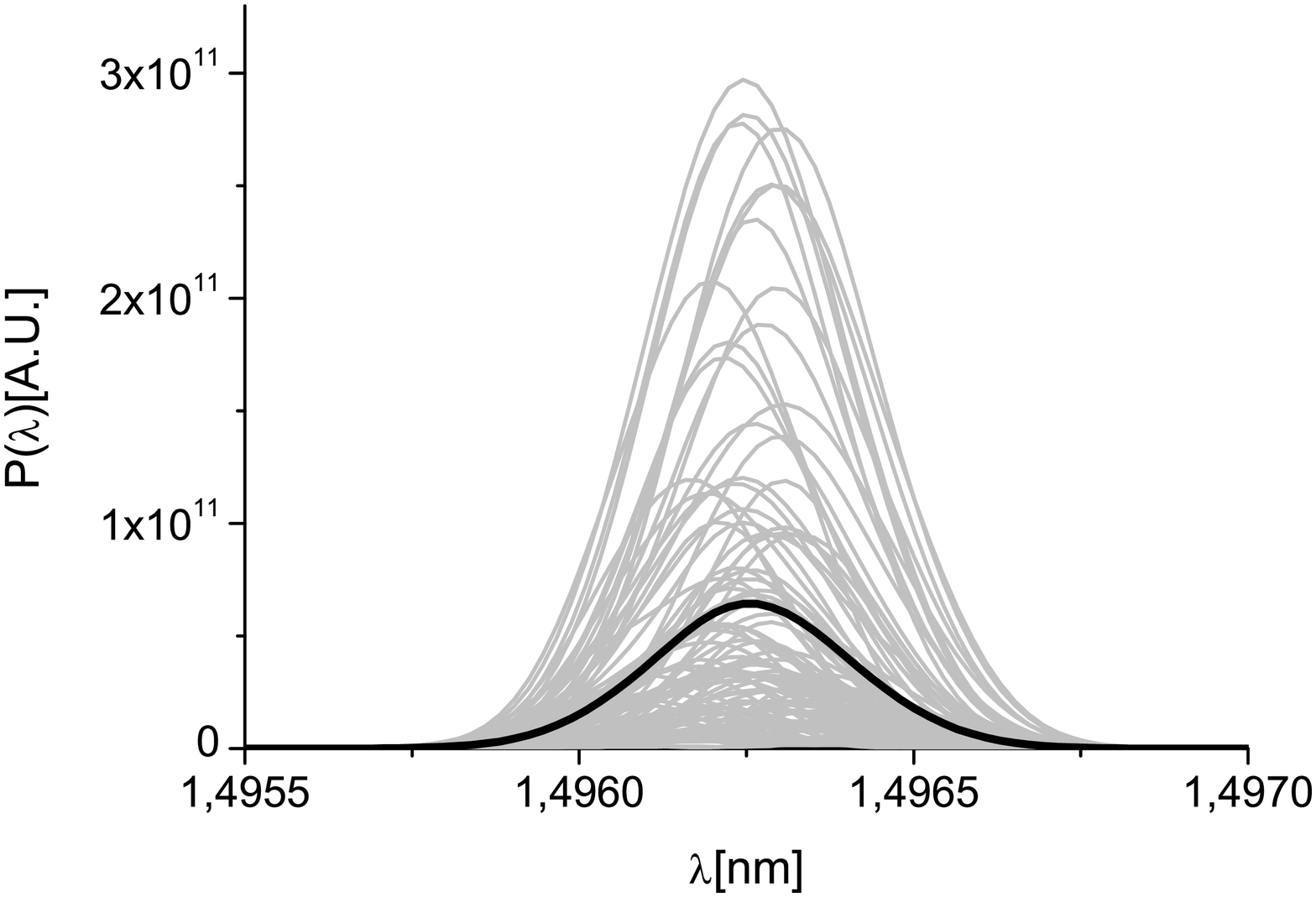}
\caption{Power and spectrum after the first magnetic chicane and
soft X-ray monochromator. Grey lines refer to single shot
realizations, the black line refers to the average over a hundred
realizations.} \label{biof18}
\end{figure}
The magnetic chicane is switched on, an the soft X-ray monochromator
is inserted. Assuming a monochromator efficiency of $10\%$, a
Gaussian line, and a resolving power of $5000$ we can filter the
incoming radiation pulse in Fig. \ref{biof17} accordingly, to obtain
the power and spectrum in Fig. \ref{biof18}. This power and spectrum
are used for seeding.

\begin{figure}[tb]
\includegraphics[width=0.5\textwidth]{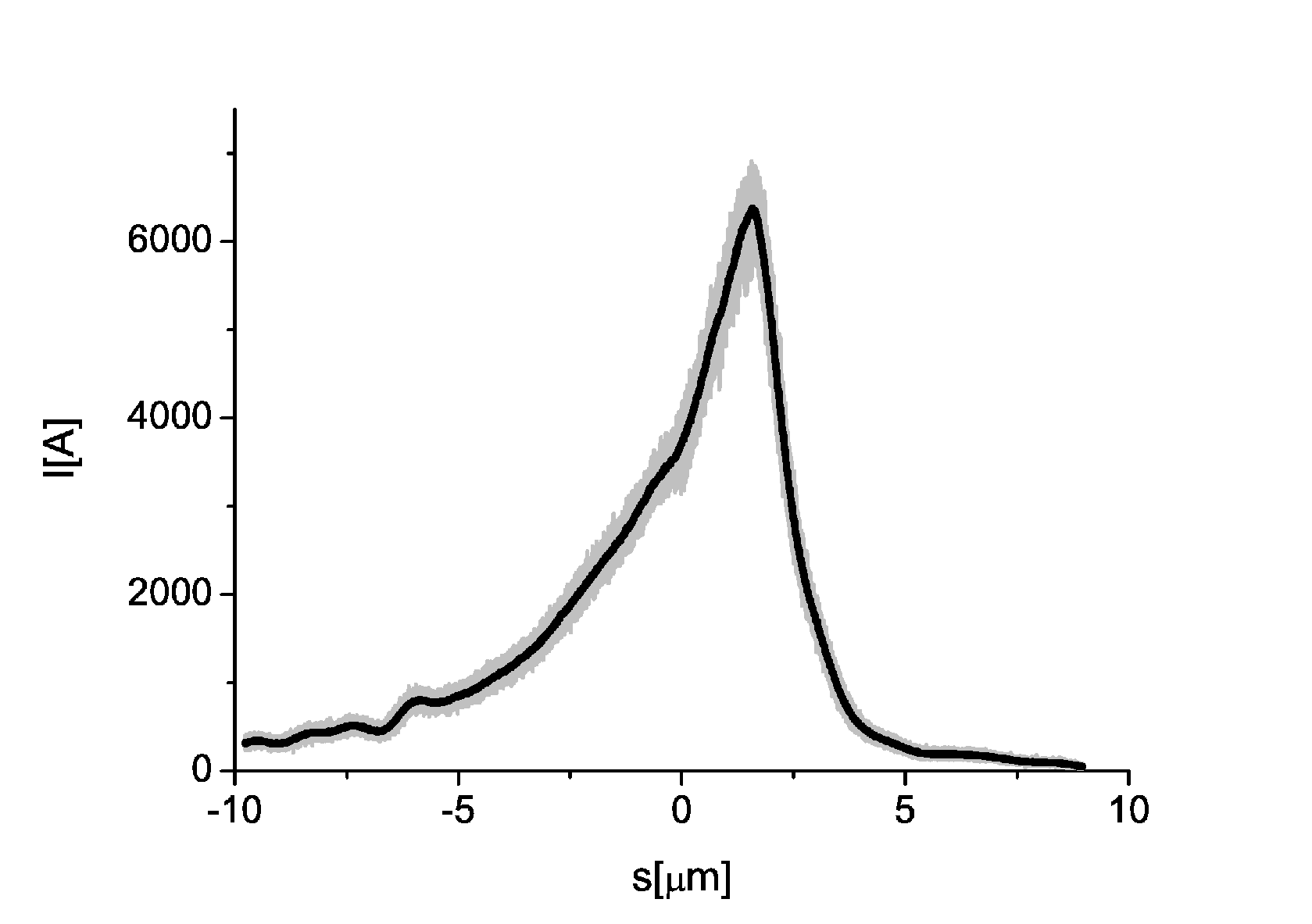}
\includegraphics[width=0.5\textwidth]{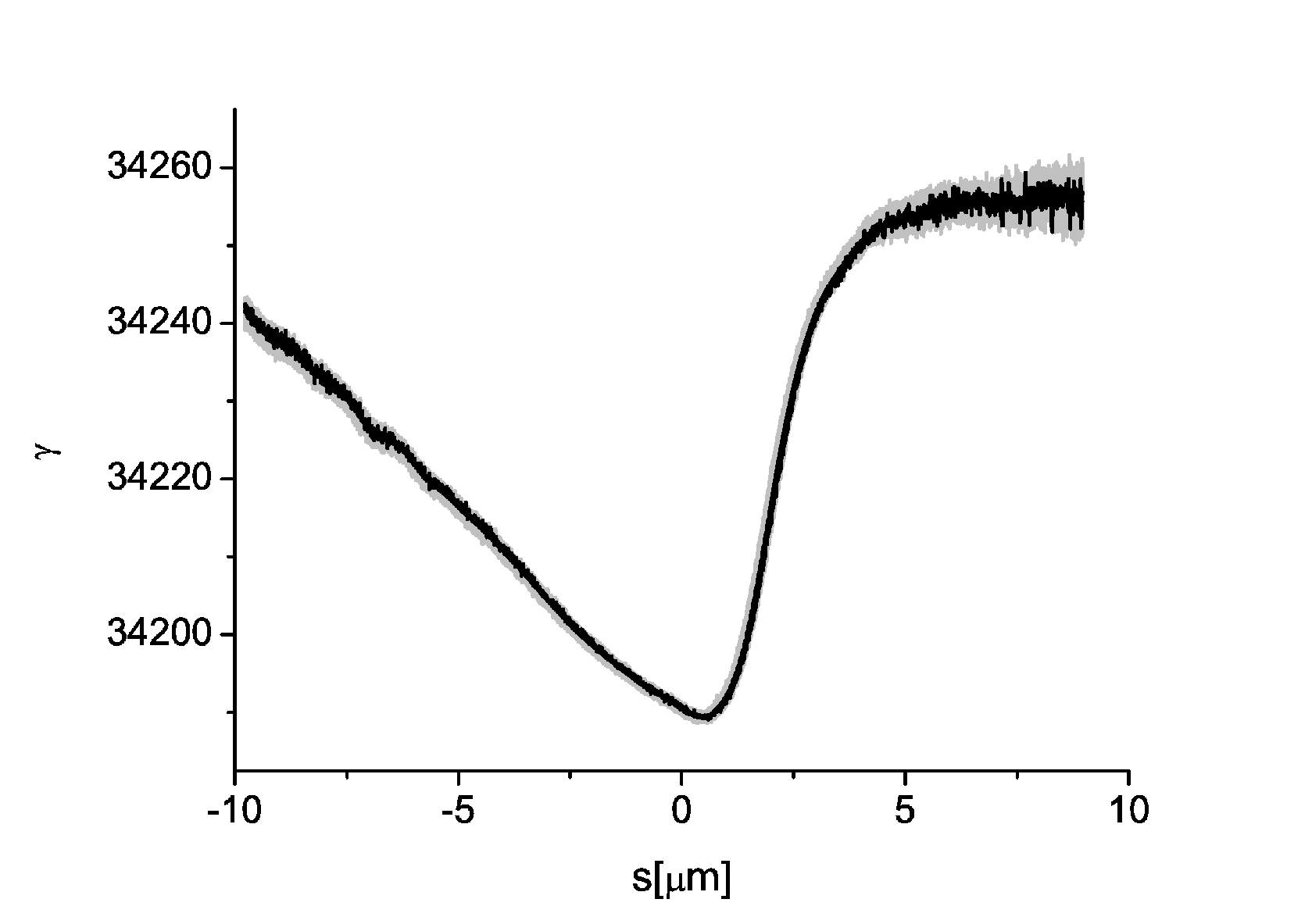}
\begin{center}
\includegraphics[width=0.5\textwidth]{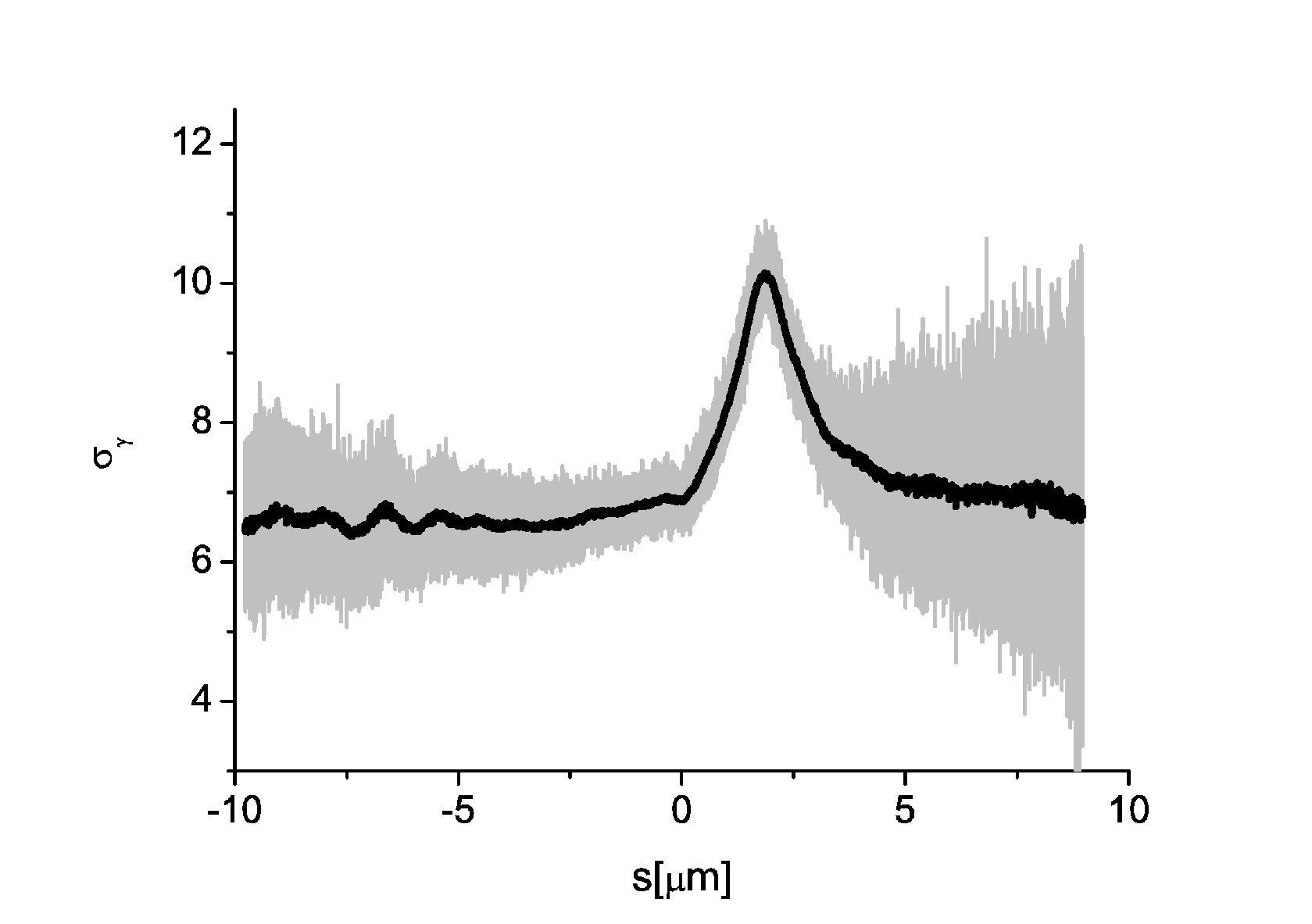}
\end{center}
\caption{Electron beam characteristics after the second magnetic
chicane. (First Row, Left) Current profile. (First Row, Right)
Energy profile along the beam. (Second Row) Electron beam energy
spread profile. } \label{biof19}
\end{figure}
Since we now deal with a conventional grating monochromator, the
photon pulse is delayed with respect to the electron pulse. In our
study case we assume a relatively large delay of about $1$ ps. In
order to compensate for such delay, one needs a chicane with a
relatively large dispersion strength $R_{56} \sim 0.6$ mm. In
principle we cannot neglect the effects of the chicane dispersion on
the electron bunch properties. We accounted for them with the help
of the code Elegant \cite{ELEG}, which was used to propagate the
electron beam distribution through the chicane. Results are shown in
Fig. \ref{biof19}.

\begin{figure}[tb]
\includegraphics[width=0.5\textwidth]{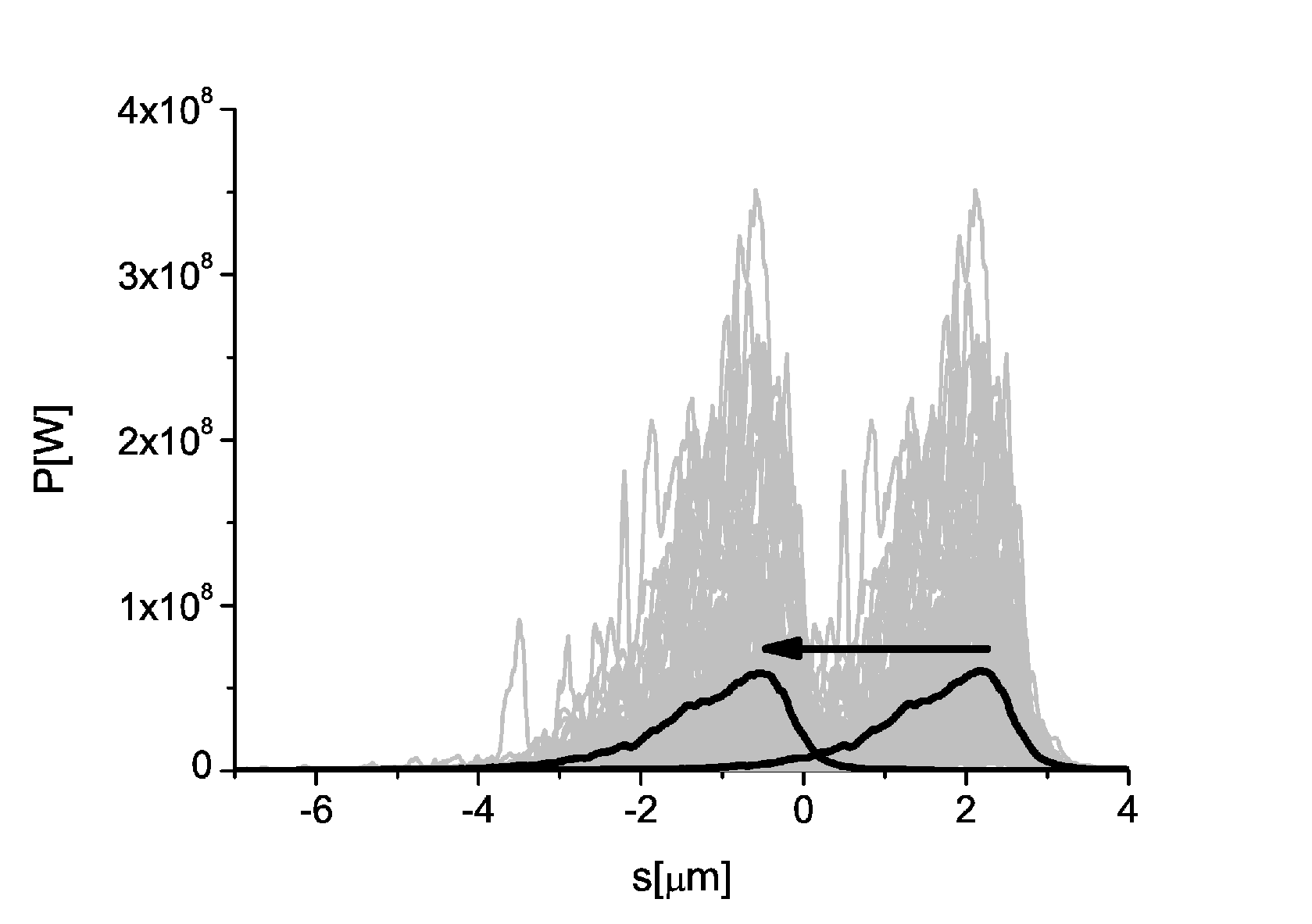}
\includegraphics[width=0.5\textwidth]{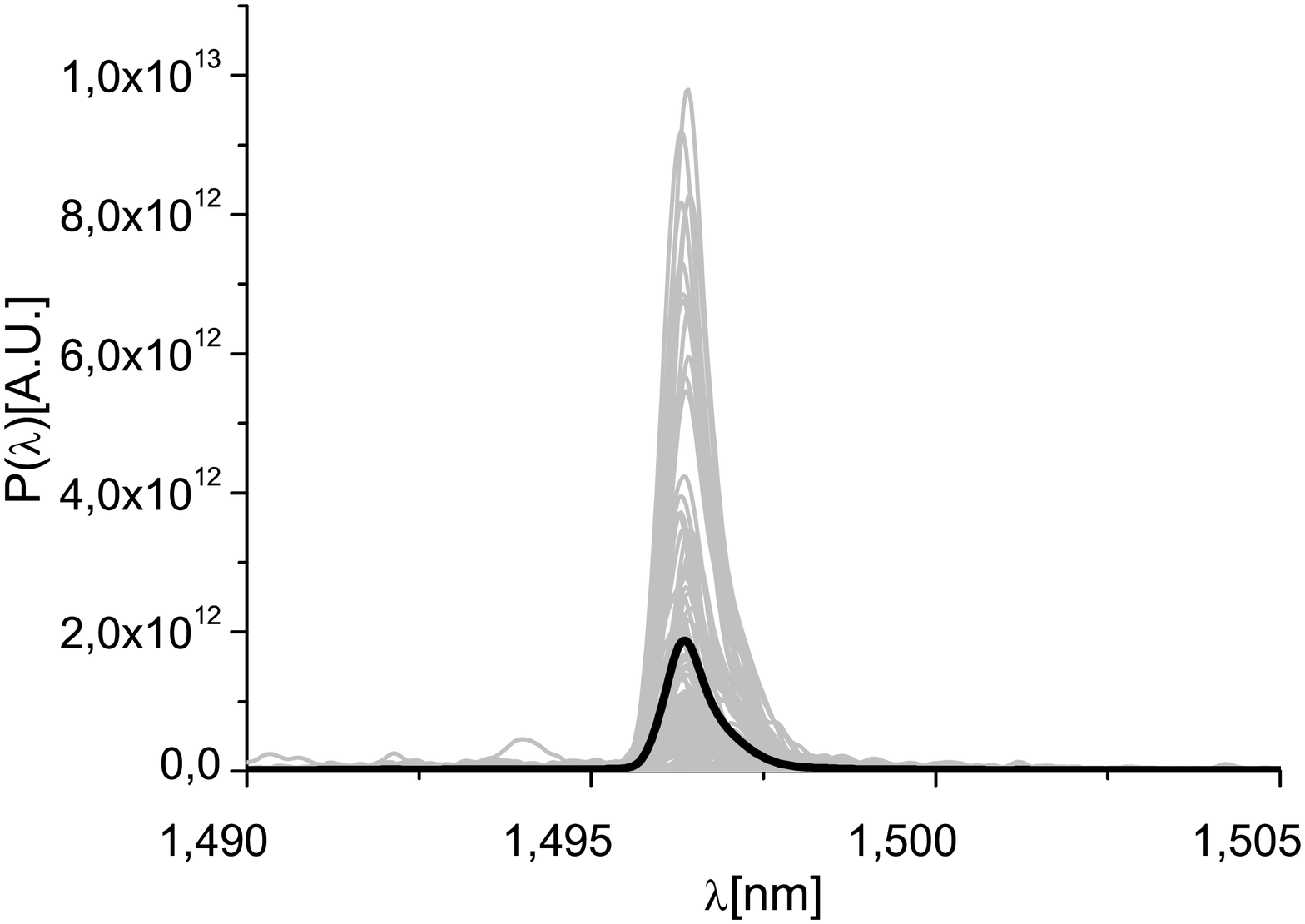}
\caption{Power and spectrum at the fundamental harmonic after the
second chicane equipped with the X-ray optical delay line, delaying
the radiation pulse with respect to the electron bunch. Grey lines
refer to single shot realizations, the black line refers to the
average over a hundred realizations.} \label{biof20}
\end{figure}
Since the $R_{56}$ is large enough to wash out the electron beam
microbunching, we assume a fresh bunch at the entrance of the
following undulator part constituted by $3$ undulator cells. This
means that the results in Fig. \ref{biof19} are taken to generate a
new beam file to be fed into GENESIS. The electron bunch is now
seeded with the monochromatized radiation pulse in Fig.
\ref{biof18}, so that the seed is amplified in the $3$ undulator
cells following the chicane. After that, the electron beam is sent
through the second chicane, while the radiation pulse goes through
the X-ray optical delay line described in Fig. \ref{biof2}, where
the radiation pulse is delayed of about $6$ fs with respect to the
electron beam, as shown in Fig. \ref{biof9}. The power and spectrum
of the radiation pulse after the optical delay line are shown in
Fig. \ref{biof20}, where the combined effect of the optical delay
and of the magnetic chicane is illustrated. This results in an
overall delay of $6$ fs. Note that the use of the mirror chicane
also allows for an $R_{56}$ in the order of ten microns. Thus, the
combination of magnetic chicane and mirror chicane also allows for
removing the electron microbunching produced in the second
undulator.

\begin{figure}[tb]
\includegraphics[width=0.5\textwidth]{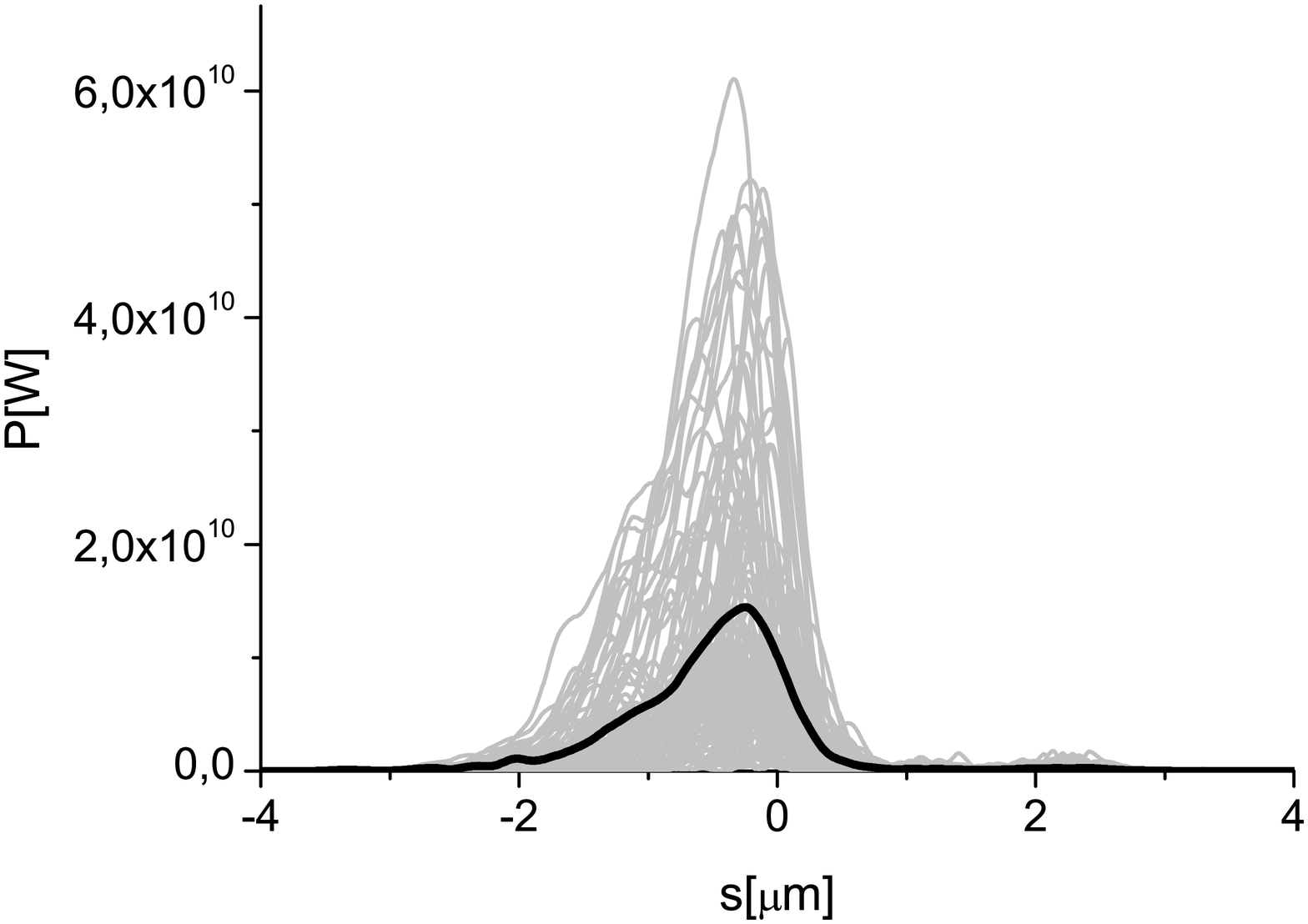}
\includegraphics[width=0.5\textwidth]{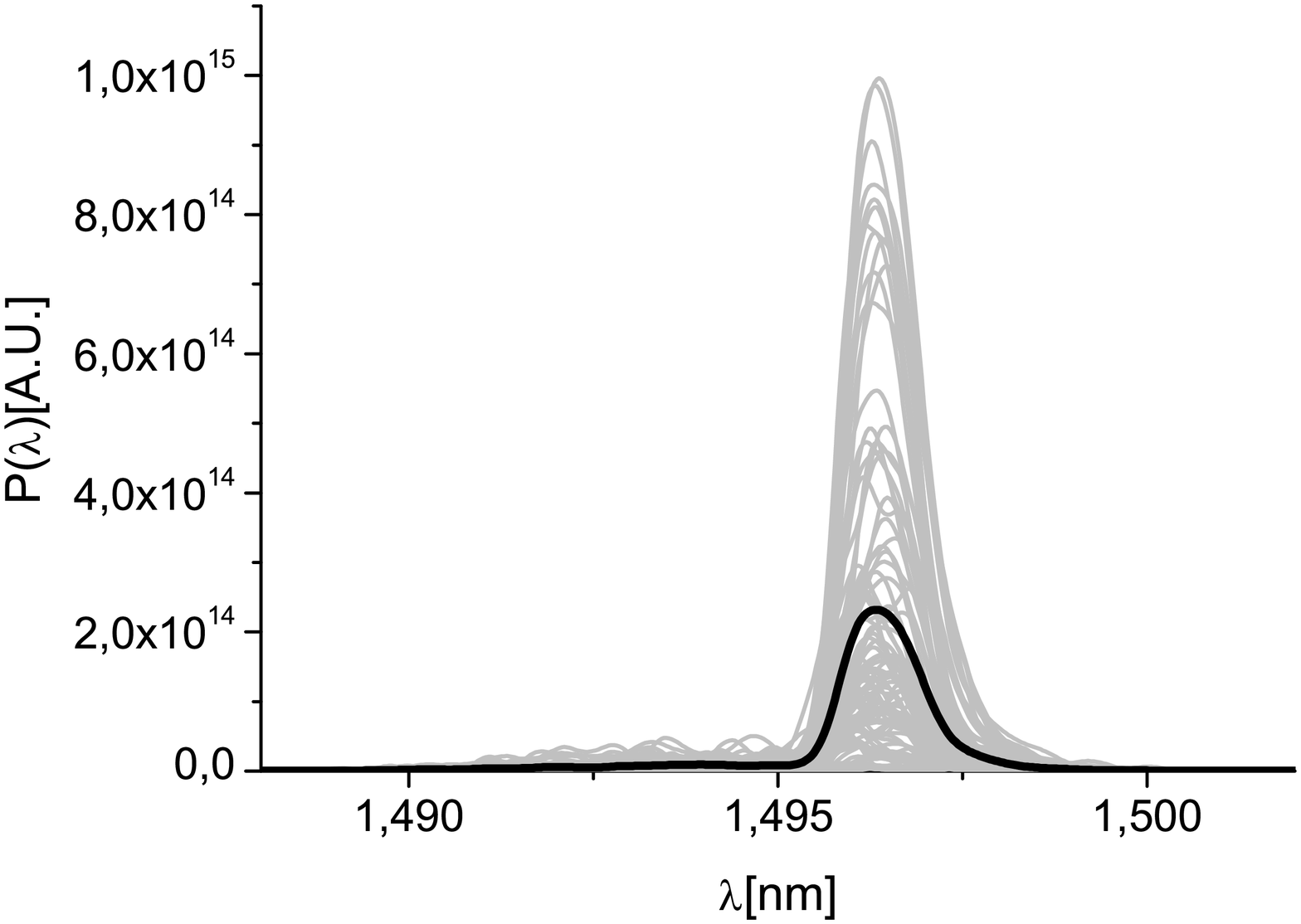}
\caption{Power and spectrum at the fundamental harmonic at the exit
of the third undulator and before the third magnetic chicane. Grey
lines refer to single shot realizations, the black line refers to
the average over a hundred realizations.} \label{biof21}
\end{figure}
Besides allowing for the installation of the optical delay line,
which delays the radiation pulse of about half of the electron bunch
size, the second chicane also smears out the microbunching in the
electron bunch. As a result, at the entrance of the third undulator
part the electron bunch can be considered as unmodulated, and half
of it is seeded with the radiation pulse. The seeded half of the
electron bunch amplifies the seed in the third undulator part,
composed by four cells. The seeded part of the electron bunch is now
spent, and its quality has deteriorated too much for further lasing.
After that, electrons and radiation are separated once more going
through the third chicane. The hard X-ray self-seeding crystal is
out, and the chicane simply acts as a delay line for the electron
beam, which also smears out the microbunching. Power and spectrum
following the third chicane are shown in Fig. \ref{biof21}.

\begin{figure}[tb]
\includegraphics[width=0.5\textwidth]{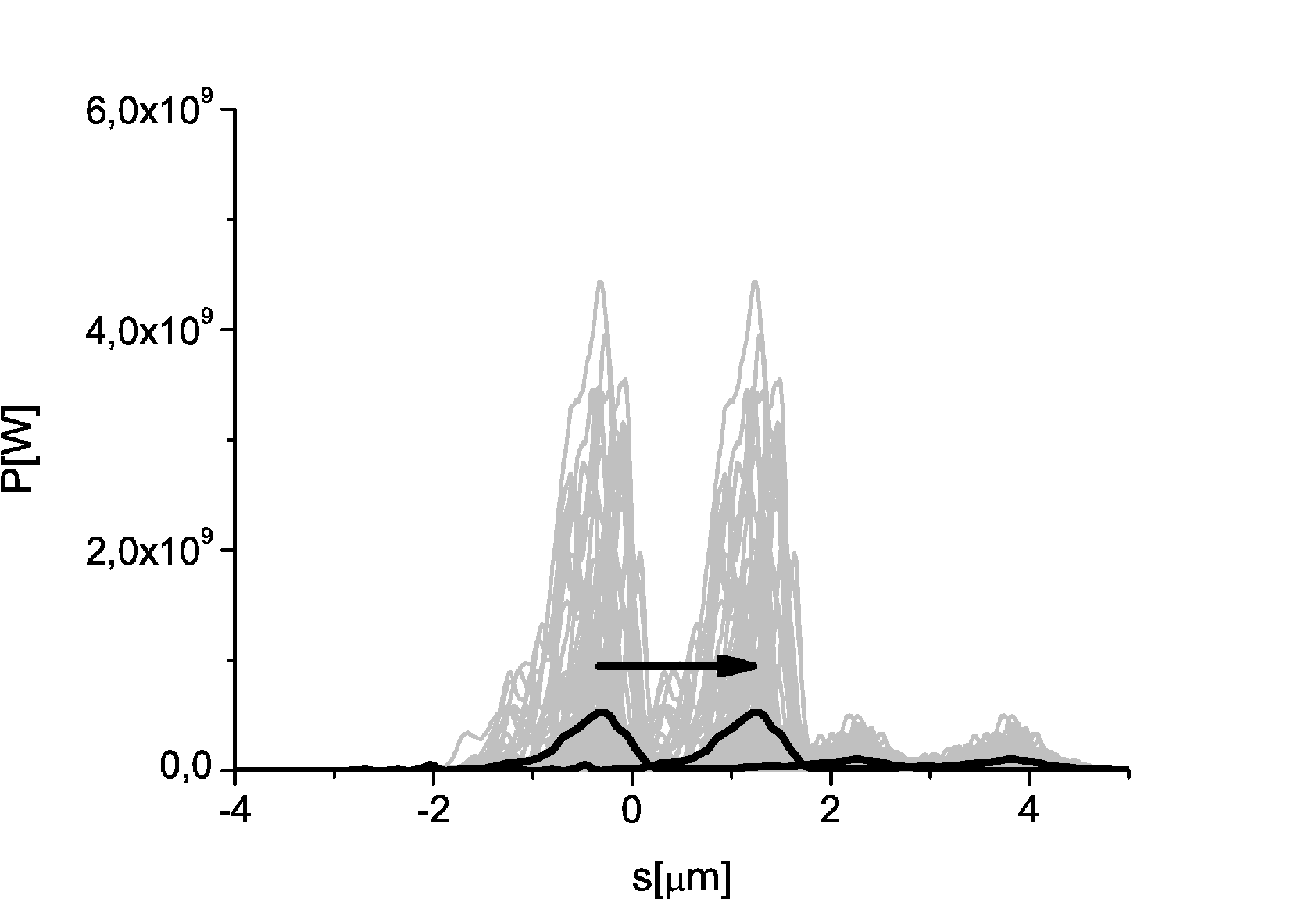}
\includegraphics[width=0.5\textwidth]{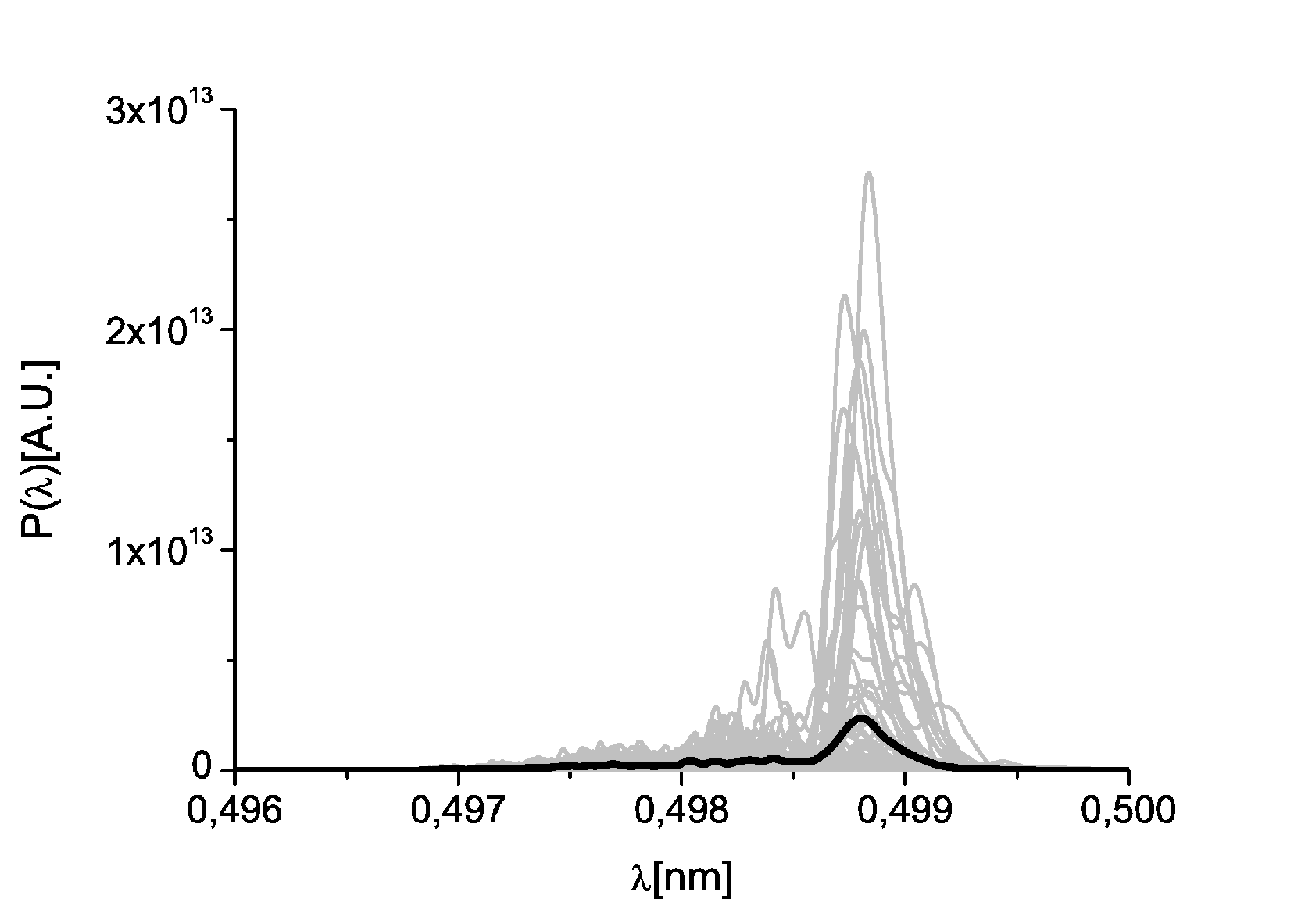}
\caption{Power and spectrum at the third harmonic after the third
magnetic chicane. Grey lines refer to single shot realizations, the
black line refers to the average over a hundred realizations.}
\label{biof21}
\end{figure}
By tuning the third chicane in the proper way, one can superimpose
the radiation beam onto that part of the electron bunch that has not
been seeded in the third undulator part. This is fresh, and can lase
again in the fourth undulator part.  Fig. \ref{biof21} shows the
effect of the magnetic chicane, which delays the electron bunch
relative to the radiation pulse in order to allow for the seeding of
the fresh part of the bunch. The radiation beam includes a relevant
third-harmonic content, just slightly below the GW level as can be
seen in Fig. \ref{biof21}, and is sufficient to act as a seed in the
last part of the undulator. The fourth undulator part is not tuned
at the fundamental harmonic, but rather at the third harmonic. This
allows to reach the photon energy range between $2$ keV and $3$ keV.

\begin{figure}[tb]
\begin{center}
\includegraphics[width=0.5\textwidth]{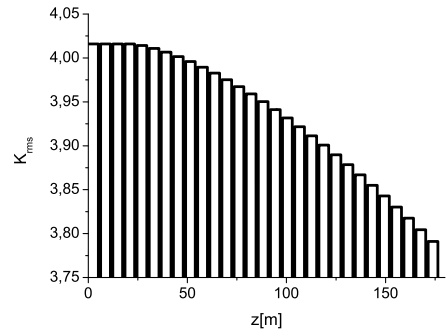}
\end{center}
\caption{Tapering law for the case $\lambda = 0.5$ nm.}
\label{biof22}
\end{figure}

\begin{figure}[tb]
\includegraphics[width=0.5\textwidth]{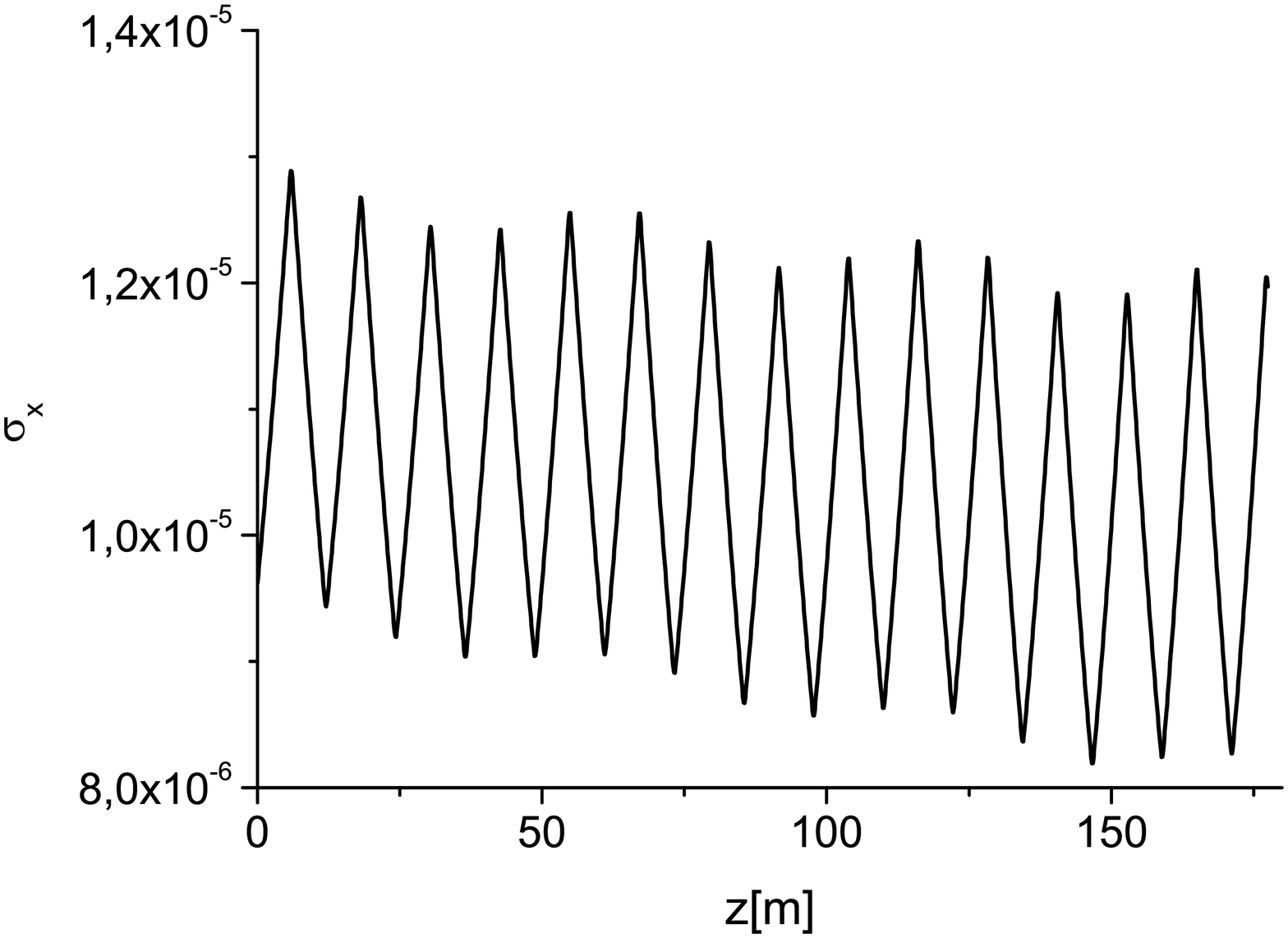}
\includegraphics[width=0.5\textwidth]{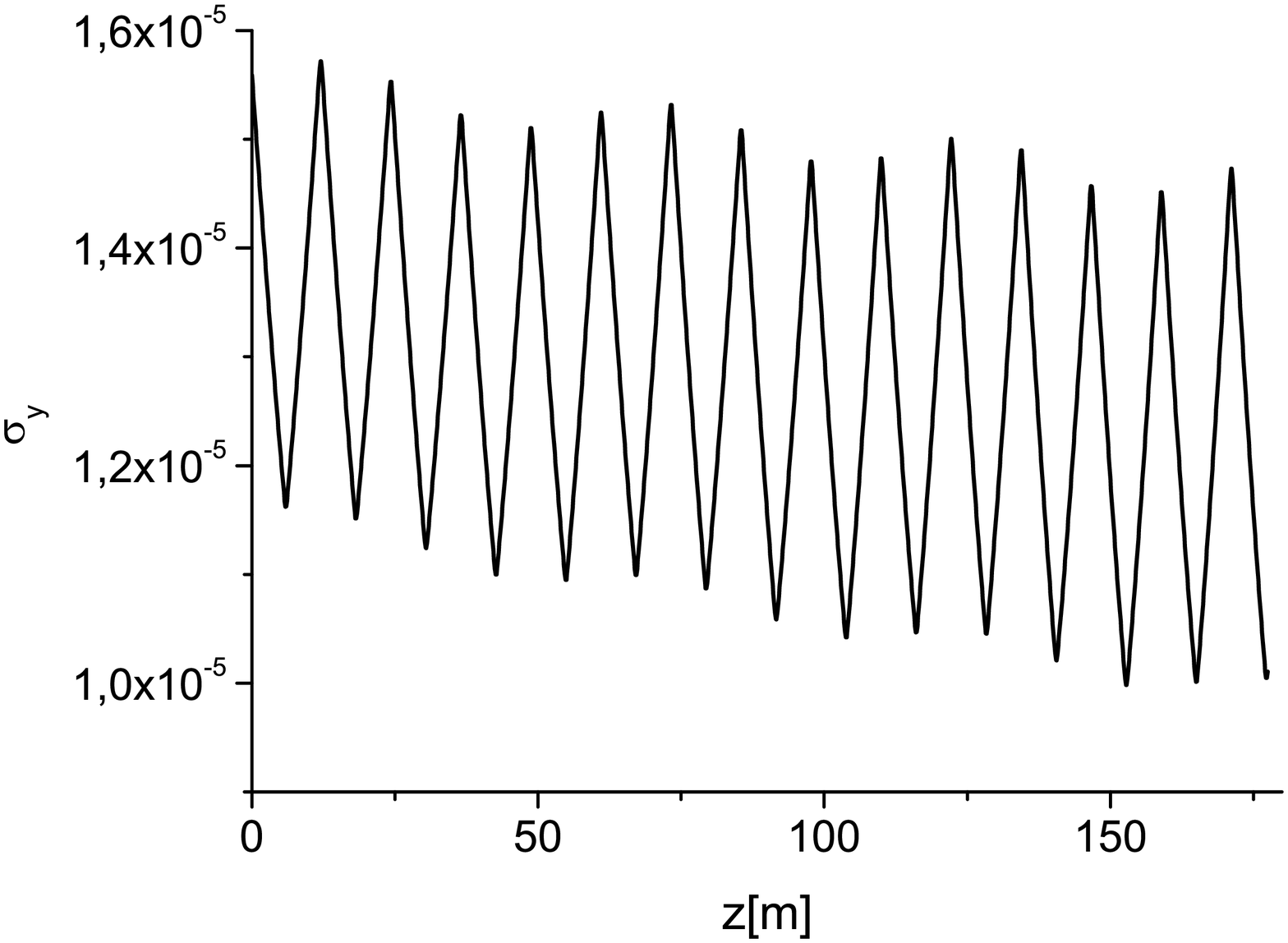}
\caption{Evolution of the horizontal (left plot) and vertical (right
plot) dimensions of the electron bunch as a function of the distance
inside the tapered part of the undulator at $\lambda = 0.5$ nm. The
plots refer to the longitudinal position inside the bunch
corresponding to the maximum current value. The quadrupole strength
varies along the undulator, and is tuned for optimum output.}
\label{sigmaaa}
\end{figure}
The fourth and last part of the undulator is composed by $29$ cells,
interrupted by a chicane which is switched off, and will be used in
different energy ranges. The last undulator part is partly tapered
post-saturation to allow for increasing the region where electrons
and radiation interact properly to the advantage of the radiation
pulse. Tapering is implemented by changing the $K$ parameter of the
undulator segment by segment according to Fig. \ref{biof22}. The
tapering law used in this work has been implemented on an empirical
basis, and the output has been optimized also by varying the
quadrupole strength as shown in Fig. \ref{sigmaaa}.

\begin{figure}[tb]
\includegraphics[width=0.5\textwidth]{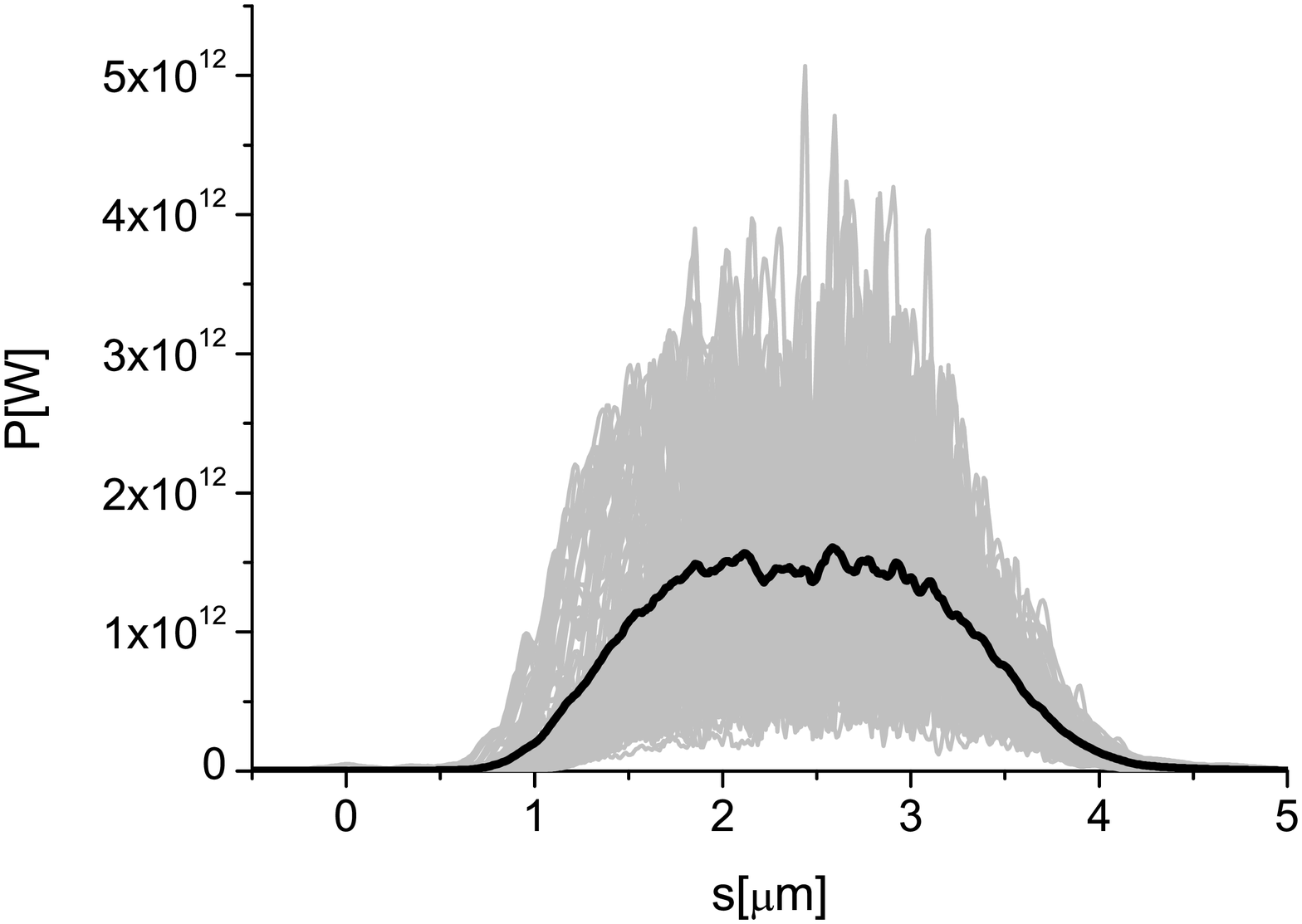}
\includegraphics[width=0.5\textwidth]{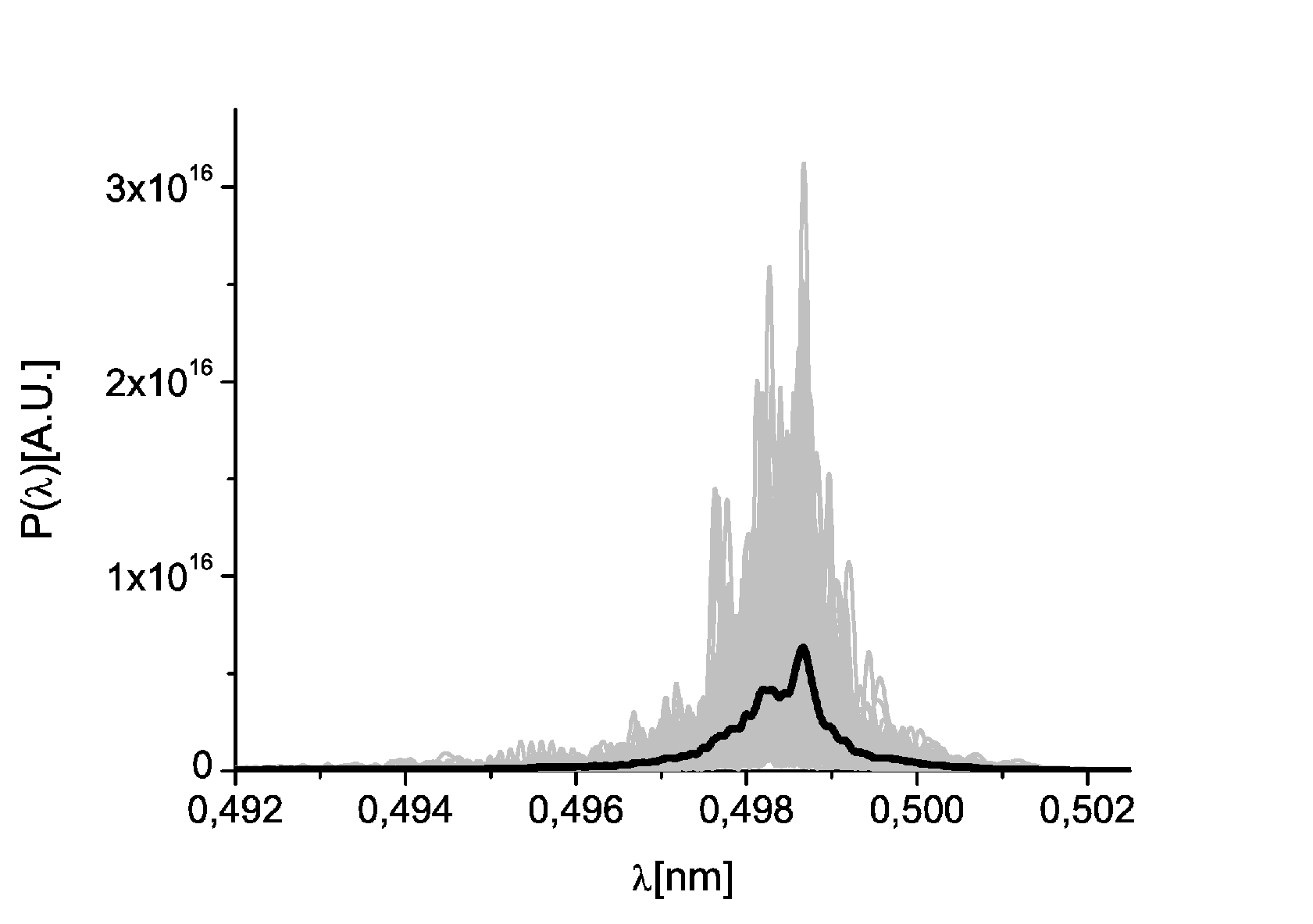}
\caption{Final output. Power and spectrum at the third harmonic
after tapering. Grey lines refer to single shot realizations, the
black line refers to the average over a hundred realizations.}
\label{biof23}
\end{figure}
The use of tapering together with monochromatic radiation is
particularly effective, since the electron beam does not experience
brisk changes of the ponderomotive potential during the slippage
process. The final output is presented in Fig. \ref{biof23} in terms
of power and spectrum. As one can see, simulations indicate an
output power of about $2$ TW.

\begin{figure}[tb]
\includegraphics[width=0.5\textwidth]{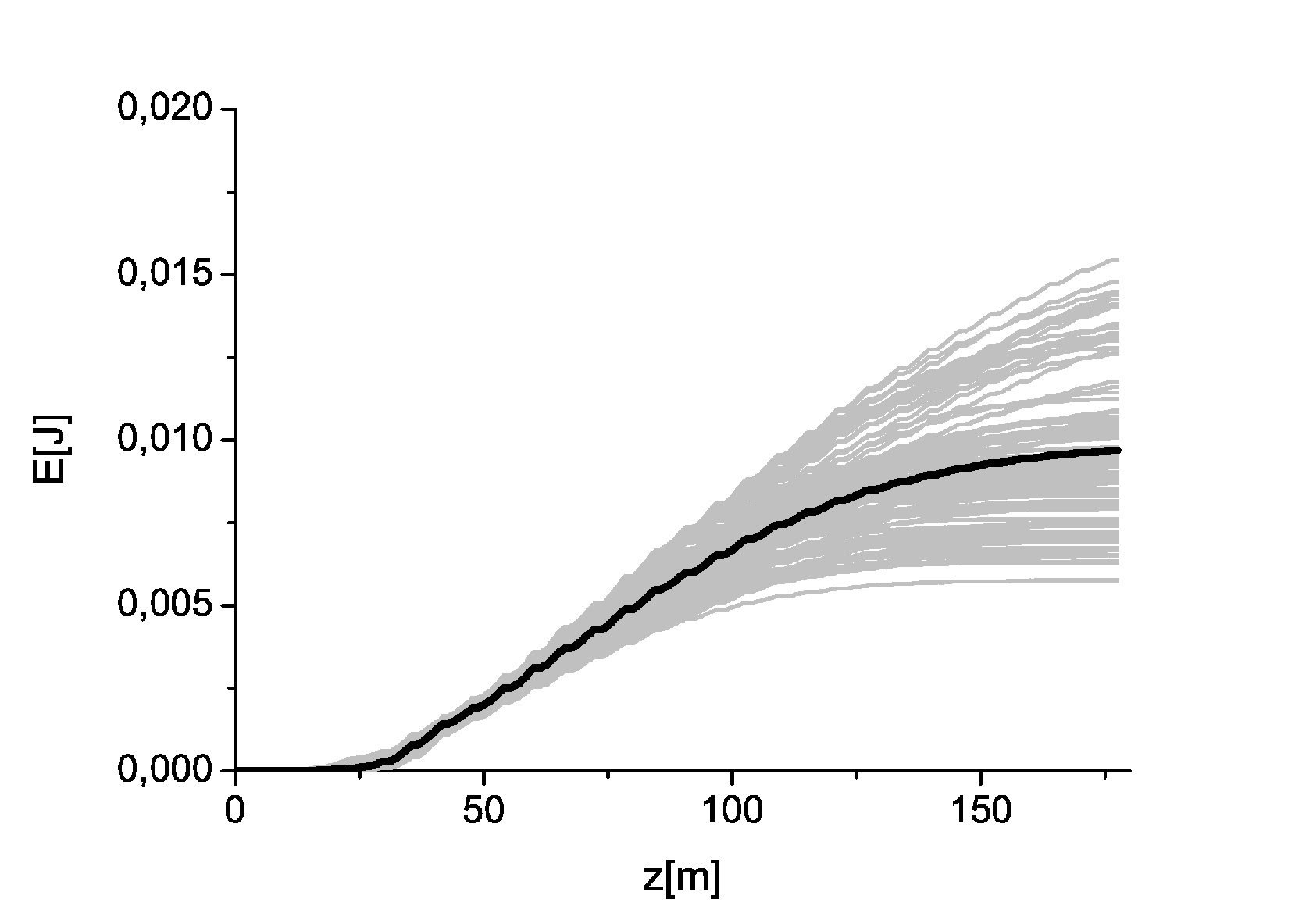}
\includegraphics[width=0.5\textwidth]{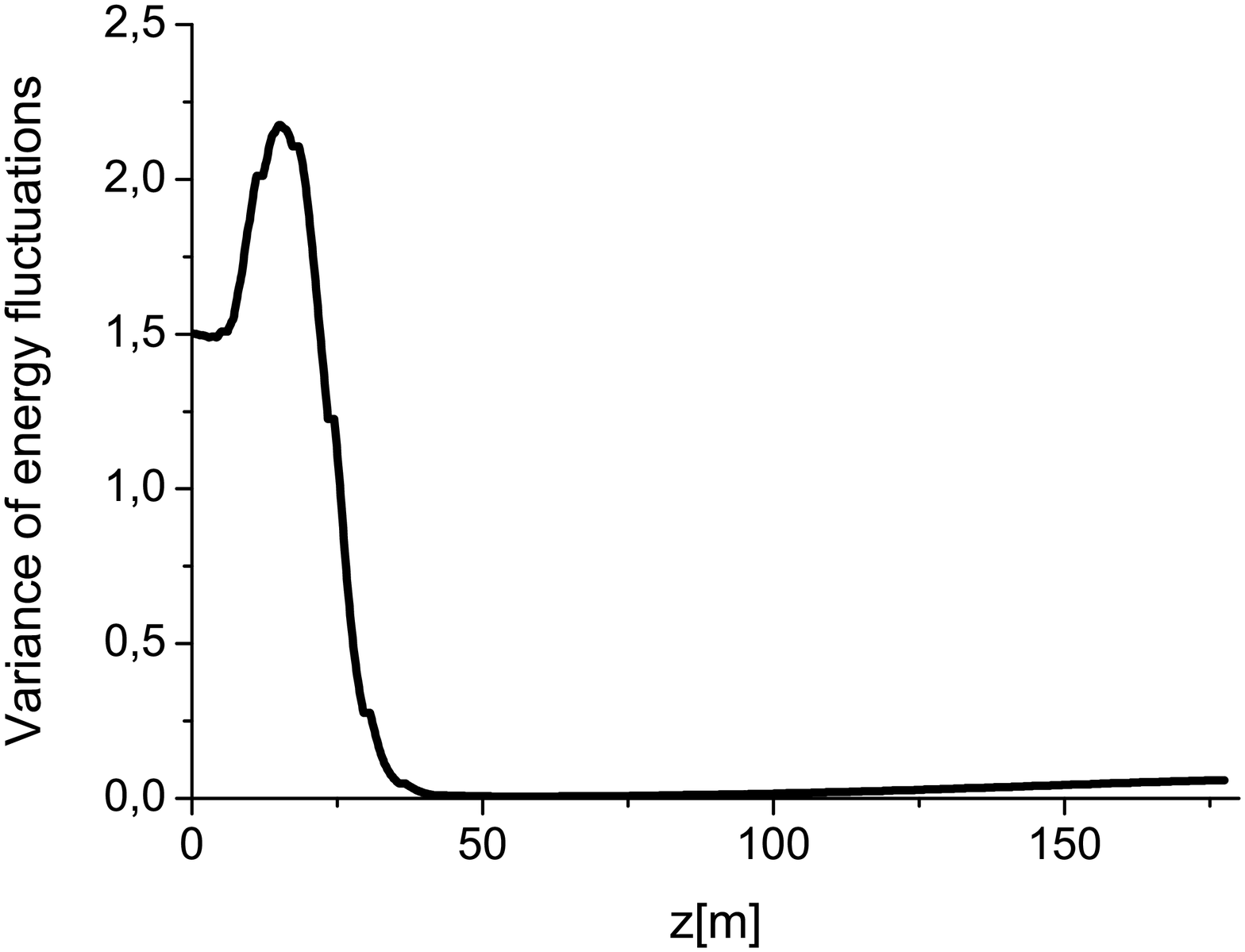}
\caption{Final output. Energy and energy variance of output pulses
for the case $\lambda = 0.5$ nm. In the left plot, grey lines refer
to single shot realizations, the black line refers to the average
over a hundred realizations.} \label{biof24}
\end{figure}

\begin{figure}[tb]
\includegraphics[width=0.5\textwidth]{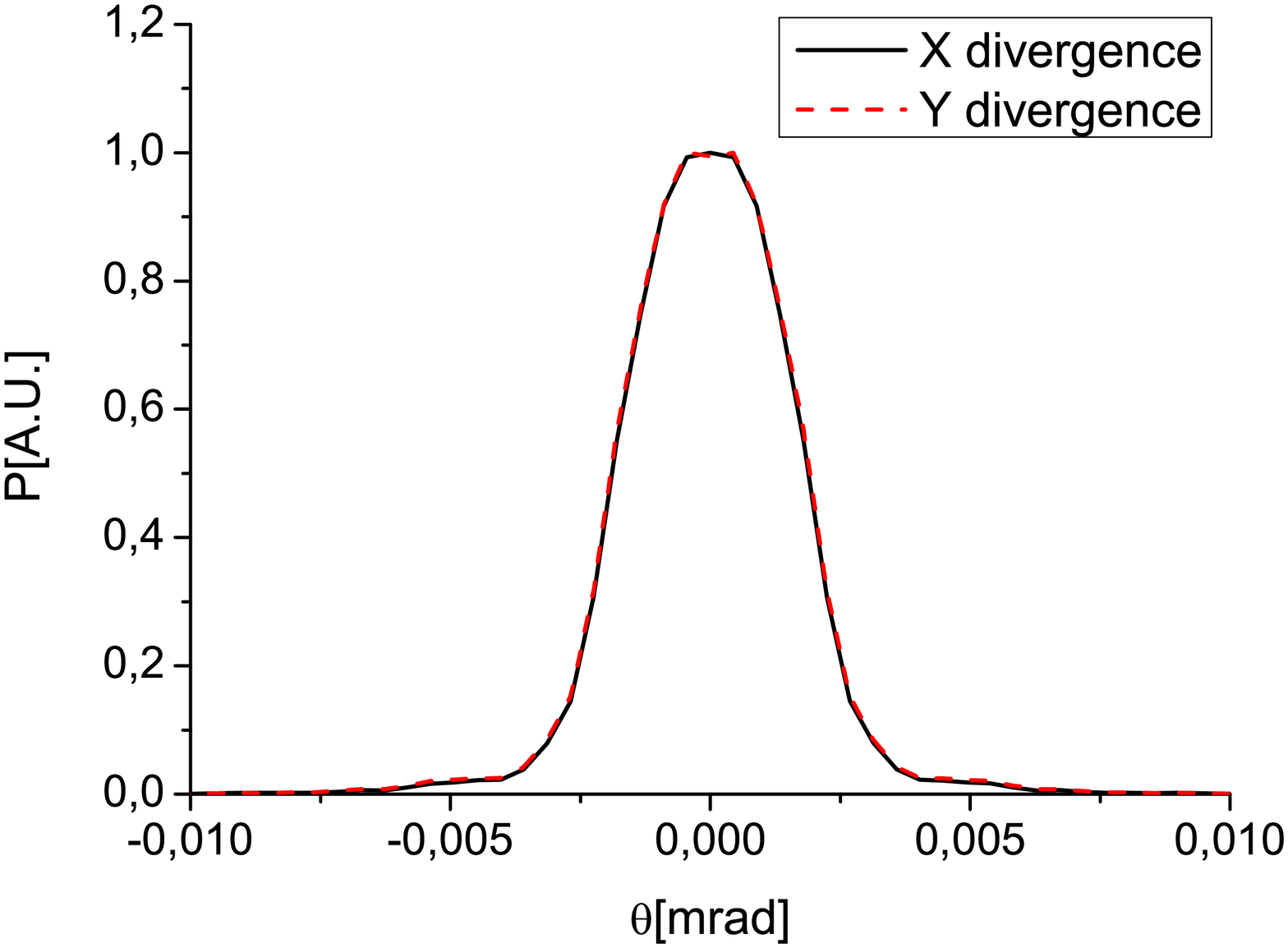}
\includegraphics[width=0.5\textwidth]{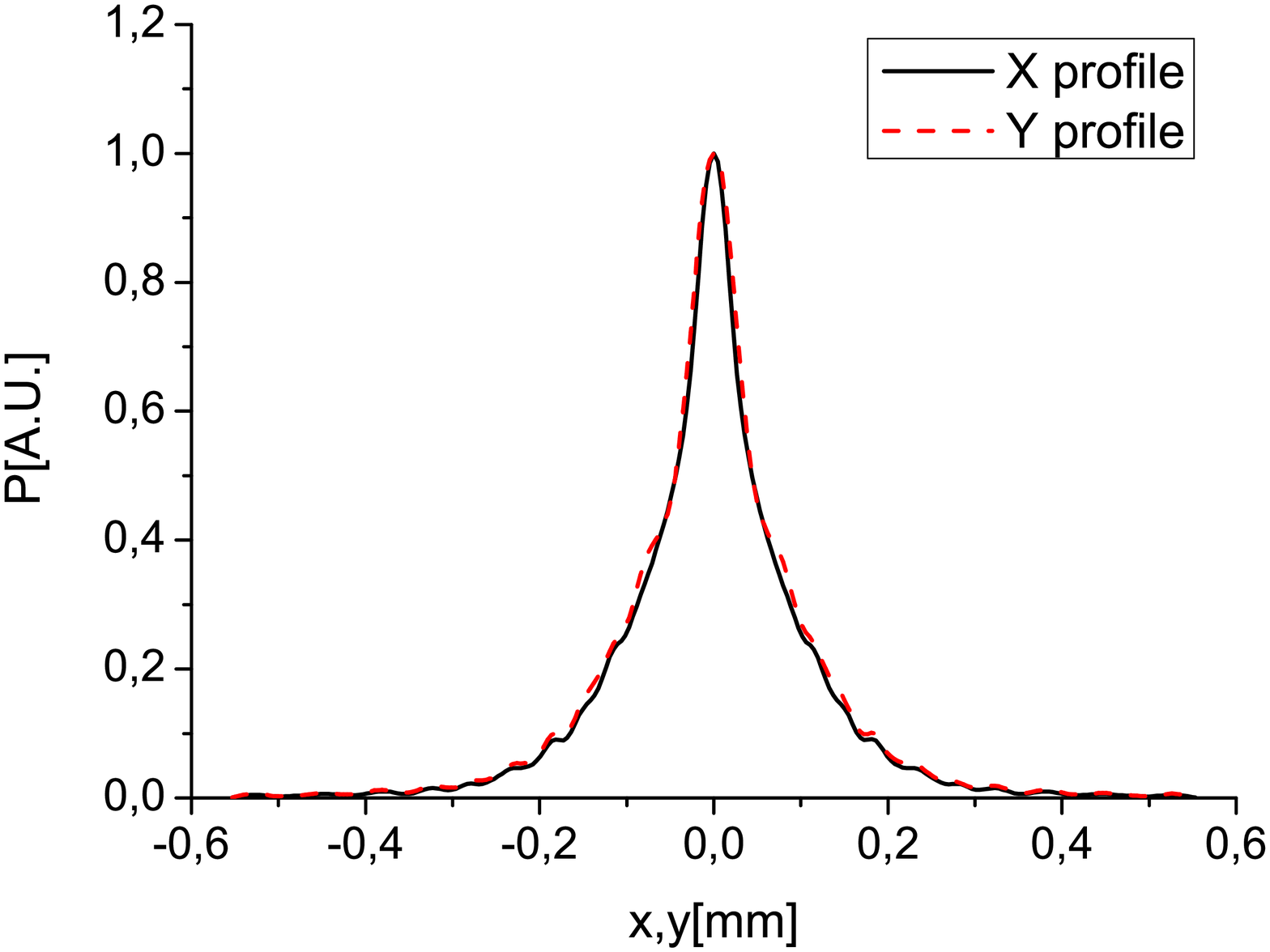}
\caption{Final output. X-ray radiation pulse energy distribution per
unit surface and angular distribution of the X-ray pulse energy at
the exit of output undulator for the case $\lambda = 0.5$ nm.}
\label{biof25}
\end{figure}
The energy of the radiation pulse and the energy variance are shown
in Fig. \ref{biof24} as a function of the position along the
undulator. The divergence and the size of the radiation pulse at the
exit of the final undulator are shown, instead, in Fig.
\ref{biof25}. In order to calculate the size, an average of the
transverse intensity profiles is taken. In order to calculate the
divergence, the spatial Fourier transform of the field is
calculated.

\subsection{Photon energy range between 3 keV and 5 keV}

We now consider generation of radiation in the photon energy between
3 keV and 5 keV, with reference to Fig. \ref{bio3f4}.

\begin{figure}[tb]
\includegraphics[width=0.5\textwidth]{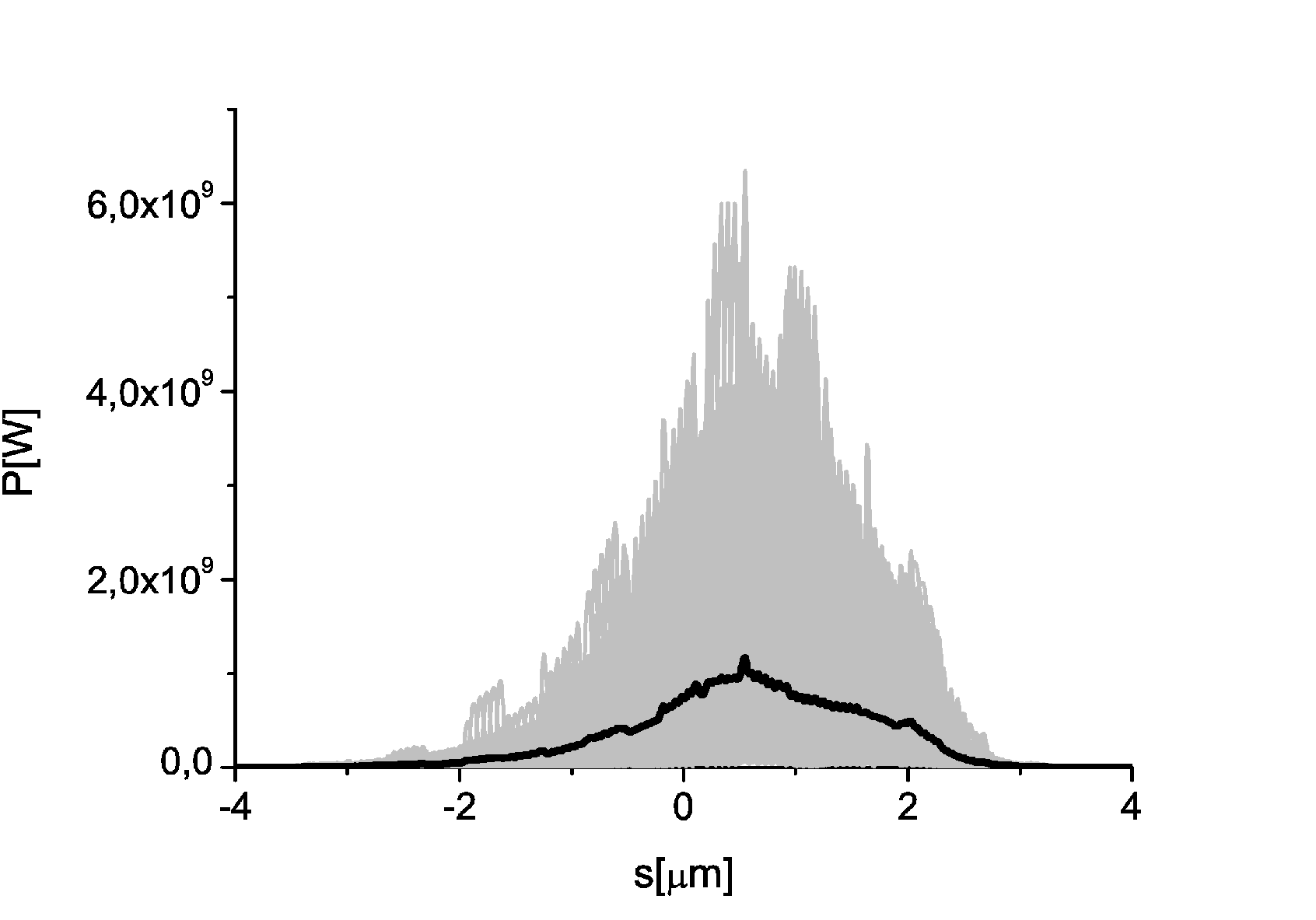}
\includegraphics[width=0.5\textwidth]{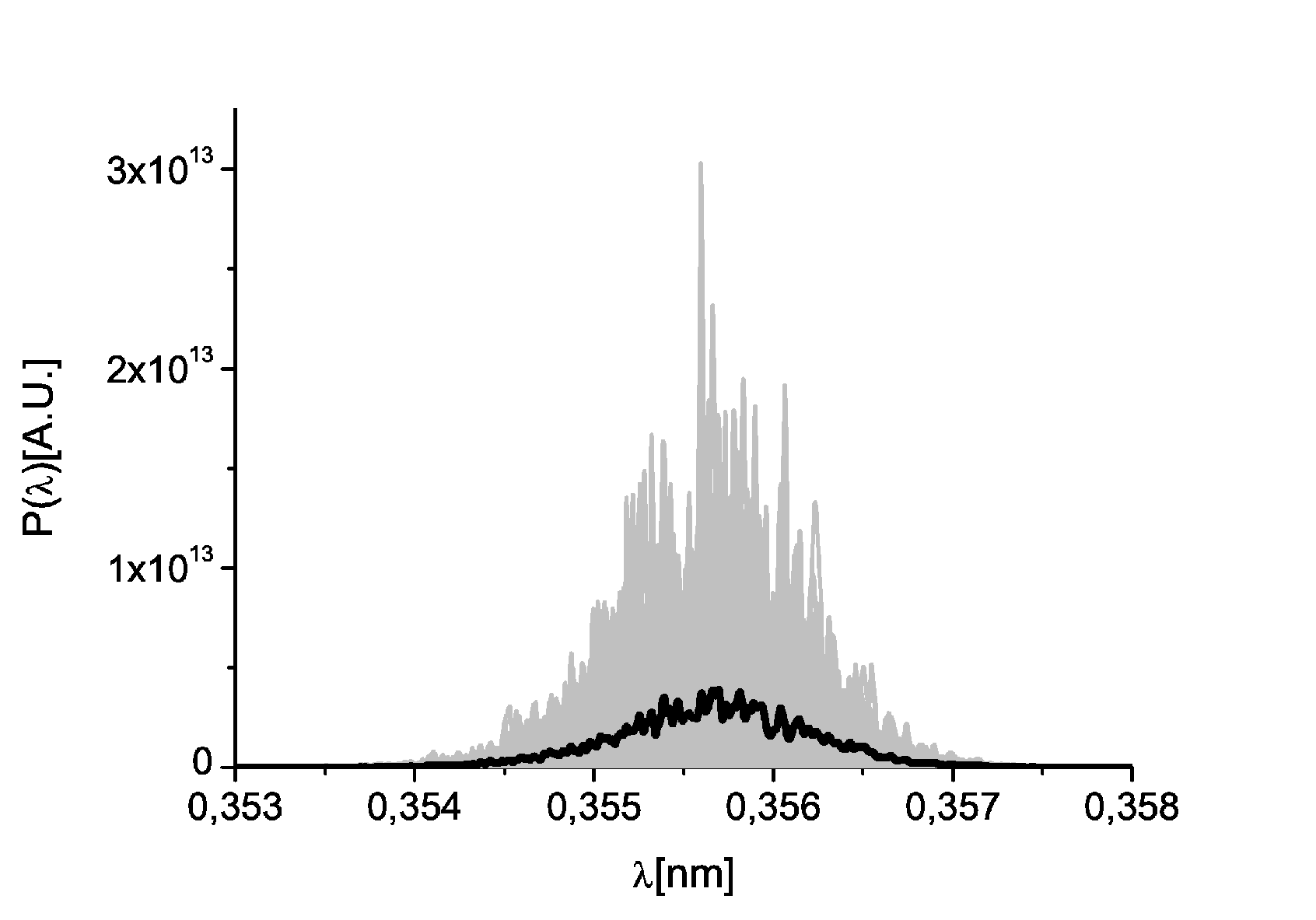}
\caption{Power and spectrum before the second magnetic chicane. Grey
lines refer to single shot realizations, the black line refers to
the average over a hundred realizations.} \label{biof173p5}
\end{figure}
For this mode of operation, the first chicane is switched off, so
that the first part of the undulator effectively consists of 7
uniform cells. We begin our investigation by simulating the SASE
power and spectrum after the first part of the undulator, that is
before the second magnetic chicane in the setup. Results are shown
in Fig. \ref{biof173p5}.

\begin{figure}[tb]
\includegraphics[width=0.5\textwidth]{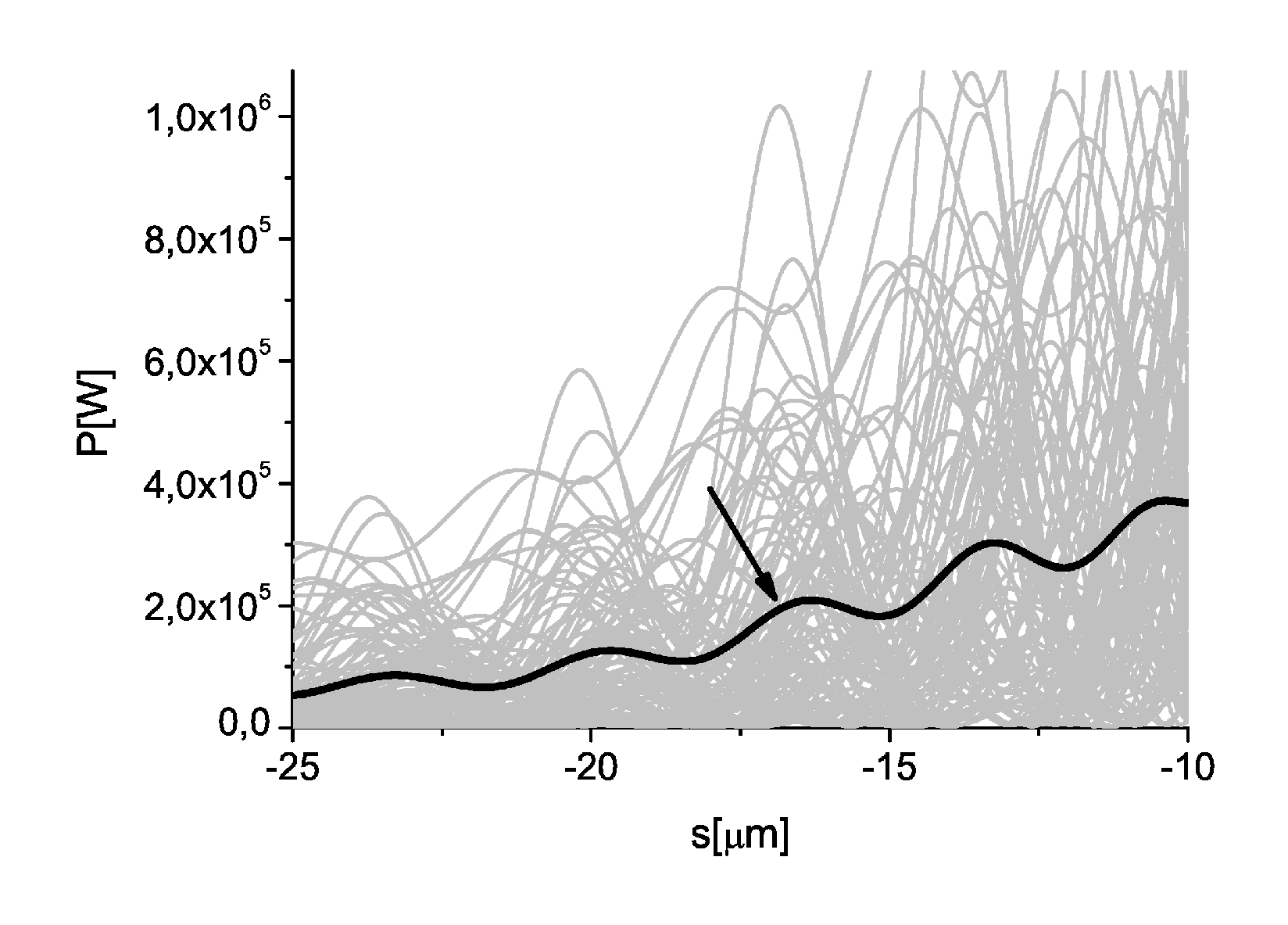}
\includegraphics[width=0.5\textwidth]{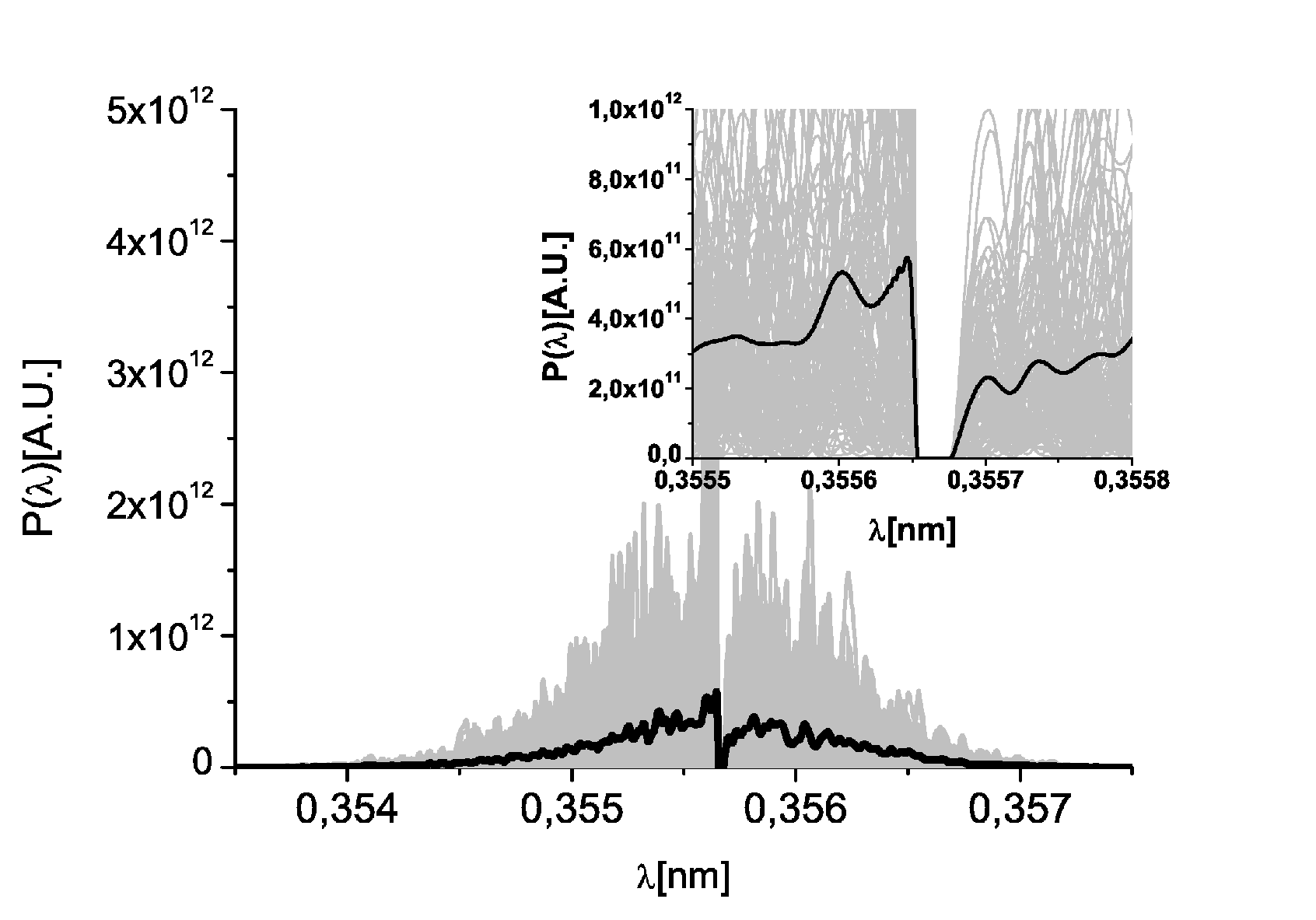}
\caption{Power and spectrum after the single crystal self-seeding
X-ray monochromator. A $100~\mu$m thick diamond crystal in Bragg
transmission geometry ( C(111) reflection, $\sigma$-polarization )
is used. Grey lines refer to single shot realizations, the black
line refers to the average over a hundred realizations. The black
arrow indicates the seeding region.} \label{biof1830p5}
\end{figure}
The second magnetic chicane is switched on, and the single-crystal
X-ray monochromator is set into the photon beam. For the 3 keV - 5
keV energy range we use a $100~\mu$m-thick diamond crystal in Bragg
transmission geometry. In particular, we take advantage of the
C(111) reflection, $\sigma$-polarization. The crystal acts as a
bandstop filter, and the output spectrum is plotted in Fig.
\ref{biof1830p5} (right). Due to the bandstop effect, the signal in
the time domain exhibits a long monochromatic tail, which is used
for seeding, Fig. \ref{biof1830p5} (left). To this purpose, the
electron bunch is slightly delayed by proper tuning of the magnetic
chicane to be superimposed to the seeding signal.

\begin{figure}[tb]
\includegraphics[width=0.5\textwidth]{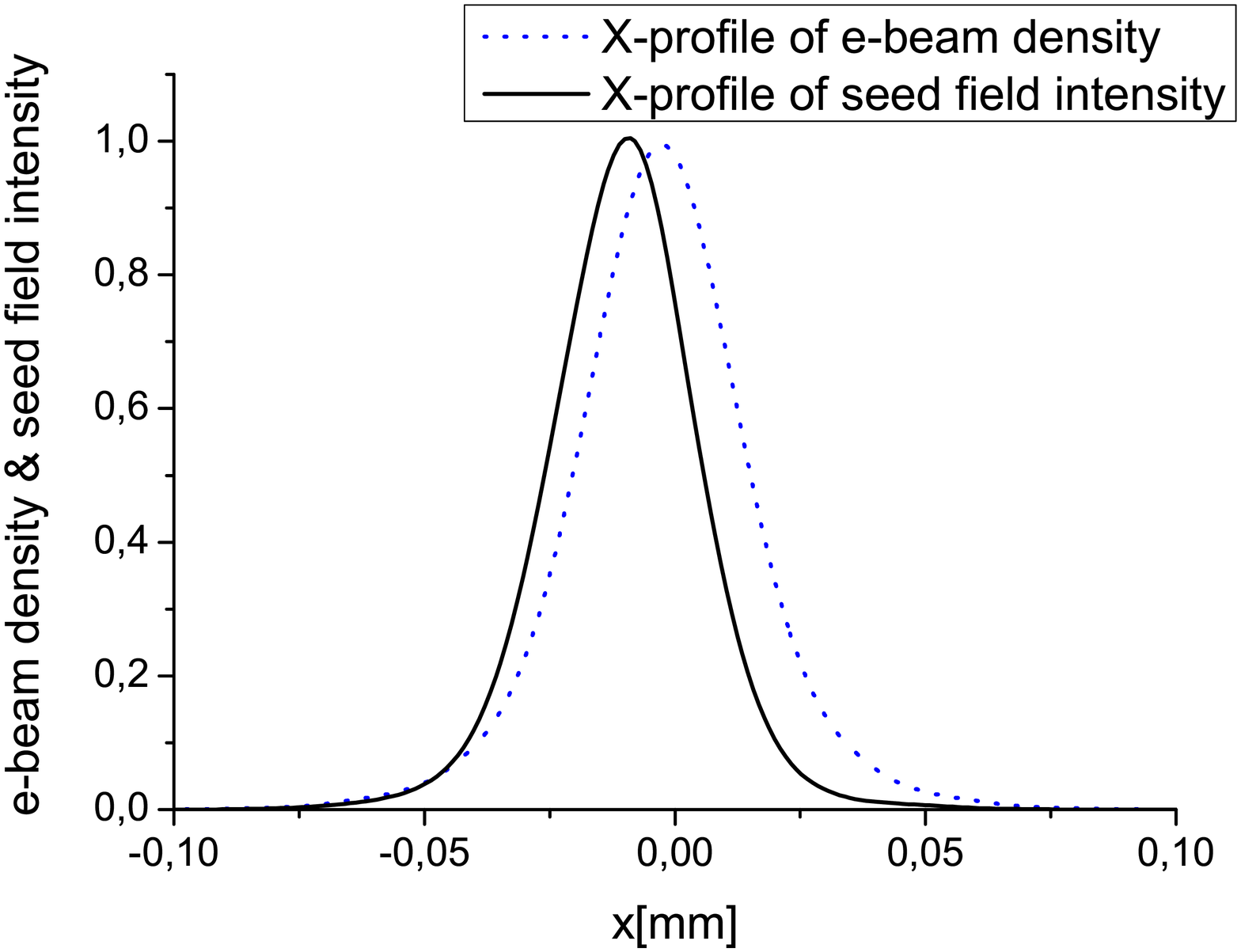}
\includegraphics[width=0.5\textwidth]{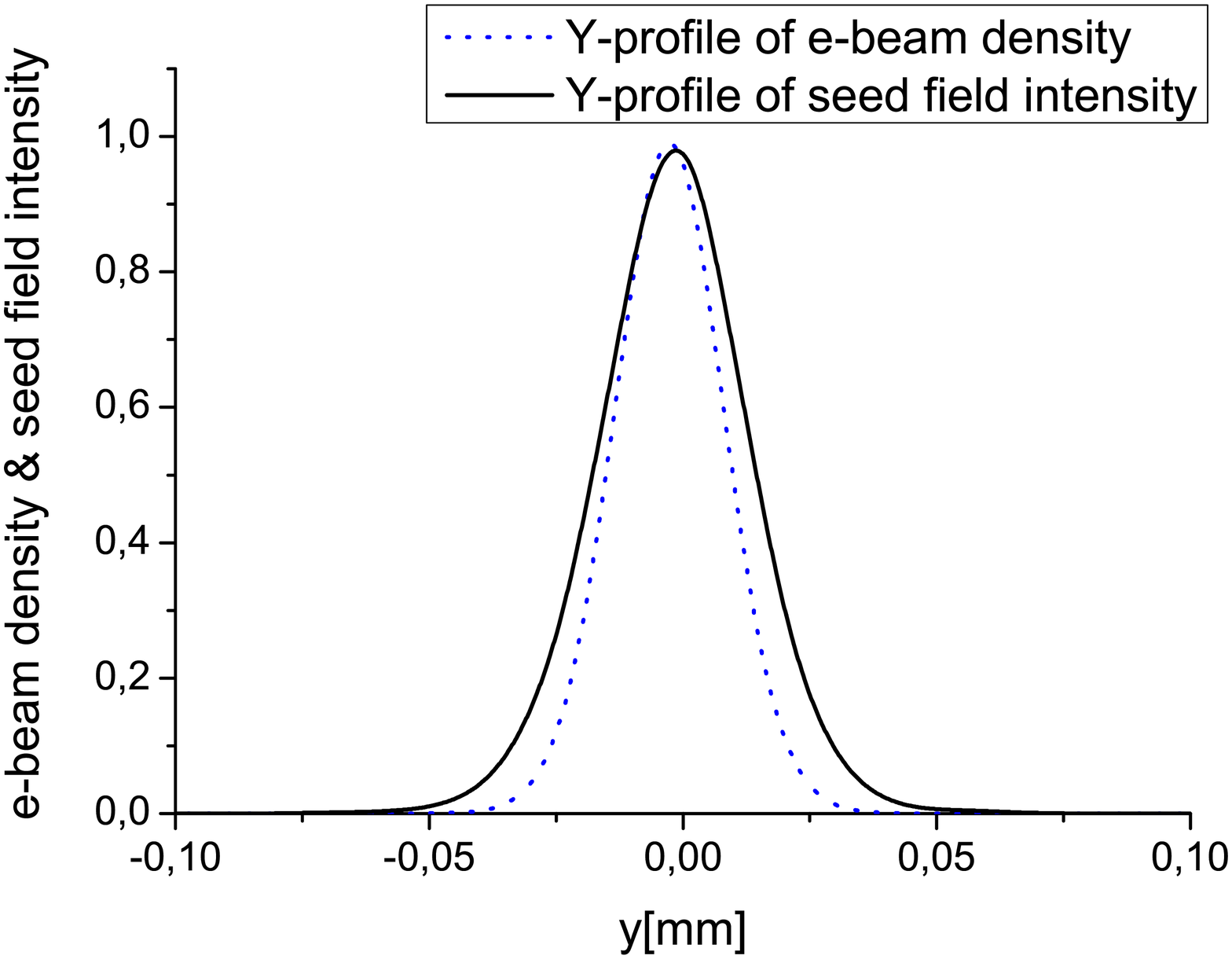}
\caption{Comparison between transverse profile of the seed field
intensity and transverse profile of the electron beam density at the
position used for seeding at $\lambda = 0.36$ nm. The plots refer to
the longitudinal position inside the bunch corresponding to the
maximum current value.} \label{compare}
\end{figure}
It should be remarked that, according to Section \ref{spatio},
spatio-temporal coupling induced by the crystal monochromator should
be accounted for in our study. In our simulations we calculate the
temporal profile of the wake by convolving the incoming radiation
pulse with the impulse response of the crystal. Since the incoming
radiation pulse has a finite length, it follows that the average
wake profile is different compared with the impulse response.

However, as concerns the inclusion of spatiotemporal coupling
effects, we based our analysis on the result discussed in Section
\ref{spatio}. This result was derived under the assumption that the
incoming radiation pulse is a Dirac $\delta$-function. In this case,
our wake in the time domain simply coincides with the impulse
response function of the crystal. The reason for such approximation
can be explained by comparing the typical delay associated with the
impulse response (about 20 $\mu$m) with the pulse duration (about 3
$\mu$m). From this comparison follows that we can neglect the length
of the incoming pulse with accuracy of about $10\% - 20\%$.

As already found in Section \ref{spatio}, the beam spot size is
independent of time, but the beam central position changes as the
pulse evolves in time. The transverse dependence of the electric
field, in fact, obeys Eq. (\ref{Eoutxt}). We account for this effect
automatically in our simulations. A comparison between the
transverse field profile used for seeding without accounting for the
spatio-temporal coupling is shown in Fig. \ref{compare}. This
feature is accounted for in calculations.

\begin{figure}[tb]
\includegraphics[width=0.5\textwidth]{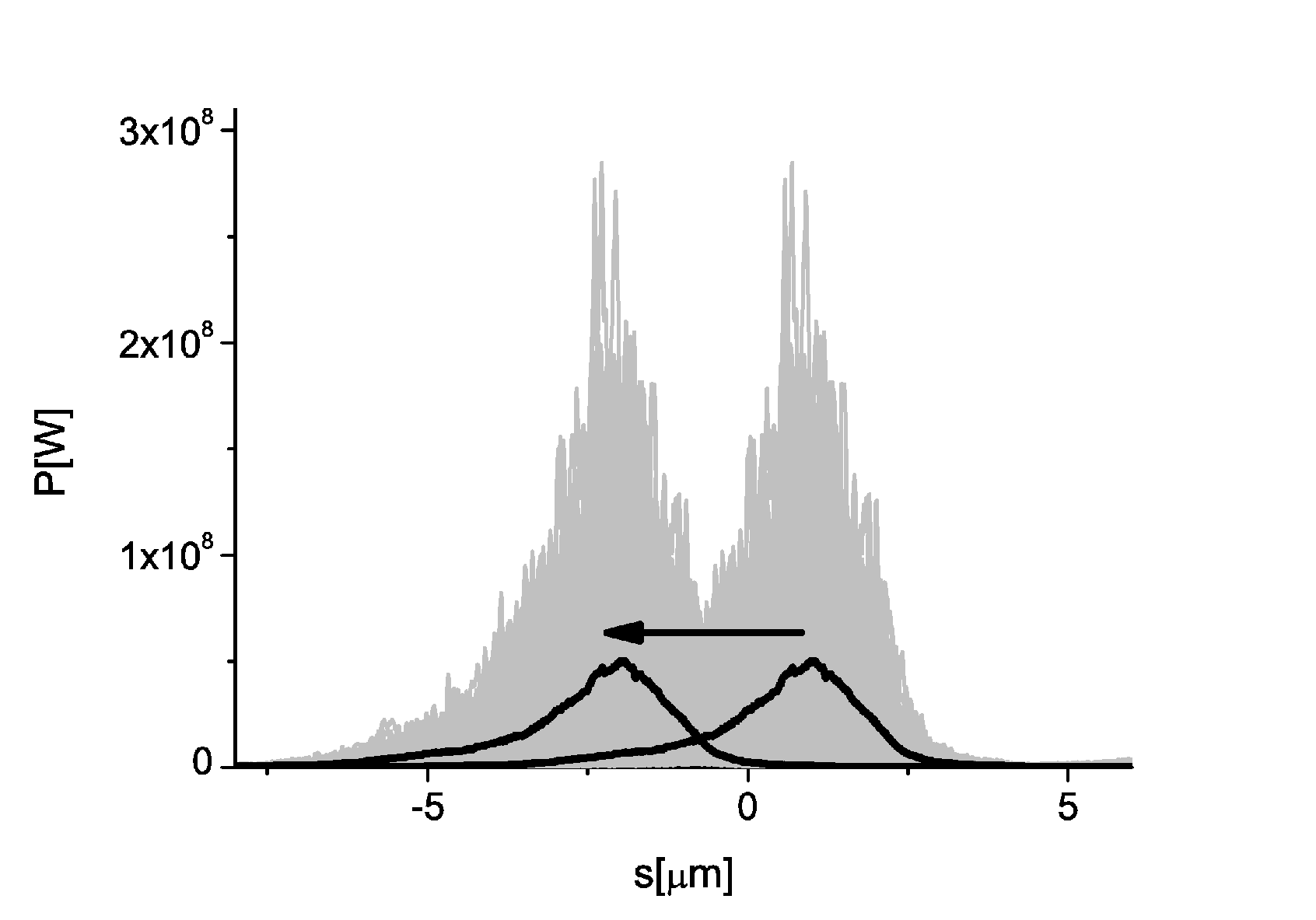}
\includegraphics[width=0.5\textwidth]{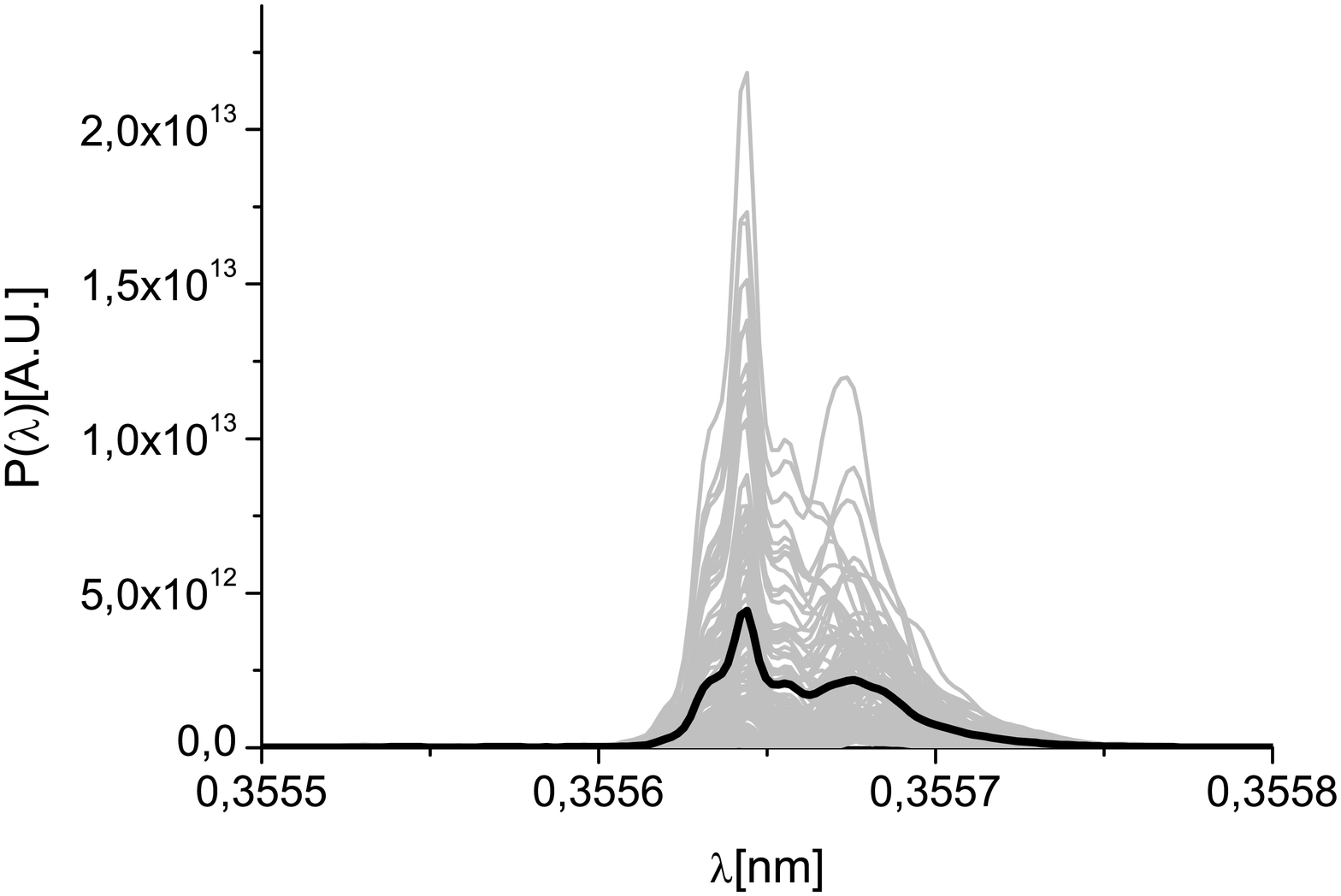}
\caption{Power and spectrum after the third chicane equipped with
the X-ray optical delay line, delaying the radiation pulse with
respect to the electron bunch. Grey lines refer to single shot
realizations, the black line refers to the average over a hundred
realizations.} \label{biof2030p5}
\end{figure}
Following the seeding setup, the electron bunch amplifies the seed
in the following 4 undulator cells. After that, a third chicane is
used to allow for the installation of an x-ray optical delay line,
which retards the radiation pulse with respect to the electron
bunch. The power and spectrum of the radiation pulse after the
optical delay line are shown in Fig. \ref{biof2030p5}, where the
combined effect of the optical delay is illustrated.

\begin{figure}[tb]
\includegraphics[width=0.5\textwidth]{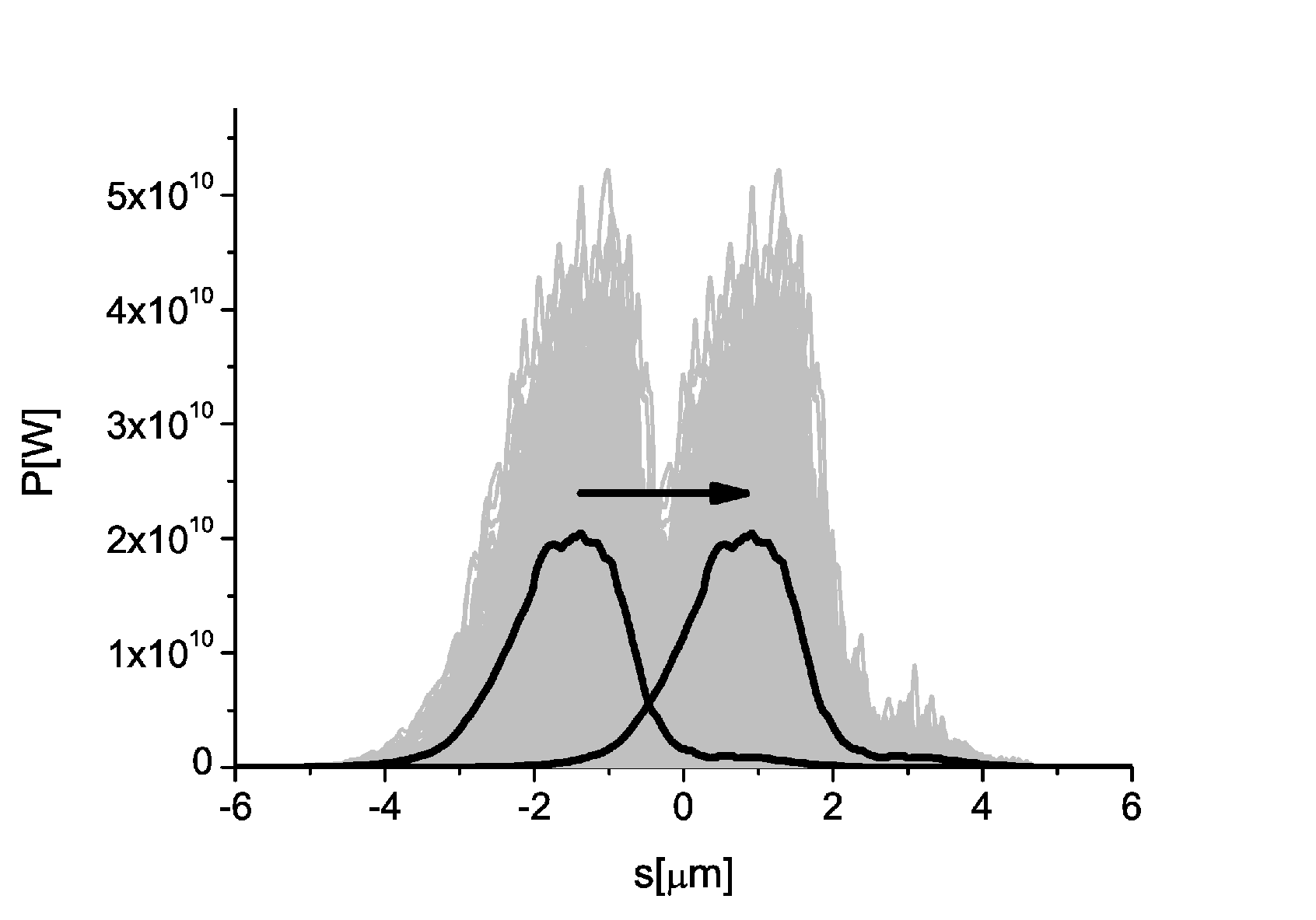}
\includegraphics[width=0.5\textwidth]{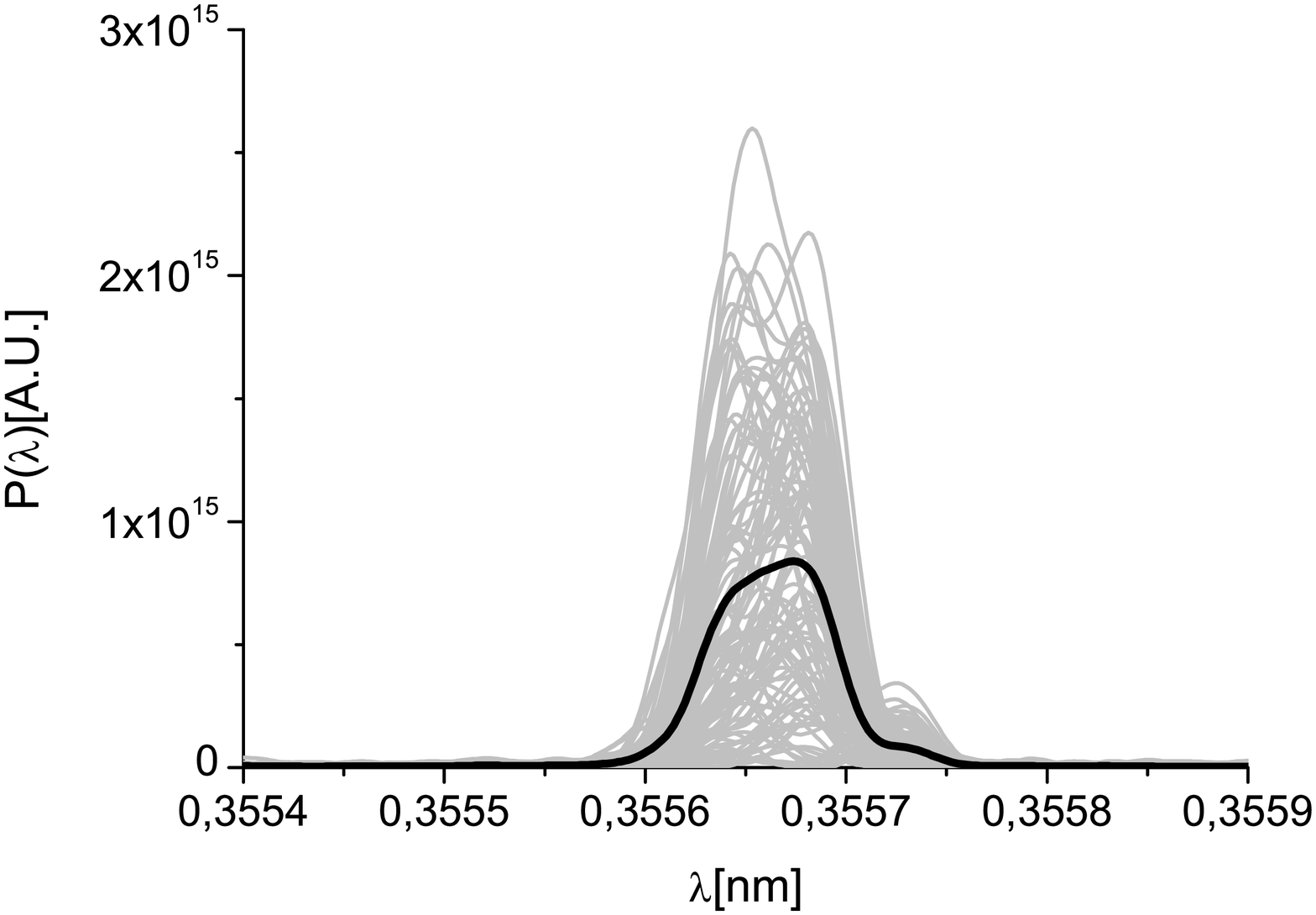}
\caption{Power and spectrum  after the last magnetic chicane. Grey
lines refer to single shot realizations, the black line refers to
the average over a hundred realizations.} \label{biof213p5}
\end{figure}
Due to the presence of the optical delay, only part of the electron
beam is used to further amplify the radiation pulse in the following
6 undulator cells. The electron beam part which has not lased is
fresh, and can be used for further lasing. In order to do so, after
amplification, the electron beam passes through the final magnetic
chicane, which delays the electron beam. The power and spectrum of
the radiation pulse after the last magnetic chicane are shown in
Fig. \ref{biof213p5}. By delaying the electron bunch, the magnetic
chicane effectively shifts forward the photon beam with respect to
the electron beam. Tunability of such shift allows the selection of
different photon pulse lengths. Moreover, an additional advantage of
brought by the use of a fresh bunch technique in this mode of
operation is the suppression of the intensity fluctuations of the
seed down to $40 \%$  in the nonlinear regime before the last
chicane. This suppression of fluctuations is useful in connection
with the application of the tapering technique in the last part of
the undulator. Fluctuations of the seed in the linear regime are, at
variance, close to $100 \%$.

\begin{figure}[tb]
\begin{center}
\includegraphics[width=0.5\textwidth]{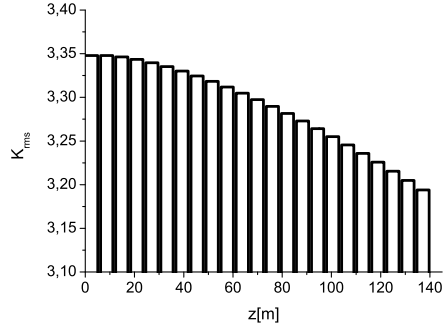}
\end{center}
\caption{Tapering law for the case $\lambda = 0.36$ nm.}
\label{biof223p5}
\end{figure}
The last part of the undulator is composed by $23$ cells. It is
partly tapered post-saturation, to increase the region where
electrons and radiation interact properly to the advantage of the
radiation pulse. Tapering is implemented by changing the $K$
parameter of the undulator segment by segment according to Fig.
\ref{biof223p5}. The tapering law used in this work has been
implemented on an empirical basis.

\begin{figure}[tb]
\includegraphics[width=0.5\textwidth]{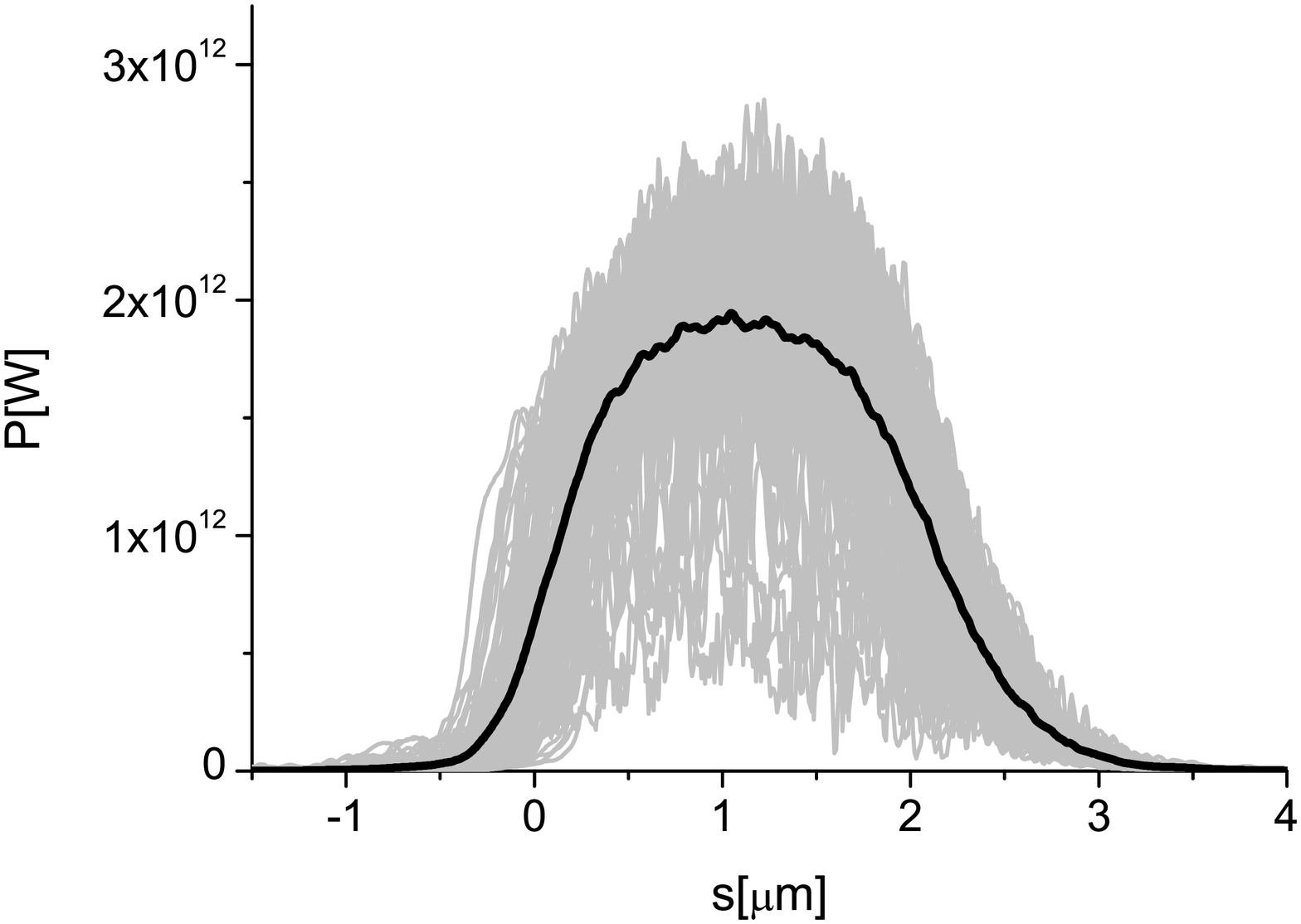}
\includegraphics[width=0.5\textwidth]{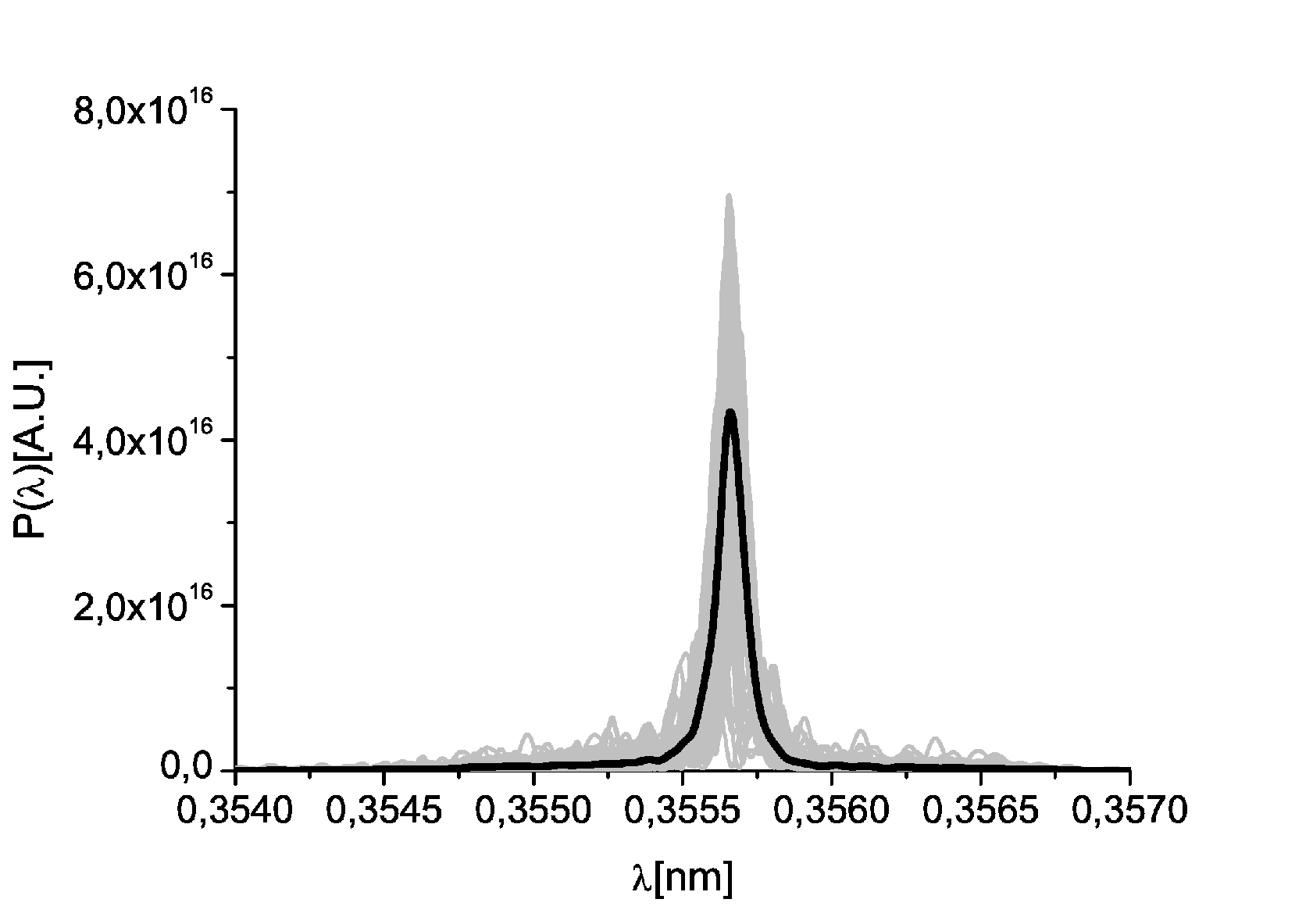}
\caption{Final output. Power and spectrum at the third harmonic
after tapering. Grey lines refer to single shot realizations, the
black line refers to the average over a hundred realizations.}
\label{biof233p5}
\end{figure}
The use of tapering together with monochromatic radiation is
particularly effective, since the electron beam does not experience
brisk changes of the ponderomotive potential during the slippage
process. The final output is presented in Fig. \ref{biof233p5} in
terms of power and spectrum. As one can see, simulations indicate an
output power of about $1.5$ TW.

\begin{figure}[tb]
\includegraphics[width=0.5\textwidth]{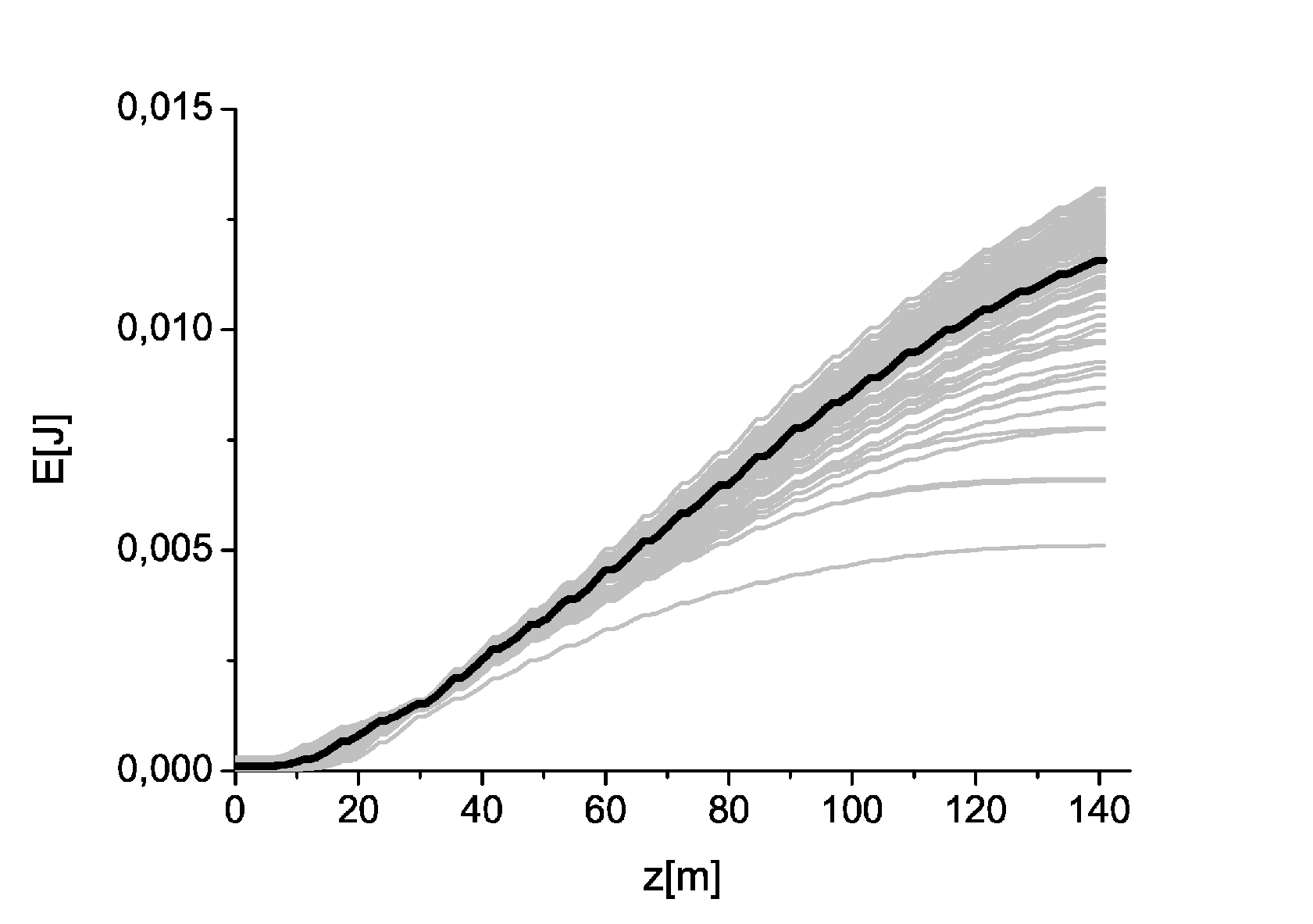}
\includegraphics[width=0.5\textwidth]{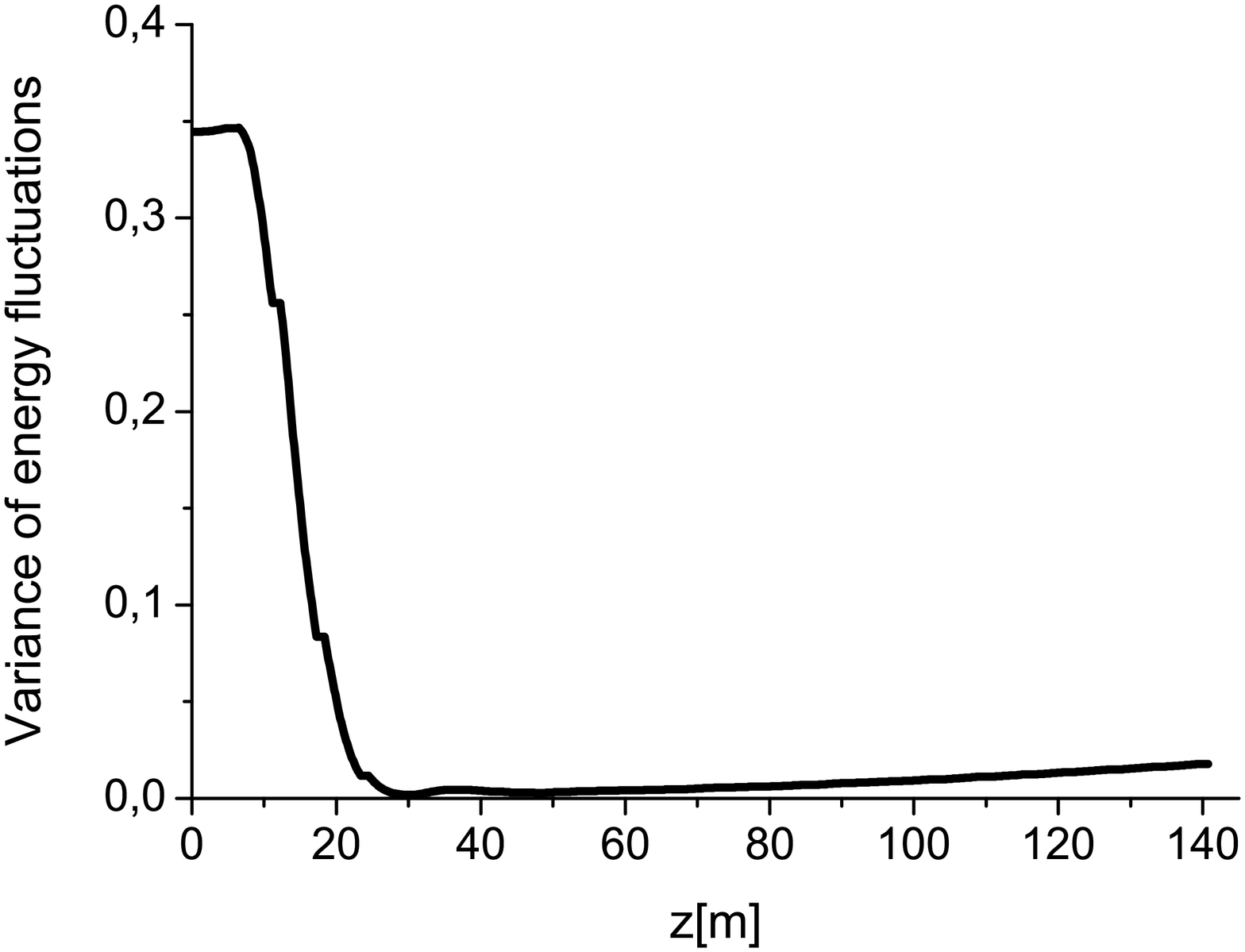}
\caption{Final output. Energy and energy variance of output pulses
for the case $\lambda = 0.36$ nm. In the left plot, grey lines refer
to single shot realizations, the black line refers to the average
over a hundred realizations.} \label{biof243p5}
\end{figure}

\begin{figure}[tb]
\includegraphics[width=0.5\textwidth]{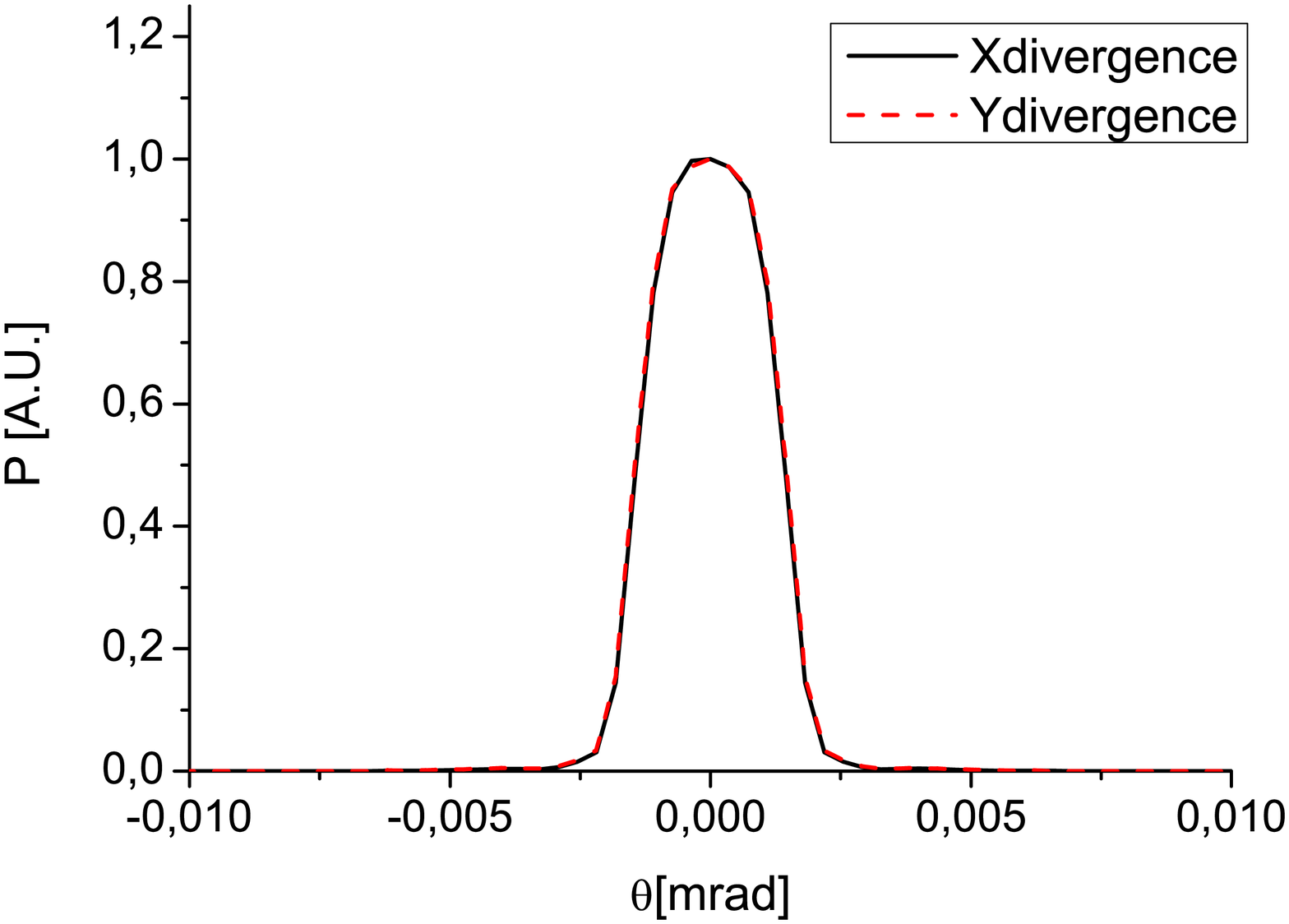}
\includegraphics[width=0.5\textwidth]{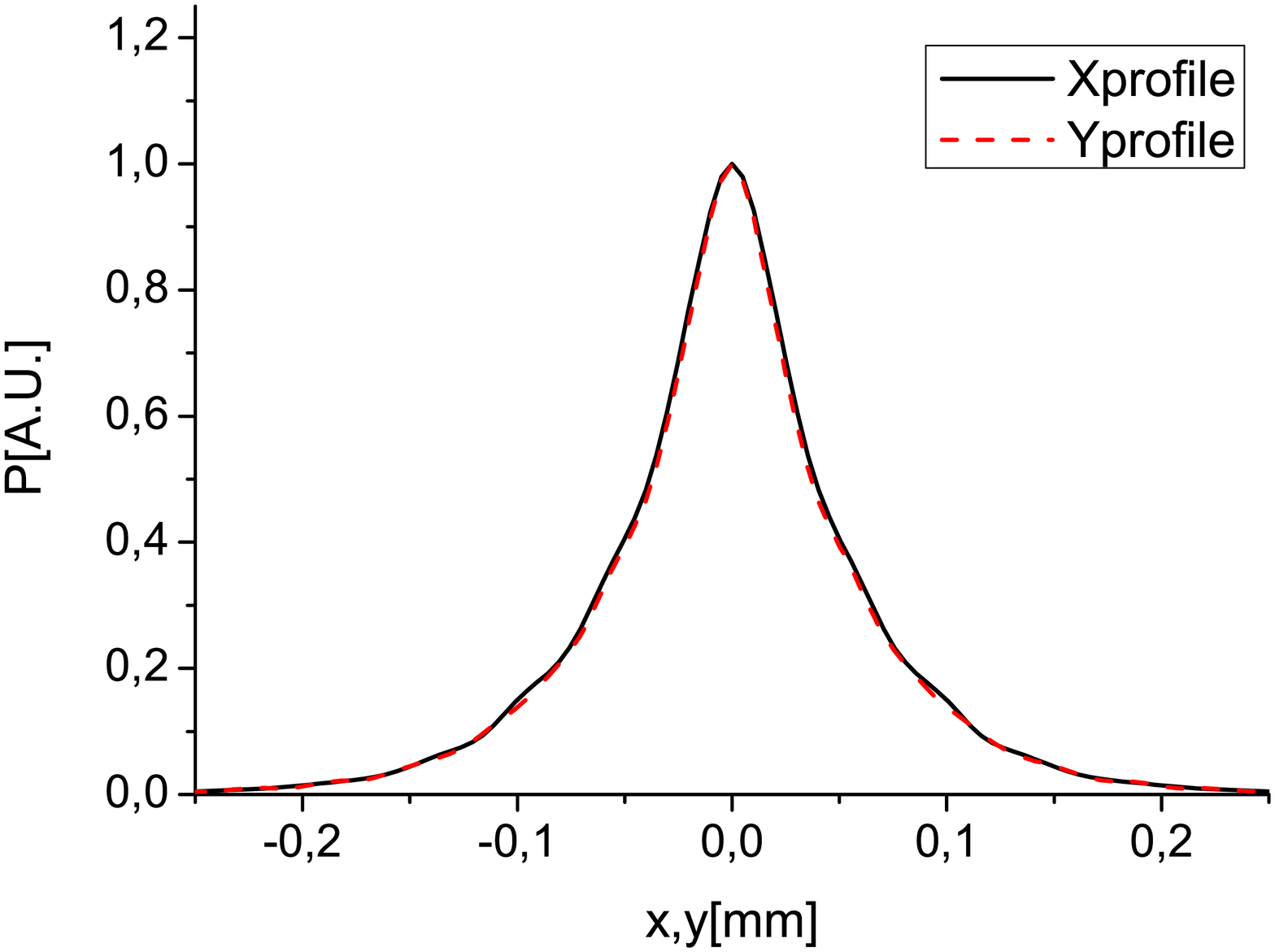}
\caption{Final output. X-ray radiation pulse energy distribution per
unit surface and angular distribution of the X-ray pulse energy at
the exit of output undulator for the case $\lambda = 0.36$ nm.}
\label{biof2530p5}
\end{figure}
The energy of the radiation pulse and the energy variance are shown
in Fig. \ref{biof243p5} as a function of the position along the
undulator. The divergence and the size of the radiation pulse at the
exit of the final undulator are shown, instead, in Fig.
\ref{biof2530p5}. In order to calculate the size, an average of the
transverse intensity profiles is taken. In order to calculate the
divergence, the spatial Fourier transform of the field is
calculated.

\subsection{Photon energy range between 5 keV and 7 keV}

Operation in the photon range between 5 keV and 7 keV will be
possible by configuring the bio-imaging beamline as in Fig.
\ref{bio3f5}. The configuration is very similar to the case for the
range between 3 keV and 5 keV. The only difference is that the
C(220) reflection is used instead of the C(111).

\begin{figure}[tb]
\includegraphics[width=0.5\textwidth]{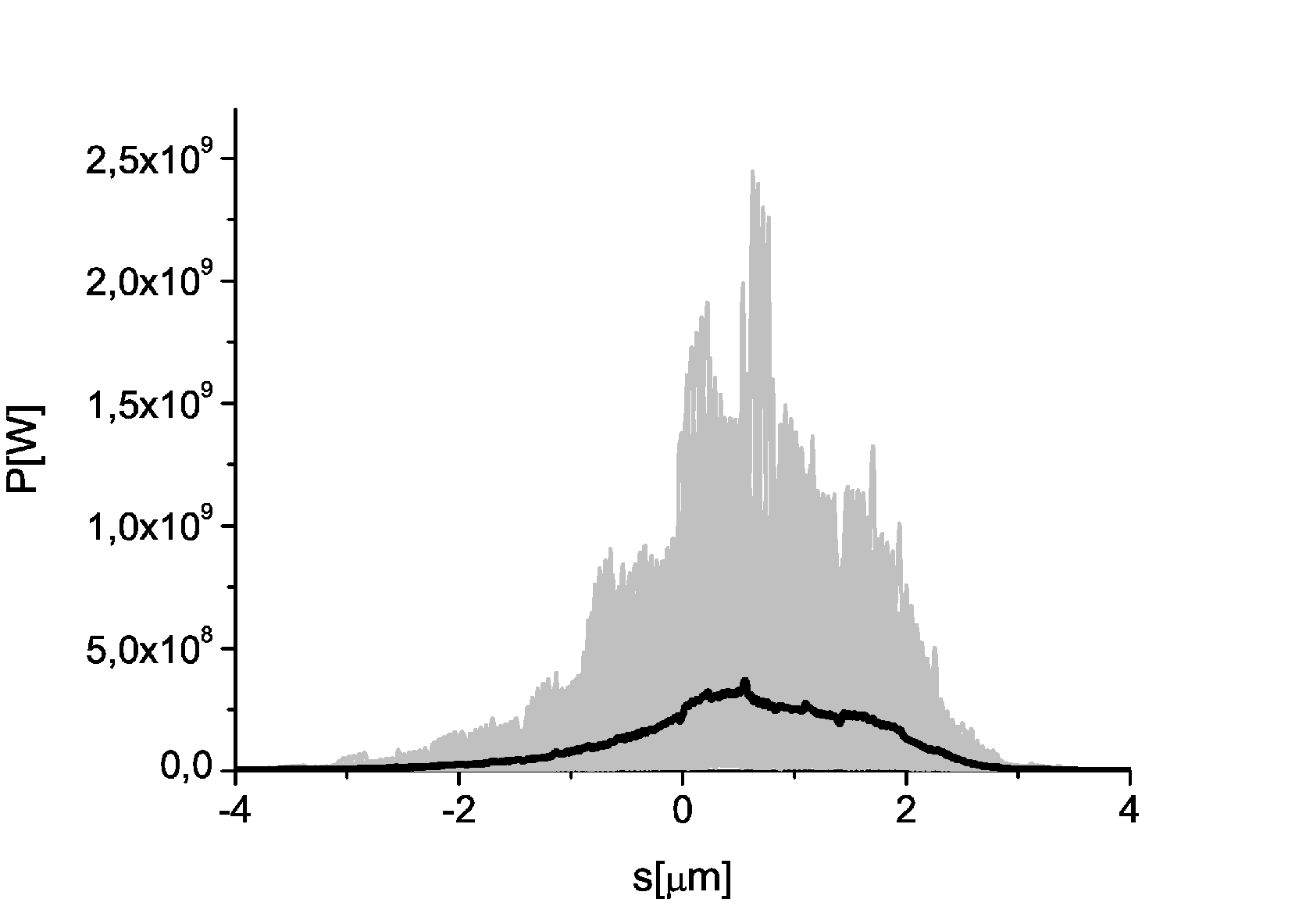}
\includegraphics[width=0.5\textwidth]{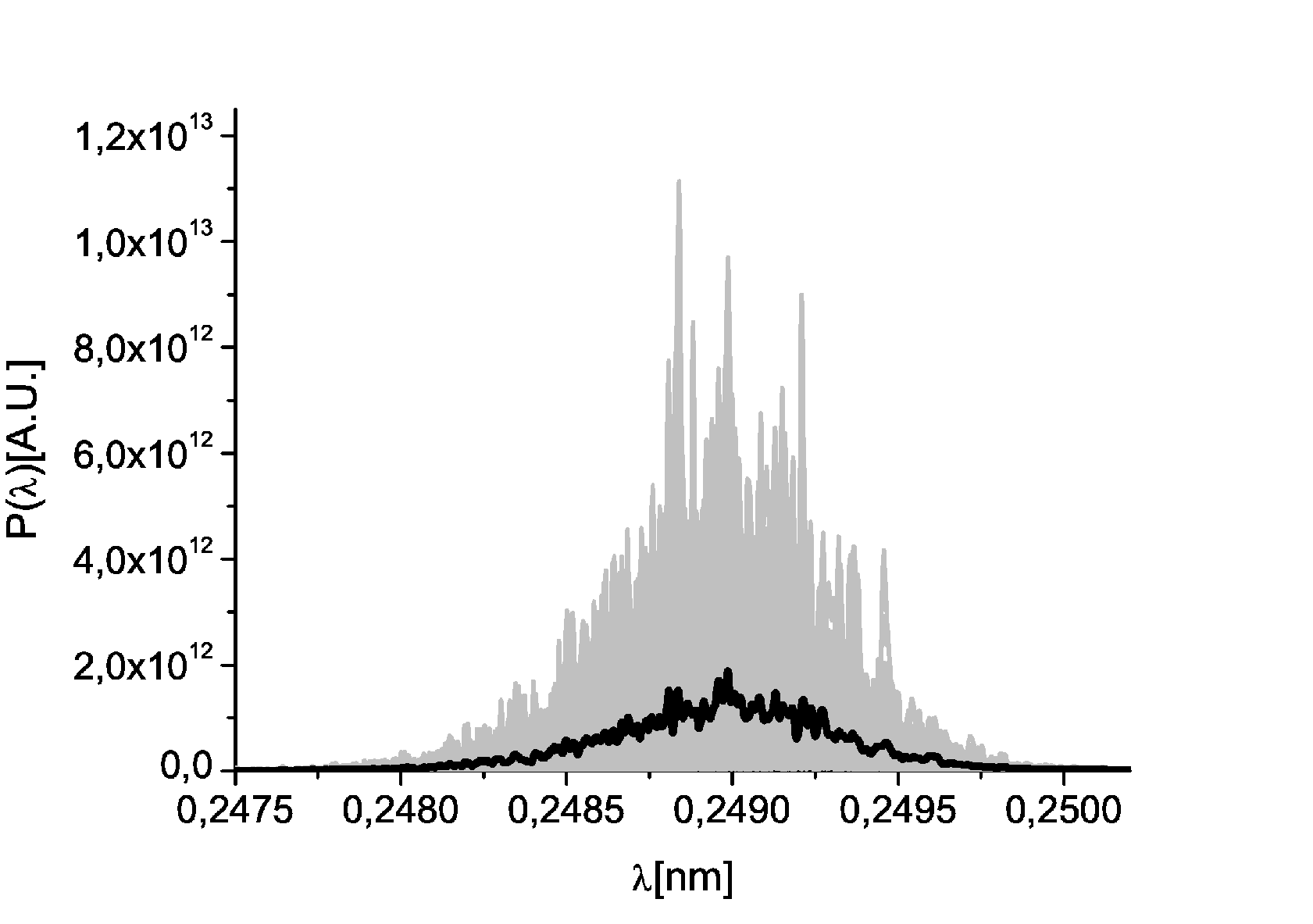}
\caption{Power and spectrum before the second magnetic chicane. Grey
lines refer to single shot realizations, the black line refers to
the average over a hundred realizations.} \label{biof175}
\end{figure}
As before,  the first chicane is switched off, so that the first
part of the undulator effectively consists of 7 uniform cells. We
begin our investigation by simulating the SASE power and spectrum
after the first part of the undulator, that is before the second
magnetic chicane in the setup. Results are shown in Fig.
\ref{biof175}.

\begin{figure}[tb]
\includegraphics[width=0.5\textwidth]{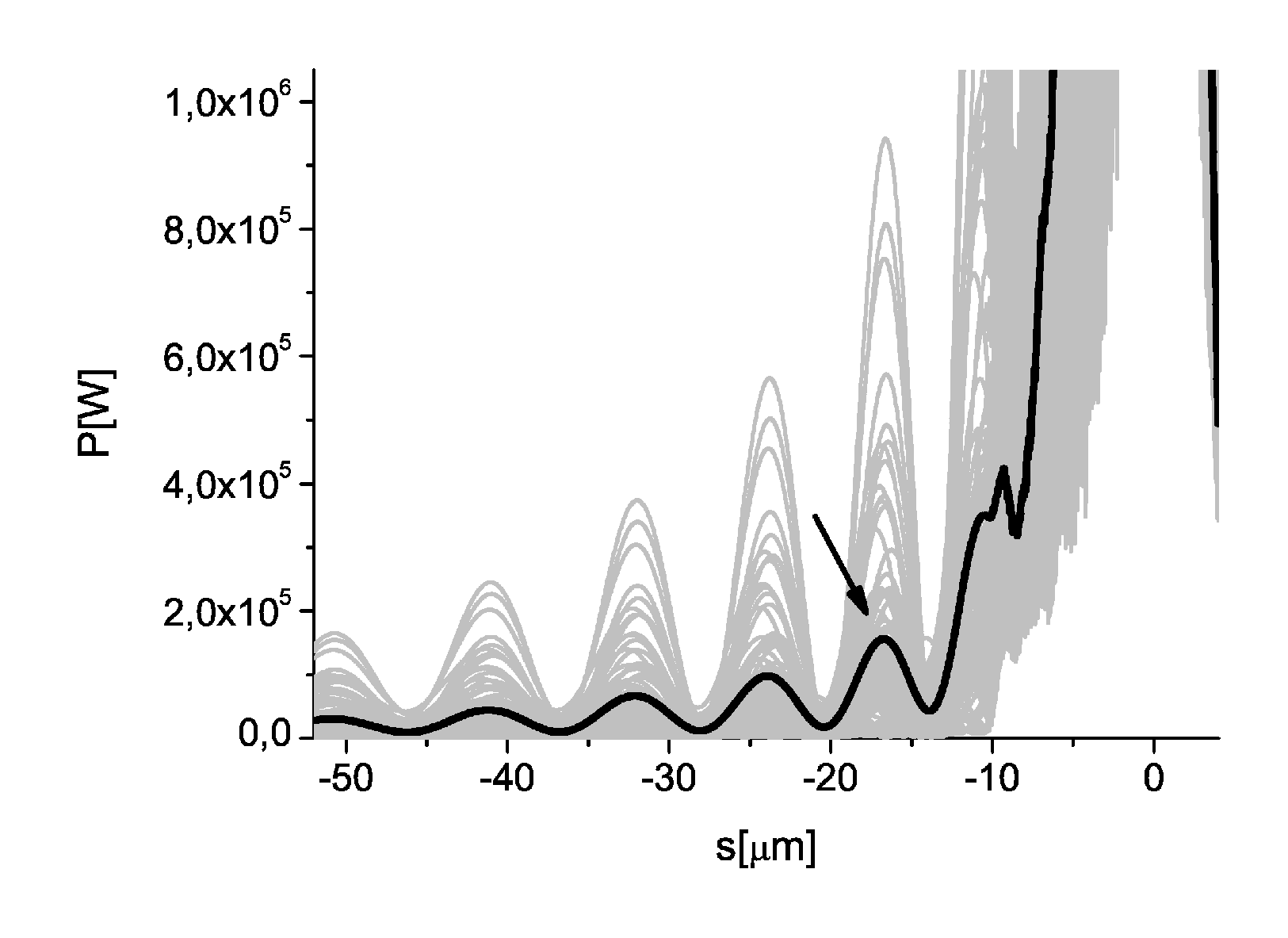}
\includegraphics[width=0.5\textwidth]{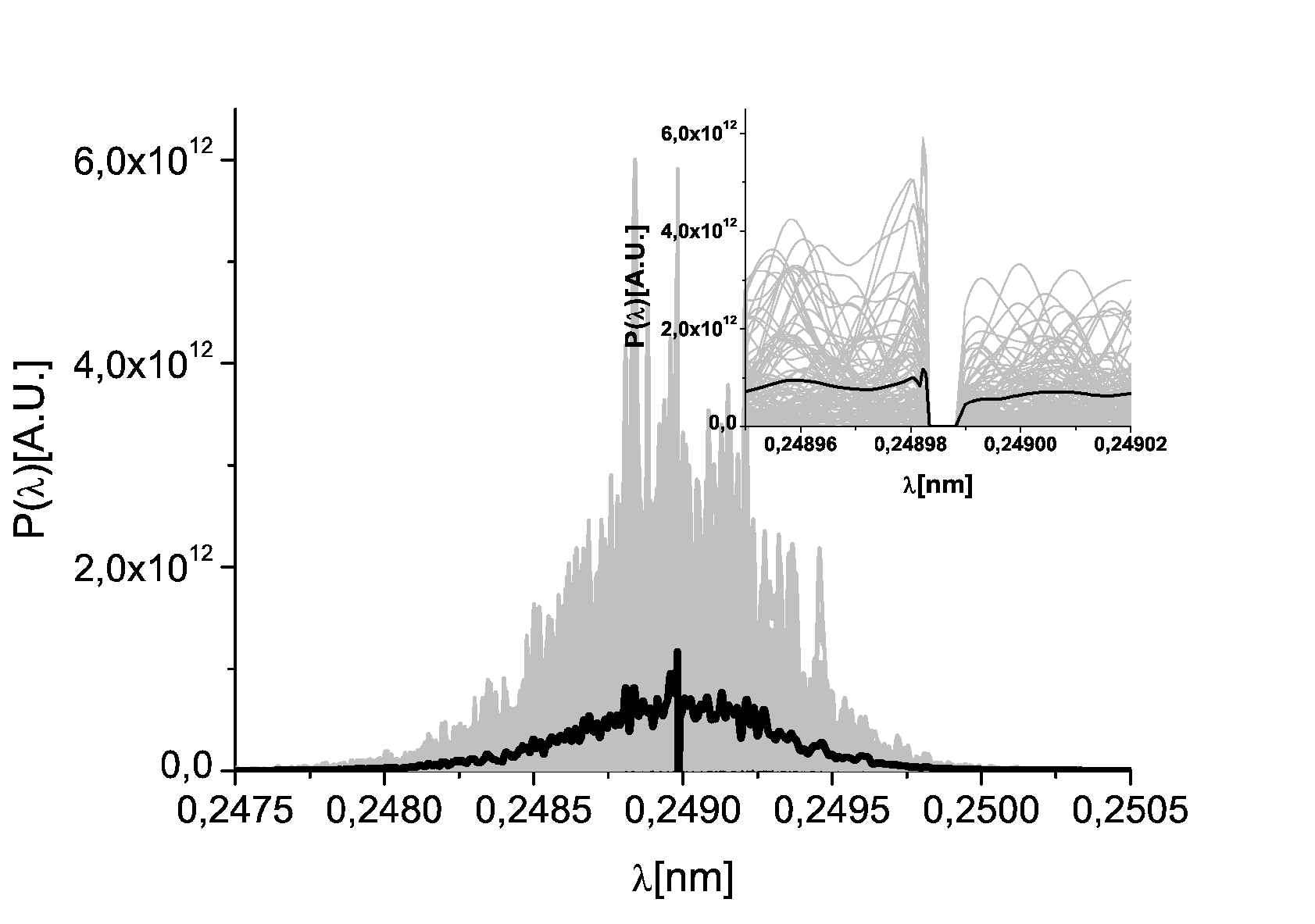}
\caption{Power and spectrum after the single crystal self-seeding
X-ray monochromator. A $100~\mu$m thick diamond crystal in Bragg
transmission geometry ( C(220) reflection, $\sigma$-polarization )
is used. Grey lines refer to single shot realizations, the black
line refers to the average over a hundred realizations. The black
arrow indicates the seeding region.} \label{biof183p5}
\end{figure}
The second magnetic chicane is switched on, and the single-crystal
X-ray monochromator is set into the photon beam. For the 5 keV - 7
keV energy range we use a $100~\mu$m-thick diamond crystal in Bragg
transmission geometry. In particular, we take advantage of the
C(220) reflection, $\sigma$-polarization, Fig. \ref{biof183p5}.

\begin{figure}[tb]
\includegraphics[width=0.5\textwidth]{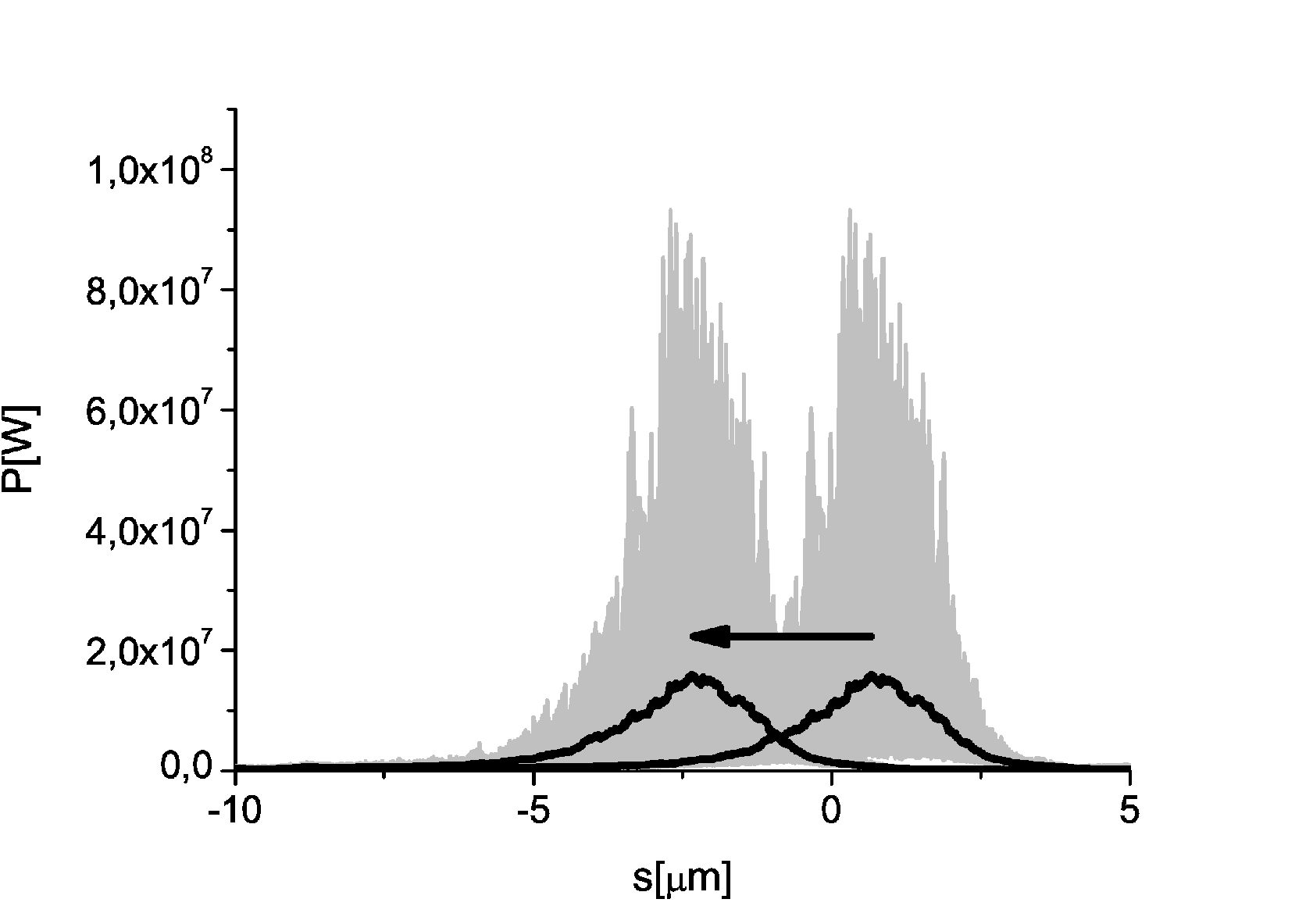}
\includegraphics[width=0.5\textwidth]{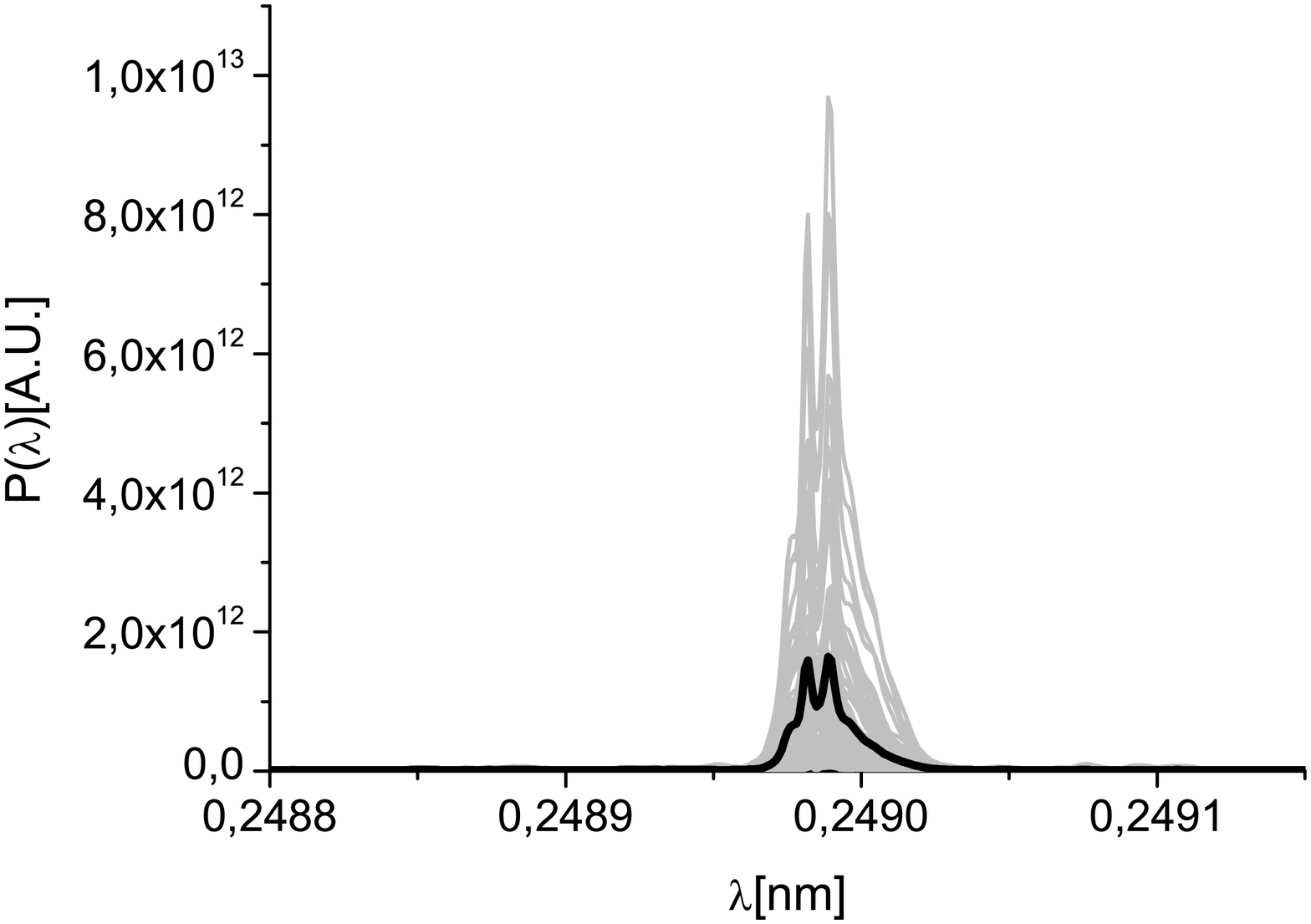}
\caption{Power and spectrum after the third chicane equipped with
the X-ray optical delay line, delaying the radiation pulse with
respect to the electron bunch. Grey lines refer to single shot
realizations, the black line refers to the average over a hundred
realizations.} \label{biof203p5}
\end{figure}
Following the seeding setup, the electron bunch amplifies the seed
in the following 4 undulator cells. After that, a third chicane is
used to allow for the installation of an x-ray optical delay line,
which delays the radiation pulse with respect to the electron bunch.
The power and spectrum of the radiation pulse after the optical
delay line are shown in Fig. \ref{biof203p5}, where the effect of
the optical delay is illustrated.

\begin{figure}[tb]
\includegraphics[width=0.5\textwidth]{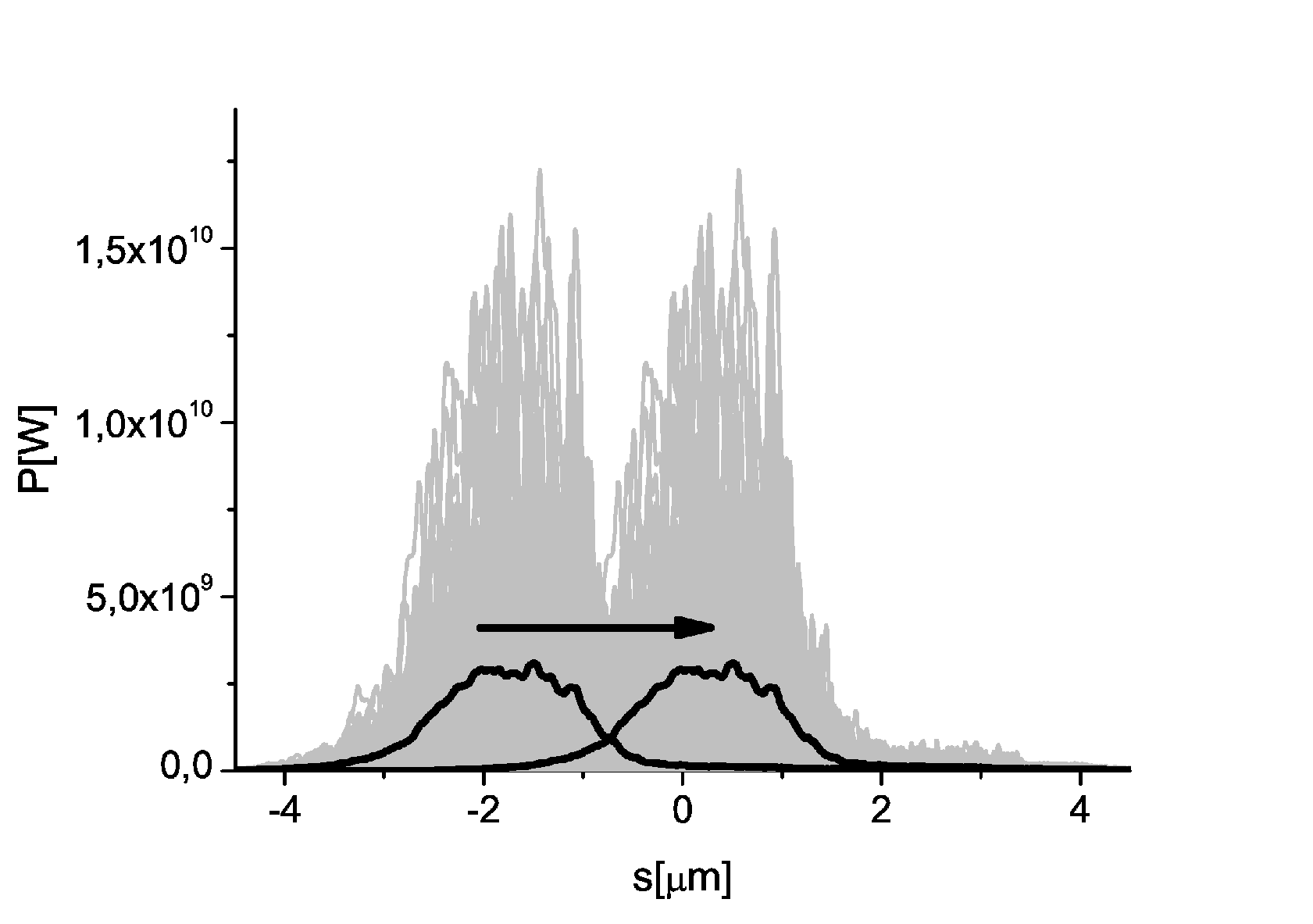}
\includegraphics[width=0.5\textwidth]{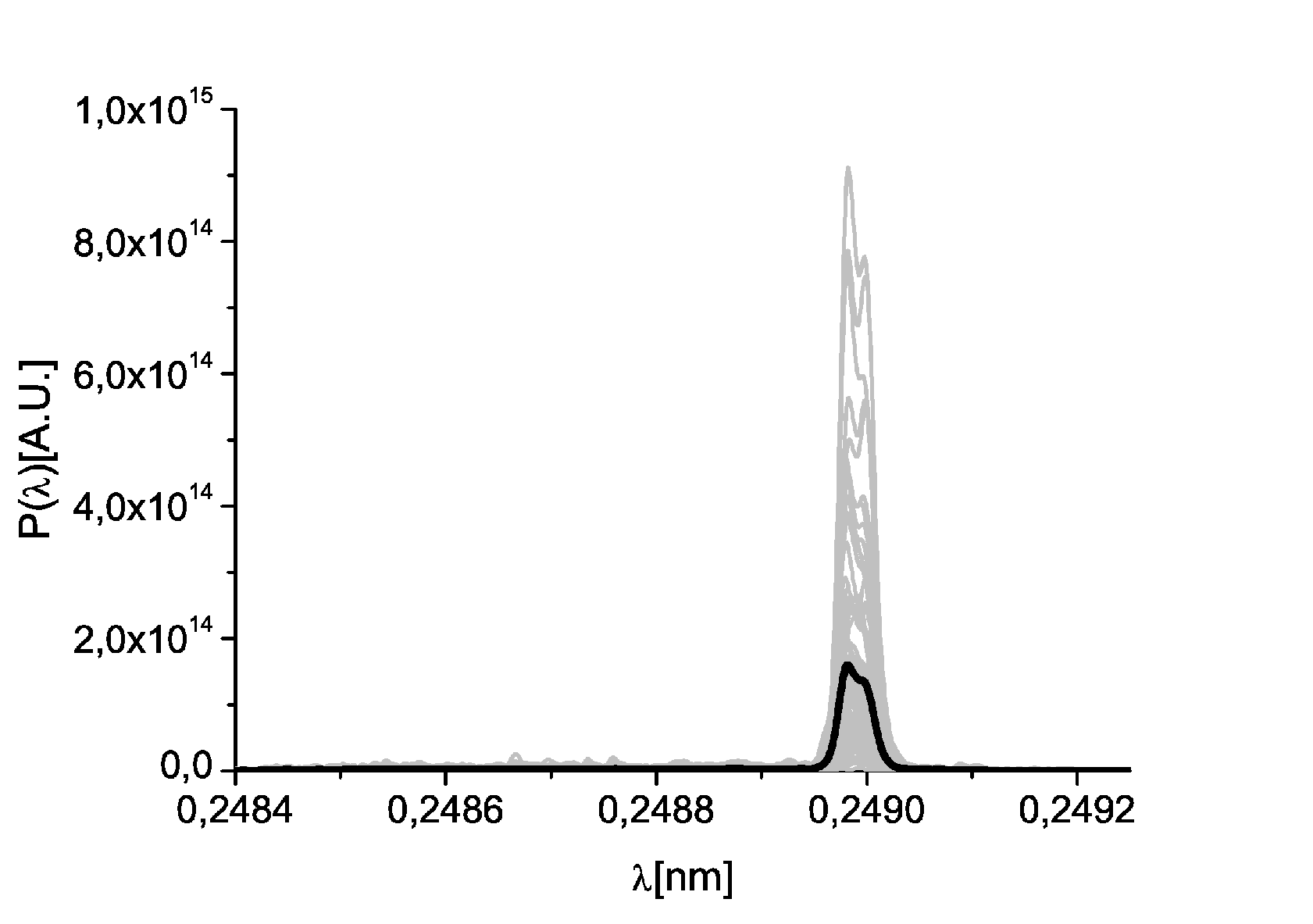}
\caption{Power and spectrum  after the last magnetic chicane. Grey
lines refer to single shot realizations, the black line refers to
the average over a hundred realizations.} \label{biof215}
\end{figure}
Due to the presence of the optical delay, only part of the electron
beam is used to further amplify the radiation pulse in the following
6 undulator cells. The electron beam part which is not used is
fresh, and can be used for further lasing. In order to do so, after
amplification, the electron beam passes through the final magnetic
chicane, which delays the electron beam. The power and spectrum of
the radiation pulse after the last magnetic chicane are shown in
Fig. \ref{biof215}. By delaying the electron bunch, the magnetic
chicane effectively shifts forward the photon beam with respect to
the electron beam. Tunability of such shift allows the selection of
different photon pulse length.

\begin{figure}[tb]
\begin{center}
\includegraphics[width=0.5\textwidth]{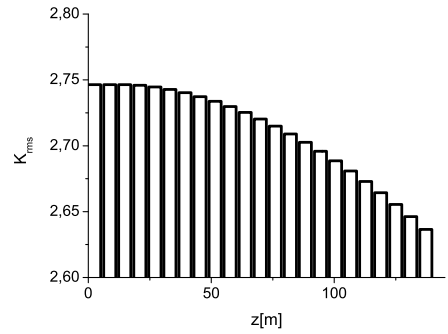}
\end{center}
\caption{Tapering law for the case $\lambda = 0.25$ nm.}
\label{biof225}
\end{figure}
The last part of the undulator is composed by $23$ cells. It is
partly tapered post-saturation, to increase the region where
electrons and radiation interact properly to the advantage of the
radiation pulse. Tapering is implemented by changing the $K$
parameter of the undulator segment by segment according to Fig.
\ref{biof225}. The tapering law used in this work has been
implemented on an empirical basis.

\begin{figure}[tb]
\includegraphics[width=0.5\textwidth]{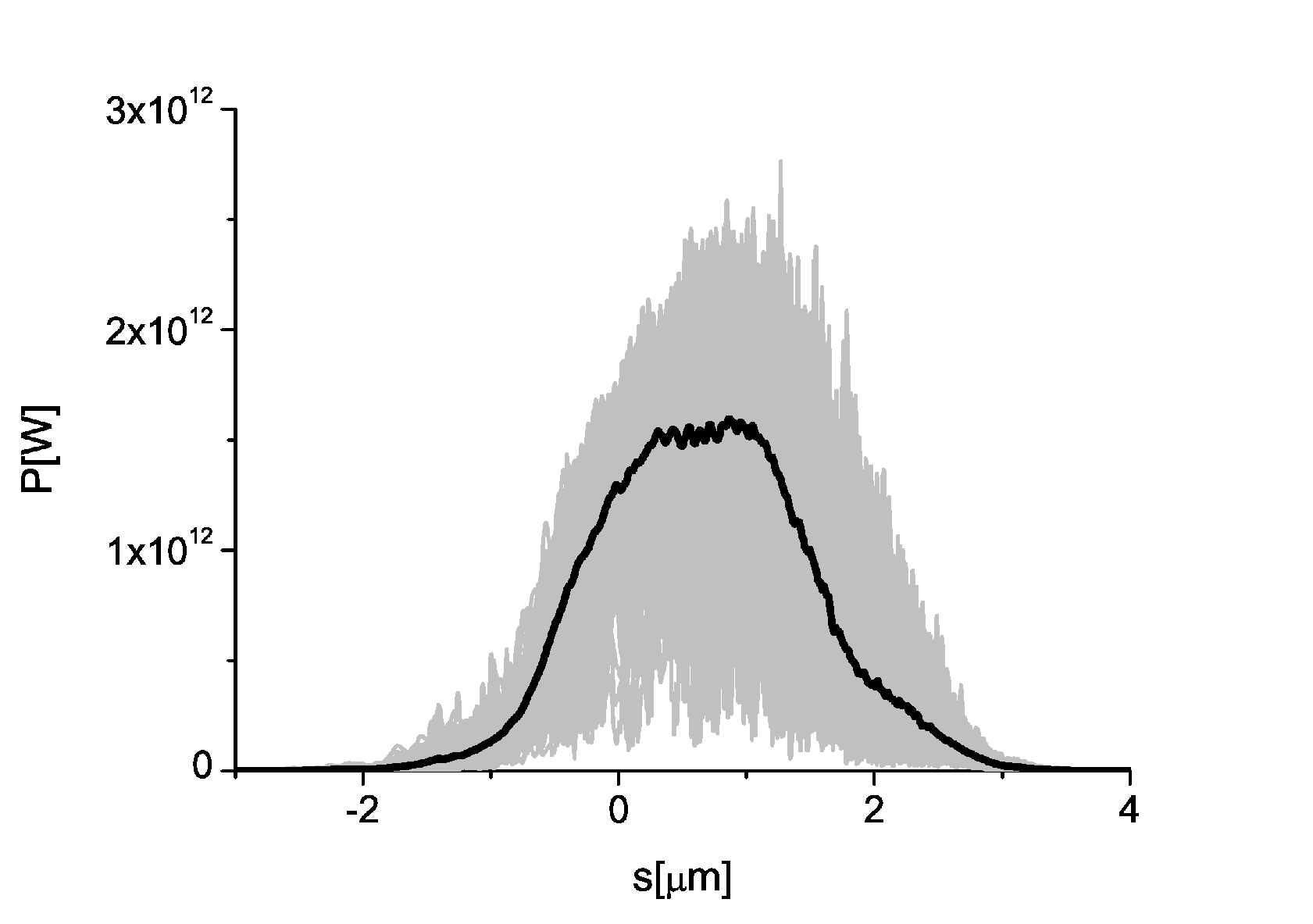}
\includegraphics[width=0.5\textwidth]{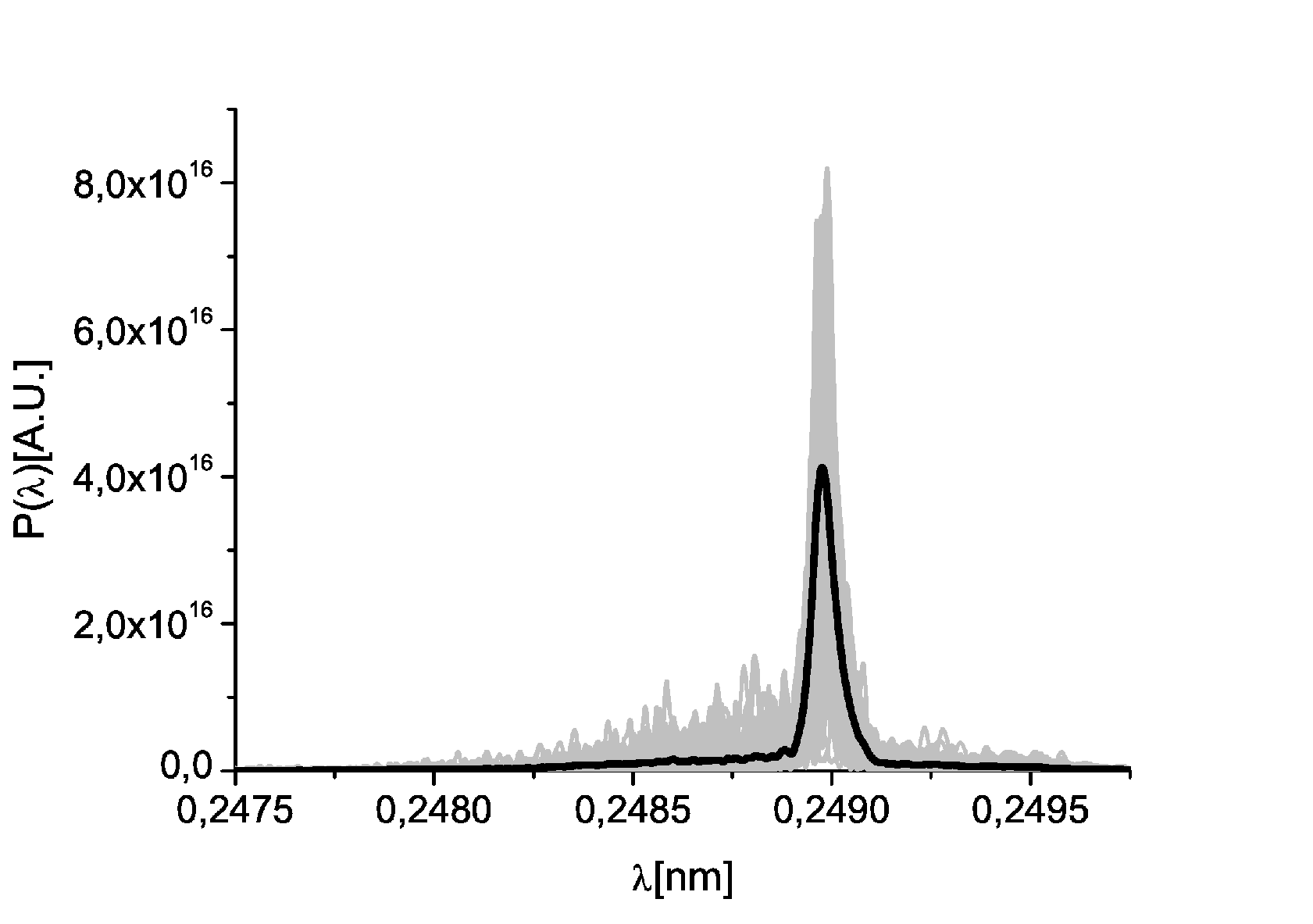}
\caption{Final output. Power and spectrum at the third harmonic
after tapering. Grey lines refer to single shot realizations, the
black line refers to the average over a hundred realizations.}
\label{biof235}
\end{figure}
The use of tapering together with monochromatic radiation is
particularly effective, since the electron beam does not experience
brisk changes of the ponderomotive potential during the slippage
process. The final output is presented in Fig. \ref{biof235} in
terms of power and spectrum. As one can see, simulations indicate an
output power of about $1.5$ TW.

\begin{figure}[tb]
\includegraphics[width=0.5\textwidth]{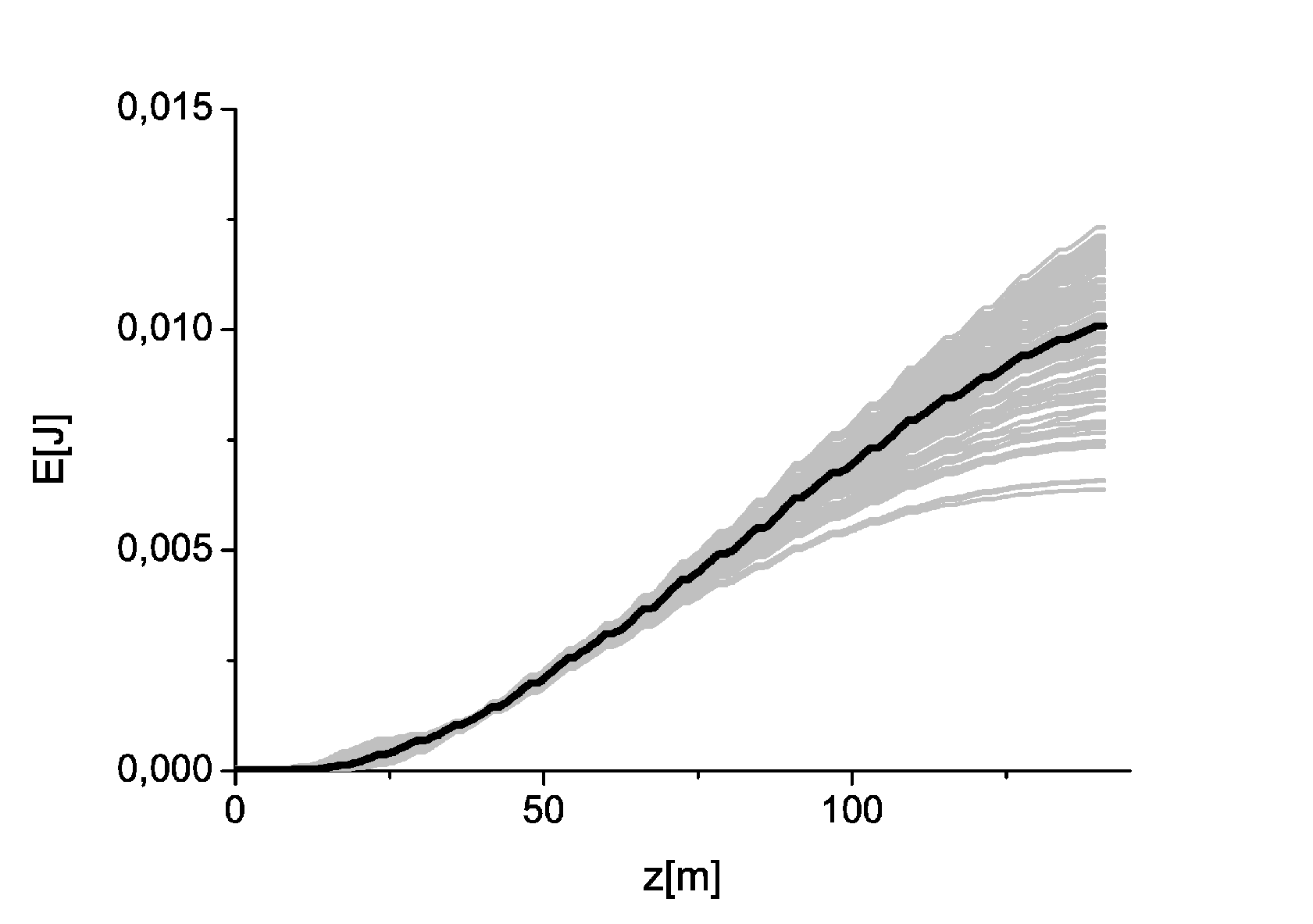}
\includegraphics[width=0.5\textwidth]{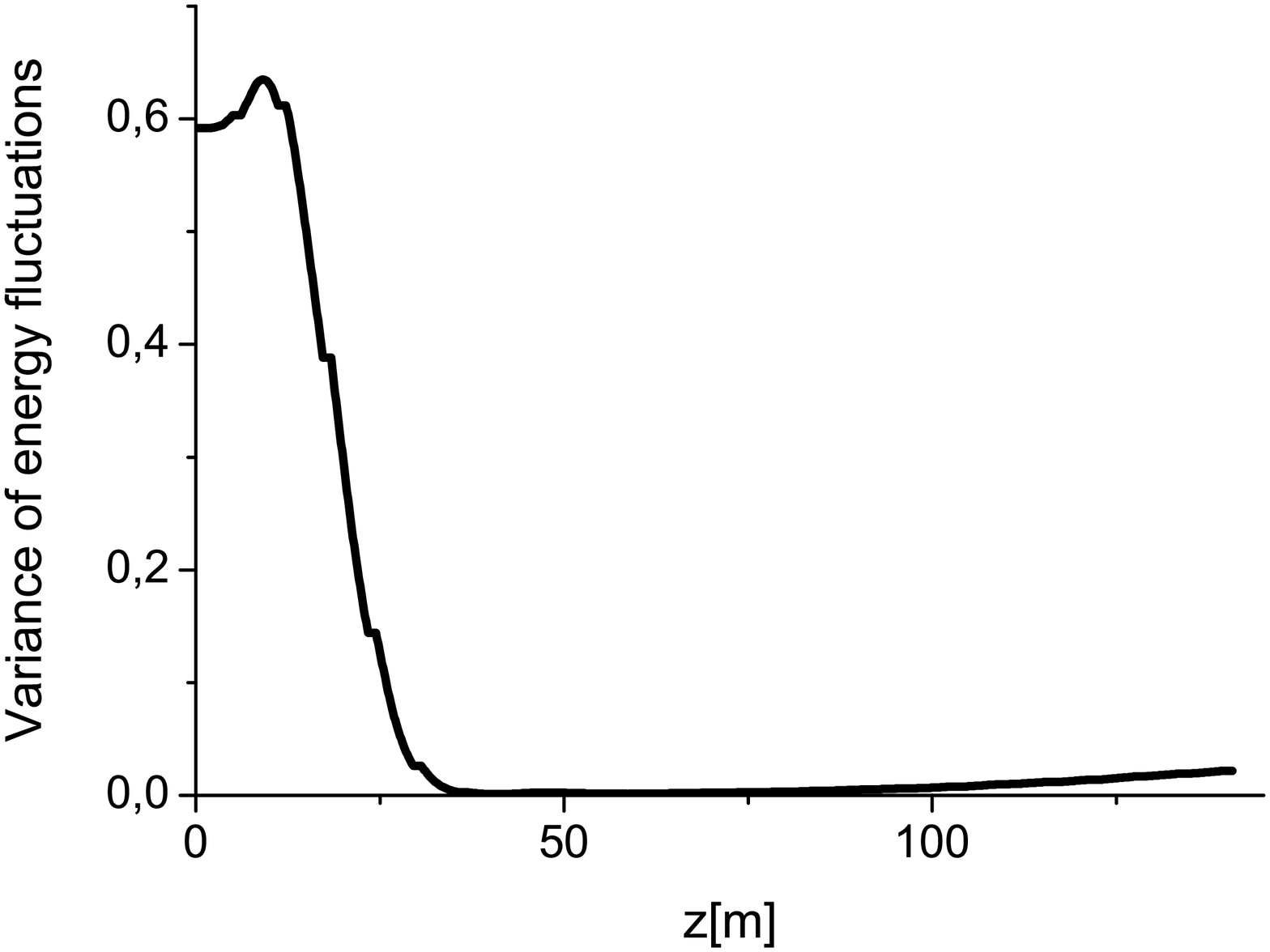}
\caption{Final output. Energy and energy variance of output pulses
for the case $\lambda = 0.25$ nm. In the left plot, grey lines refer
to single shot realizations, the black line refers to the average
over a hundred realizations.} \label{biof245}
\end{figure}

\begin{figure}[tb]
\includegraphics[width=0.5\textwidth]{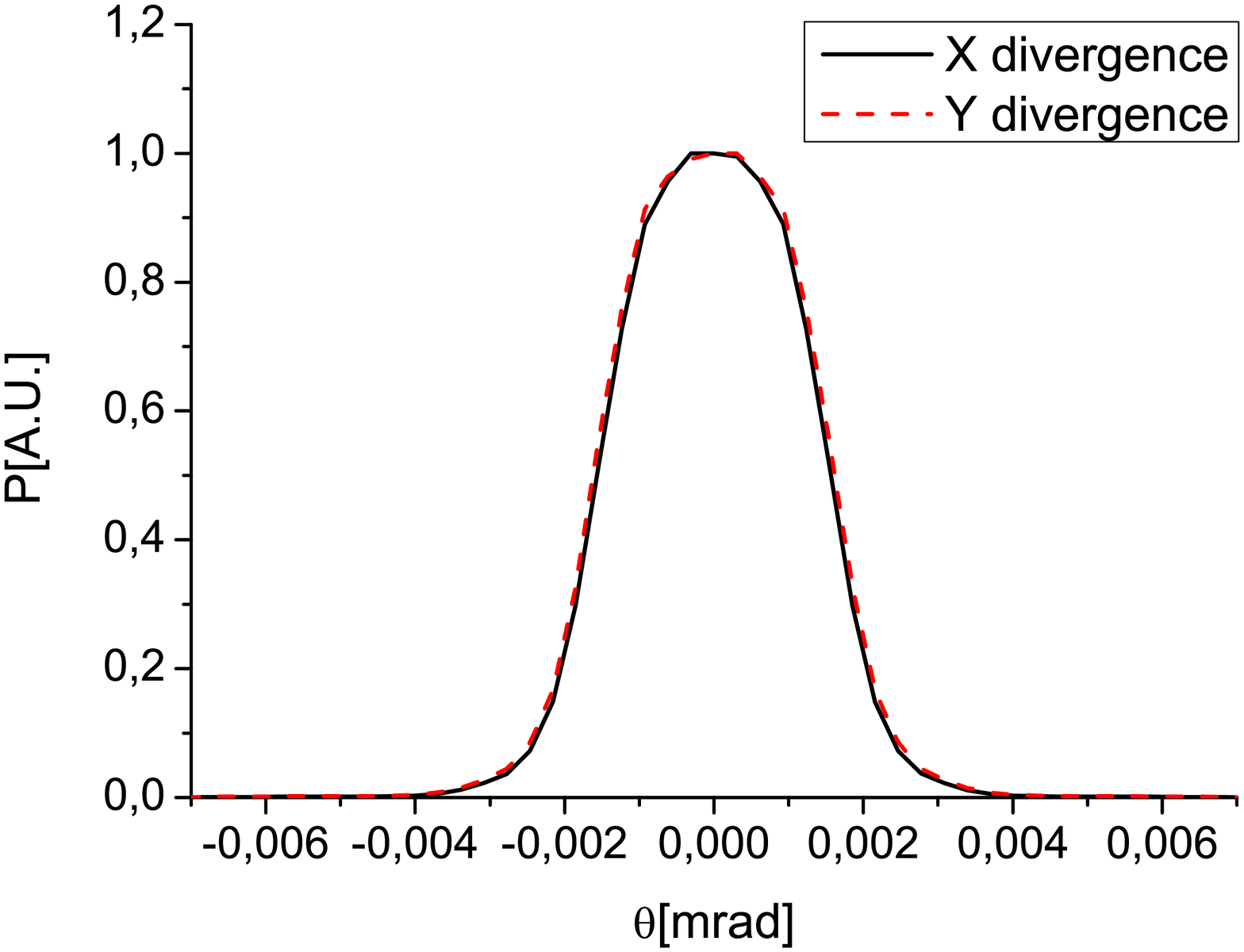}
\includegraphics[width=0.5\textwidth]{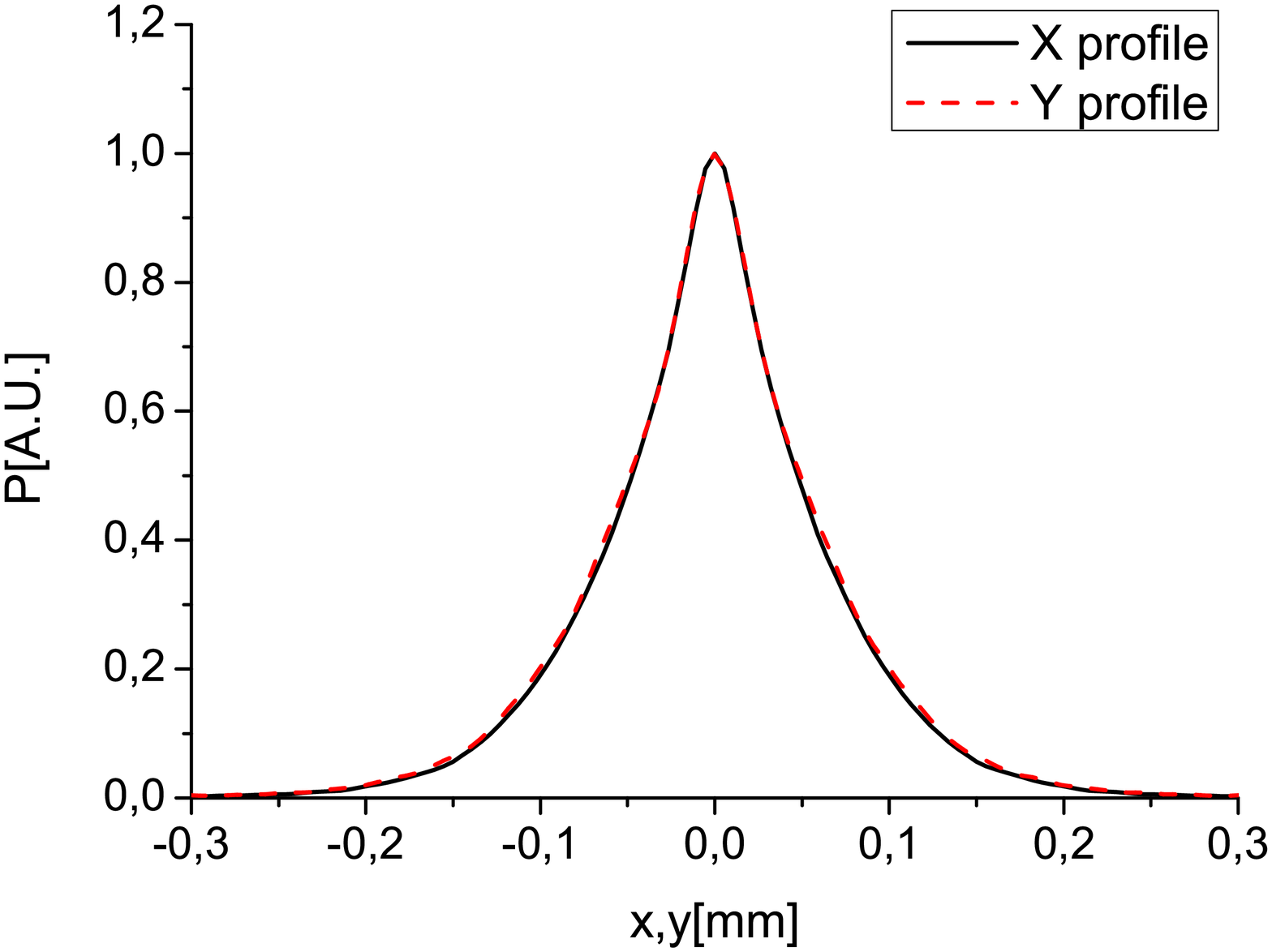}
\caption{Final output. X-ray radiation pulse energy distribution per
unit surface and angular distribution of the X-ray pulse energy at
the exit of output undulator for the case $\lambda = 0.25$ nm.}
\label{biof253p5}
\end{figure}
The energy of the radiation pulse and the energy variance are shown
in Fig. \ref{biof245} as a function of the position along the
undulator. The divergence and the size of the radiation pulse at the
exit of the final undulator are shown, instead, in Fig.
\ref{biof253p5}. In order to calculate the size, an average of the
transverse intensity profiles is taken. In order to calculate the
divergence, the spatial Fourier transform of the field is
calculated.

\subsection{Photon energy range between 7 keV and 9 keV}

We now consider the photon energy range between 7 keV and 9 keV. In
this case the beamline will be configured as in Fig. \ref{bio3f6}. A
feasibility study dealing with this energy range can be found in
\cite{OURCC}.

\subsection{Photon energy range between 9 keV and 13 keV}

Finally, we consider generation of radiation in the photon energy
between 9 keV and 13 keV. The undulator line will be configured as
in Fig. \ref{bio3f4}, the only difference now being that the final
$23$ cells will be tuned at the third harmonic of the fundamental,
thus allowing to reach the photon energy between 9 keV and 13 keV.

\begin{figure}[tb]
\includegraphics[width=0.5\textwidth]{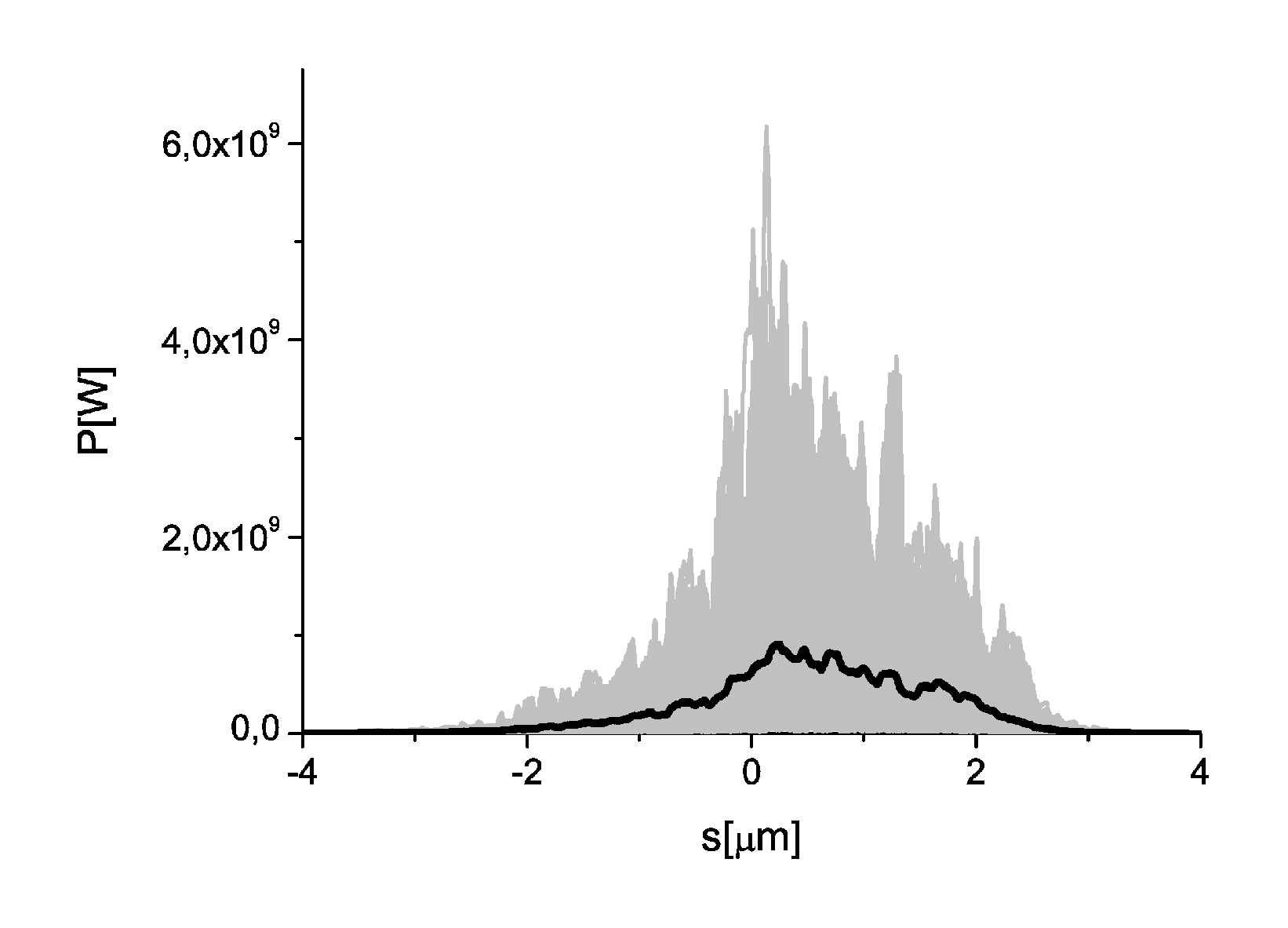}
\includegraphics[width=0.5\textwidth]{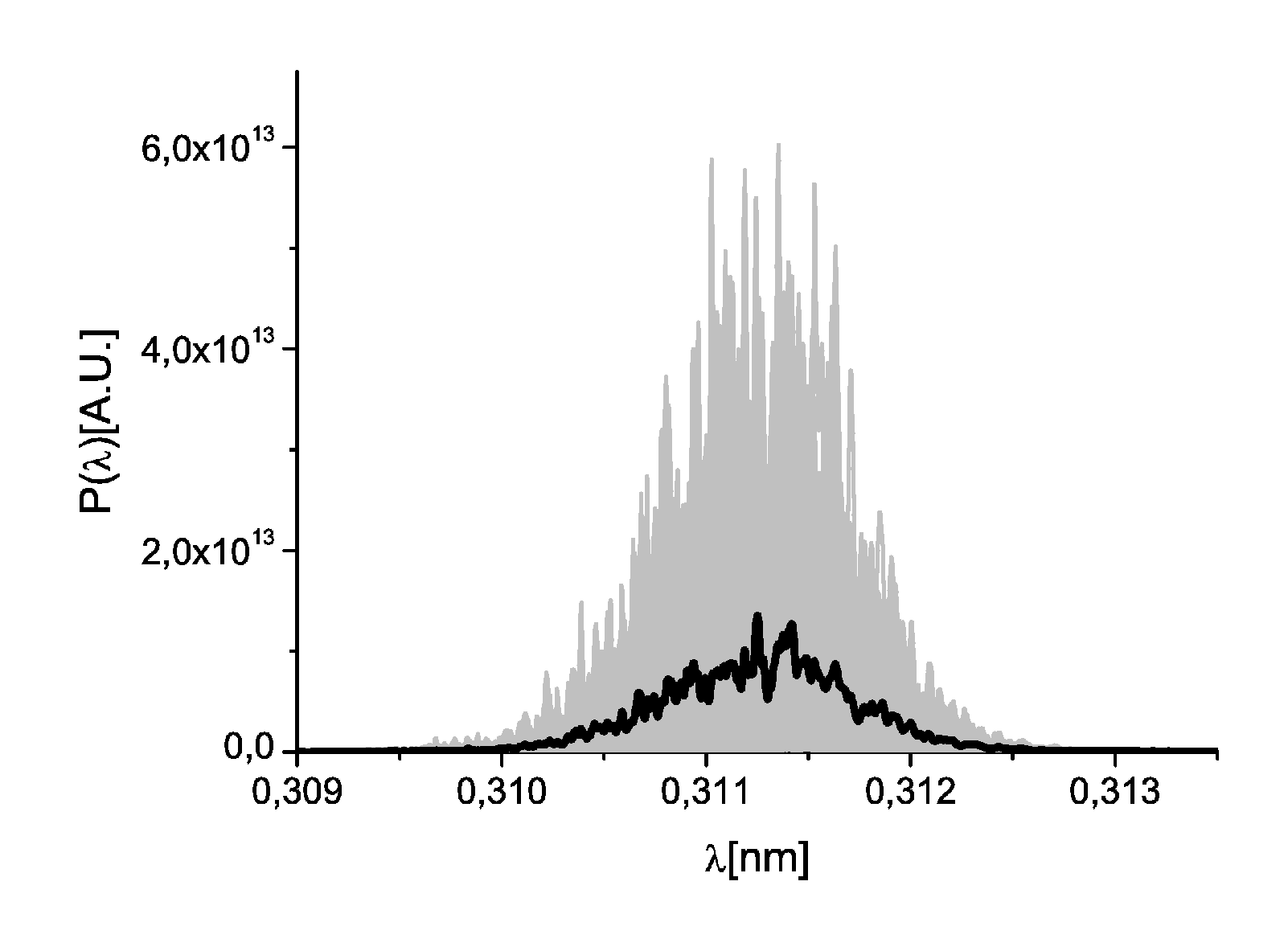}
\caption{Power and spectrum before the second magnetic chicane. Grey
lines refer to single shot realizations, the black line refers to
the average over a hundred realizations.} \label{12biof173p5}
\end{figure}
For this mode of operation, the first chicane is switched off, so
that the first part of the undulator effectively consists of 7
uniform cells. We begin our investigation by simulating the SASE
power and spectrum after the first part of the undulator, that is
before the second magnetic chicane in the setup. Results are shown
in Fig. \ref{12biof173p5}.

\begin{figure}[tb]
\includegraphics[width=0.5\textwidth]{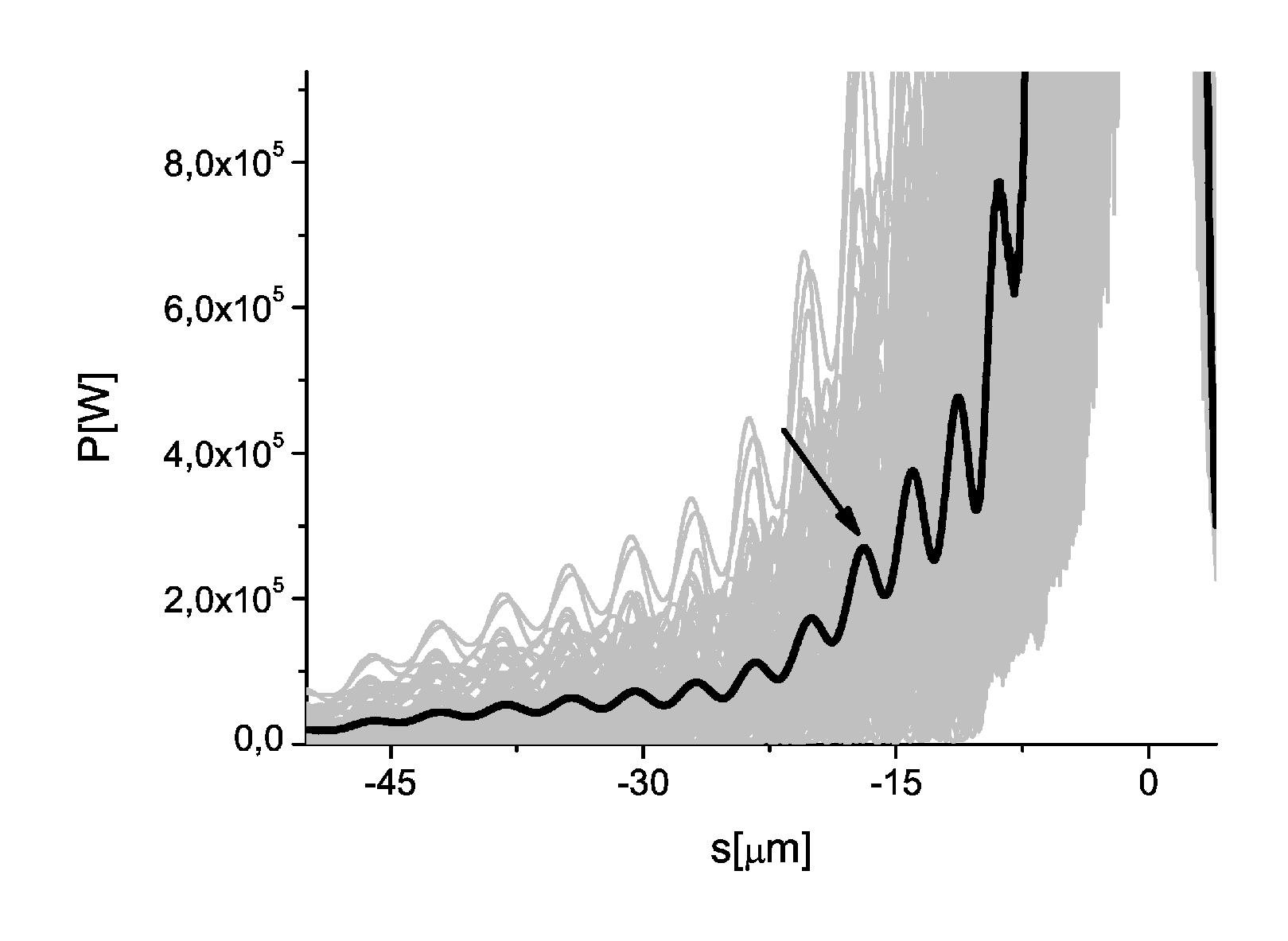}
\includegraphics[width=0.5\textwidth]{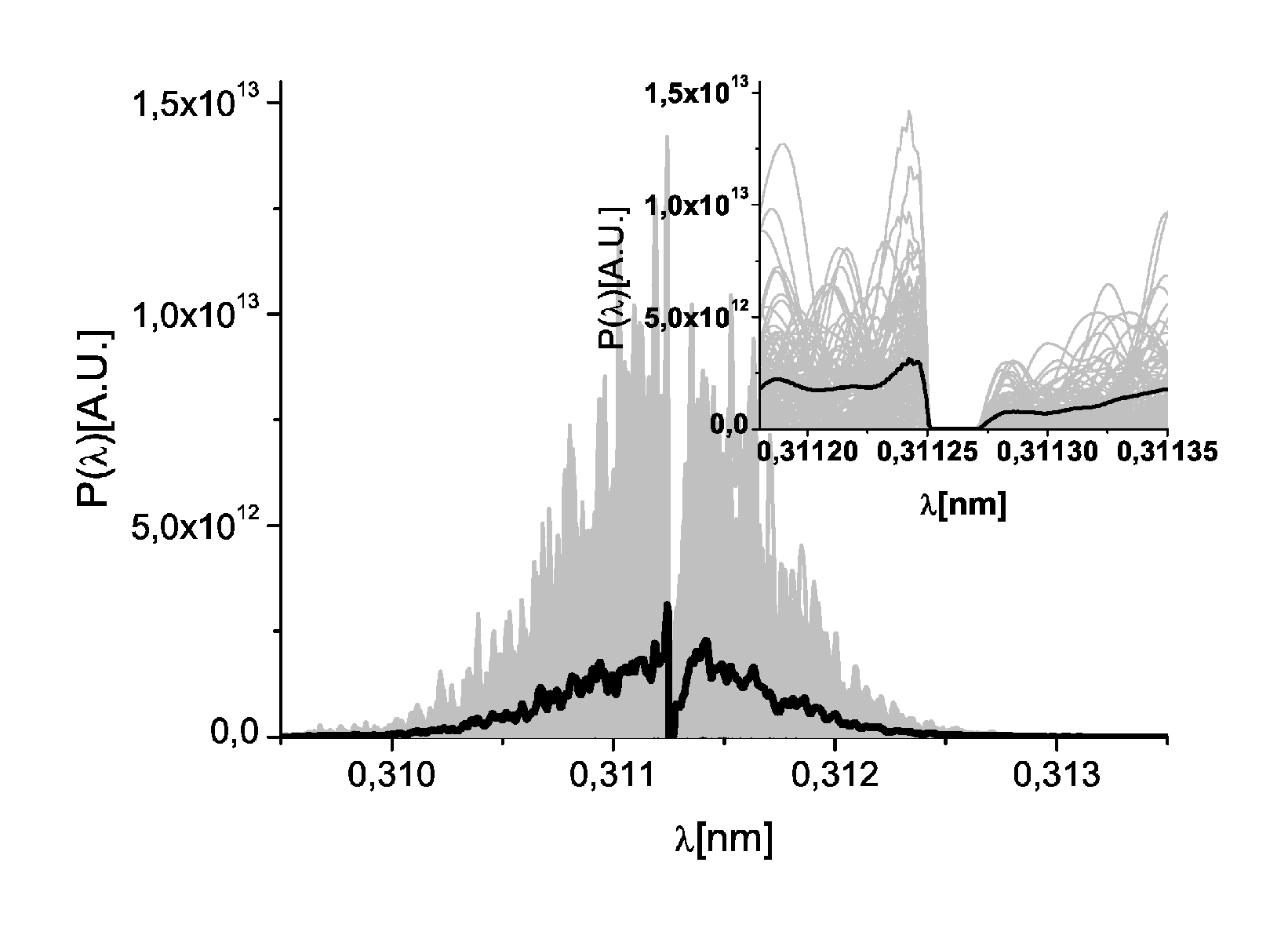}
\caption{Power and spectrum after the single crystal self-seeding
X-ray monochromator. A $100~\mu$m thick diamond crystal in Bragg
transmission geometry ( C(111) reflection, $\sigma$-polarization )
is used. Grey lines refer to single shot realizations, the black
line refers to the average over a hundred realizations. The black
arrow indicates the seeding region.} \label{12biof1830p5}
\end{figure}
The second magnetic chicane is switched on, and the single-crystal
X-ray monochromator is set into the photon beam. Since we want to
generate radiation in the 3 keV - 4 keV energy range, we use a
$100~\mu$m-thick diamond crystal in Bragg transmission geometry. In
particular, we take advantage of the C(111) reflection,
$\sigma$-polarization. The crystal acts as a bandstop filter, and
the output spectrum is plotted in Fig. \ref{12biof1830p5} (right).
Due to the bandstop effect, the signal in the time domain exhibits a
long monochromatic tail, which is used for seeding, Fig.
\ref{12biof1830p5} (left). To this purpose, the electron bunch is
slightly delayed by proper tuning of the magnetic chicane to be
superimposed to the seeding signal. The difference with respect to
Fig. \ref{biof1830p5} is in a slightly different frequency, since in
this Section we want to study the feasibility of production of $12$
keV radiation.

\begin{figure}[tb]
\includegraphics[width=0.5\textwidth]{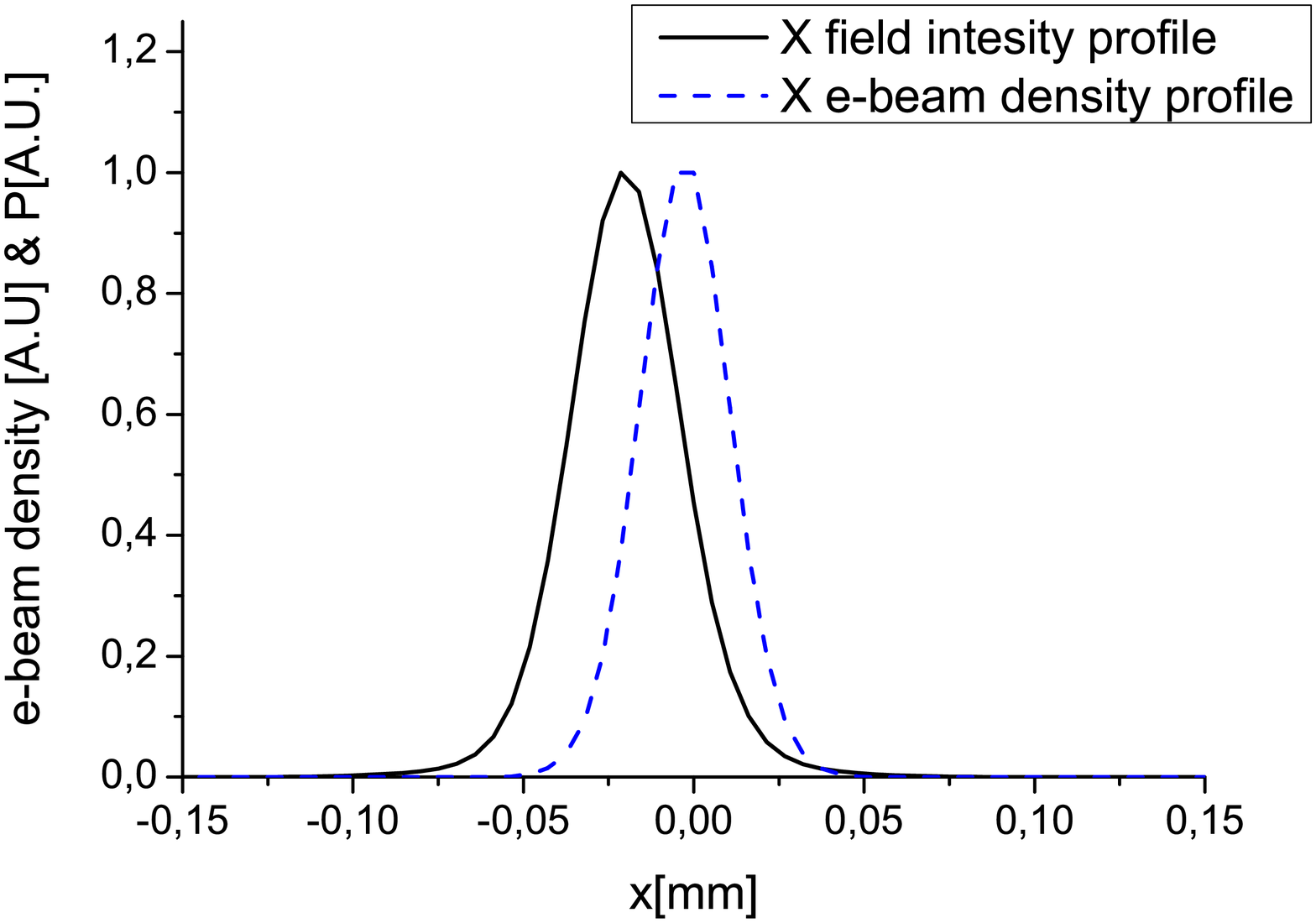}
\includegraphics[width=0.5\textwidth]{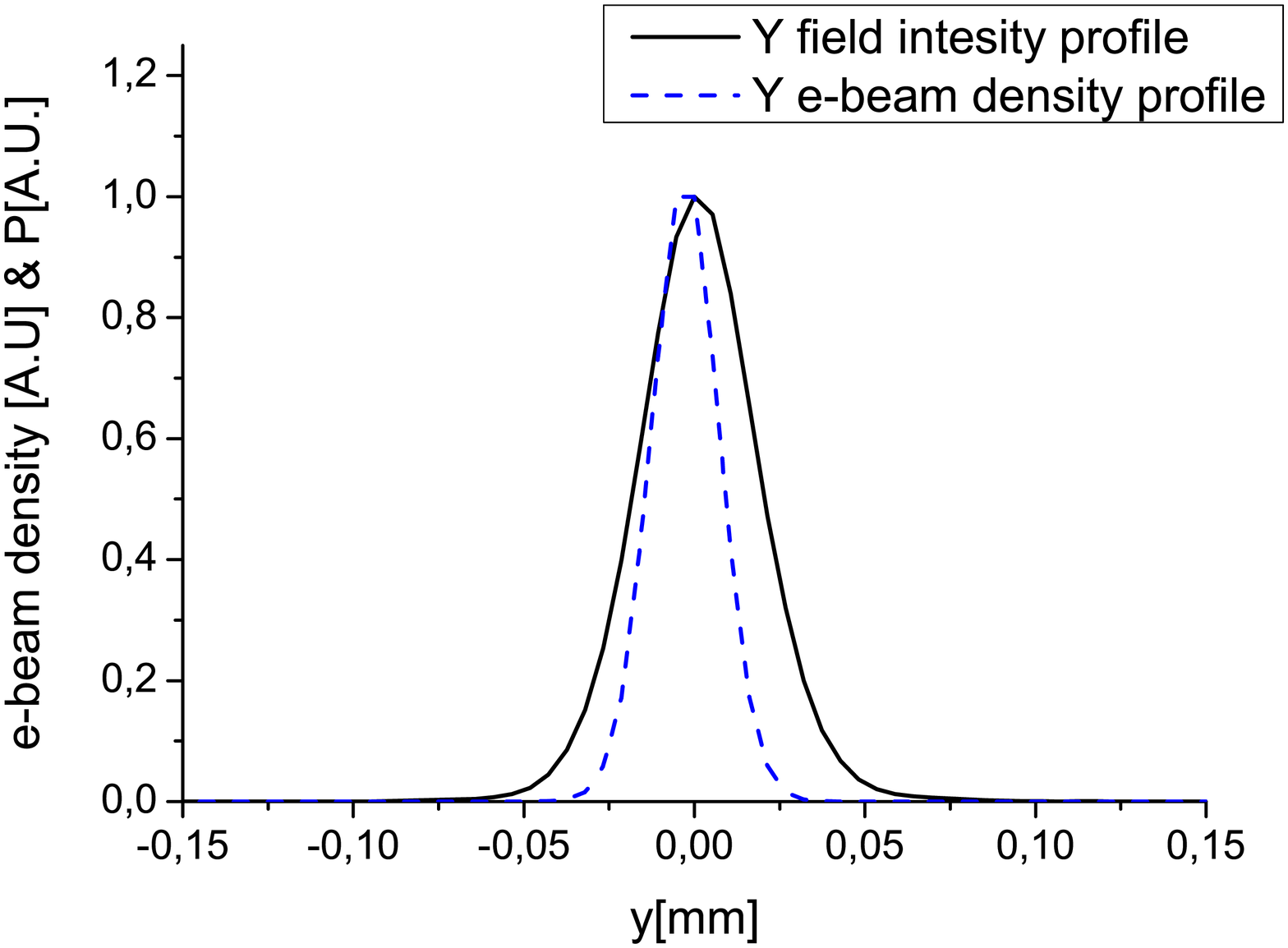}
\caption{Comparison between transverse profile of the seed field
intensity and transverse profile of the electron beam density at the
position used for seeding at $\lambda = 0.3$ nm. The plots refer to
the longitudinal position inside the bunch corresponding to the
maximum current value.} \label{12compare}
\end{figure}
%

As before it should be remarked that, according to Section
\ref{spatio}, spatio-temporal coupling induced by the crystal
monochromator should be accounted for in our study. A comparison
between the electron beam position and the photon beam position
after the crystal is shown in Fig. \ref{12compare}. The shift
difference is due to the spatio-temporal coupling induced by the
crystal, and must be accounted for in calculations. The longitudinal
position at which Fig. \ref{12compare} refers is that of the seeding
peak, compared to the lasing part of the bunch.

\begin{figure}[tb]
\includegraphics[width=0.5\textwidth]{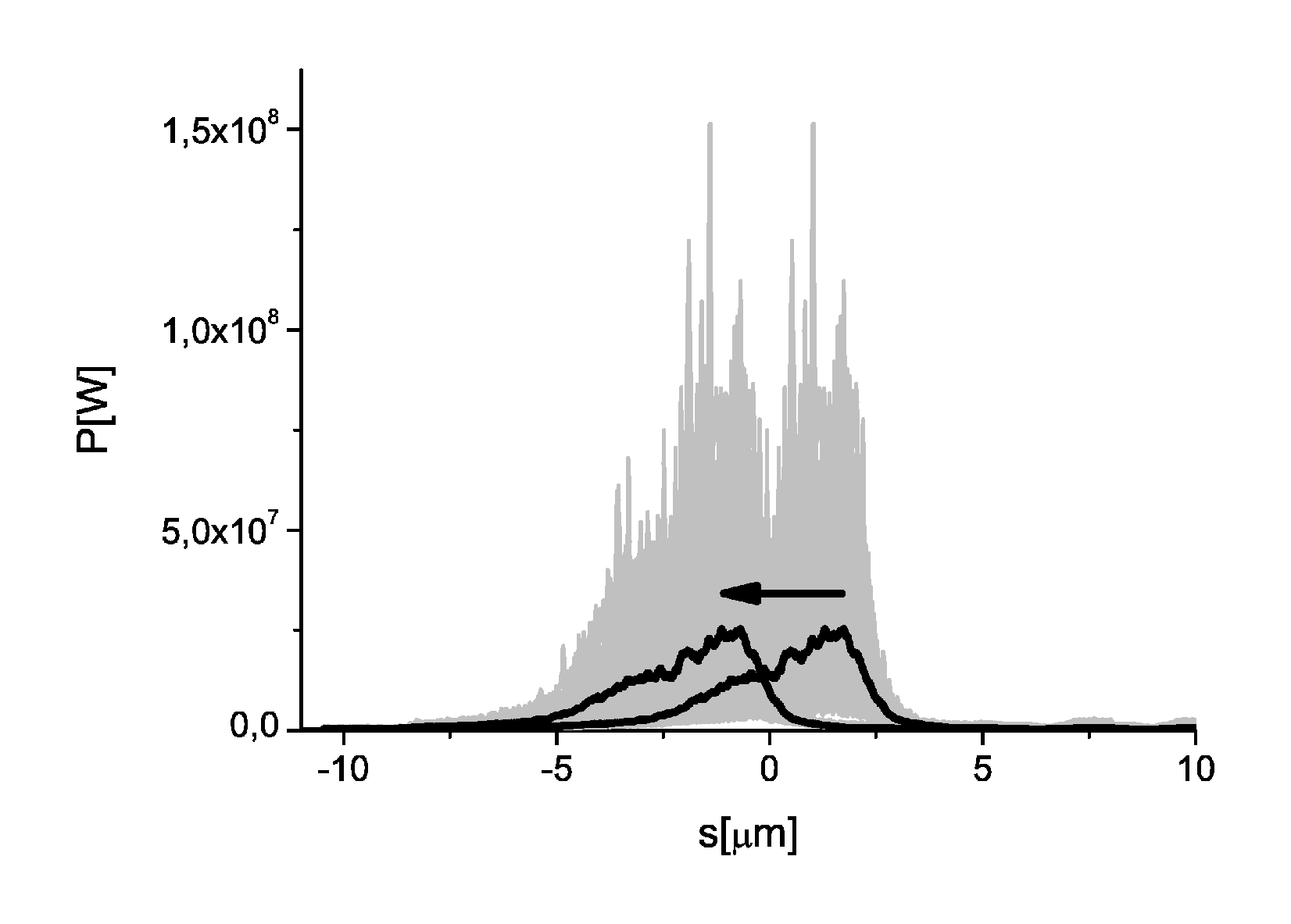}
\includegraphics[width=0.5\textwidth]{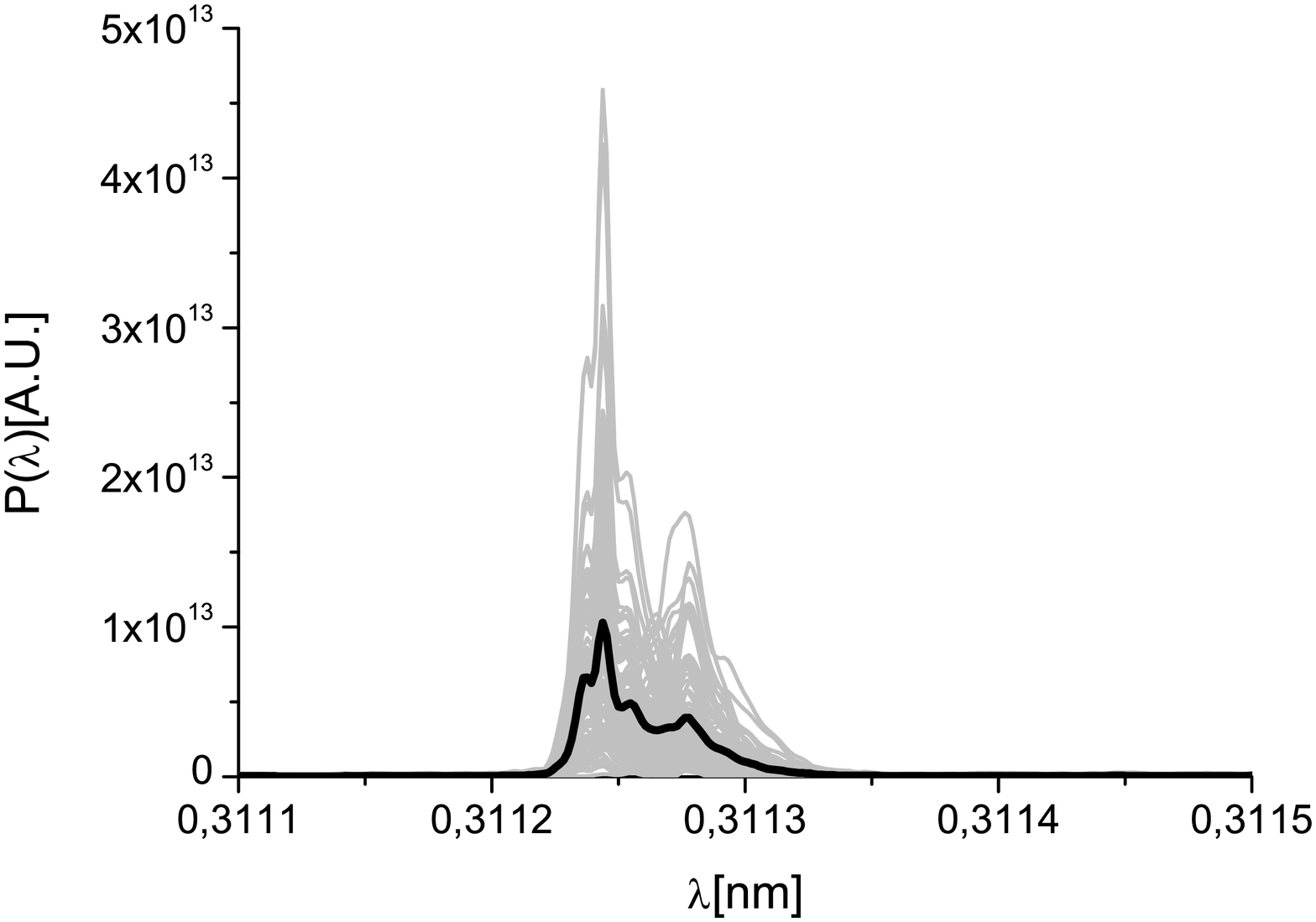}
\caption{Power and spectrum after the third chicane equipped with
the X-ray optical delay line, delaying the radiation pulse with
respect to the electron bunch. Grey lines refer to single shot
realizations, the black line refers to the average over a hundred
realizations.} \label{12biof2030p5}
\end{figure}
Following the seeding setup, the electron bunch amplifies the seed
in the following 4 undulator cells. After that, a third chicane is
used to allow for the installation of an x-ray optical delay line,
which retards the radiation pulse with respect to the electron
bunch. The power and spectrum of the radiation pulse after the
optical delay line are shown in Fig. \ref{12biof2030p5}, where the
effect of the optical delay is illustrated.

\begin{figure}[tb]
\includegraphics[width=0.5\textwidth]{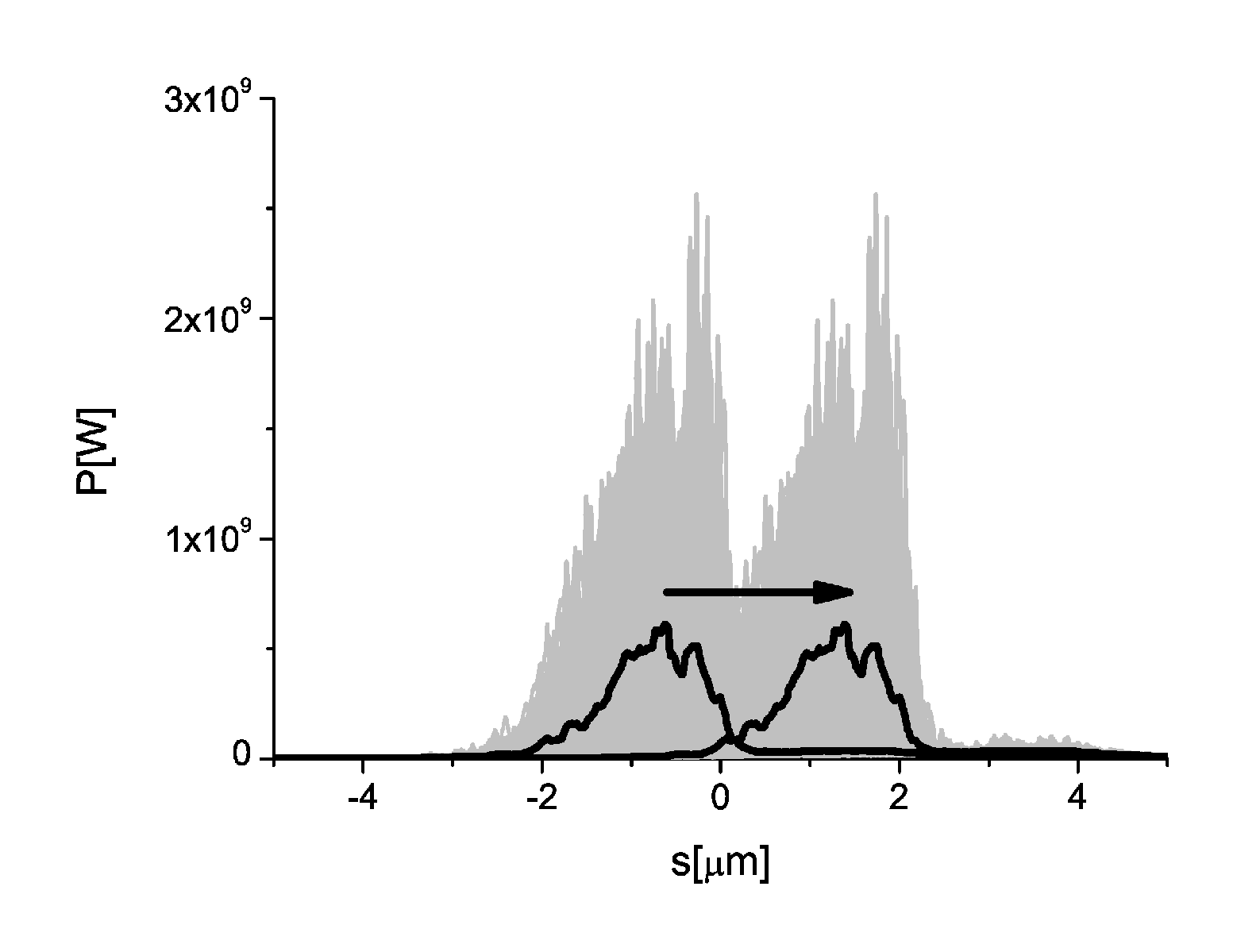}
\includegraphics[width=0.5\textwidth]{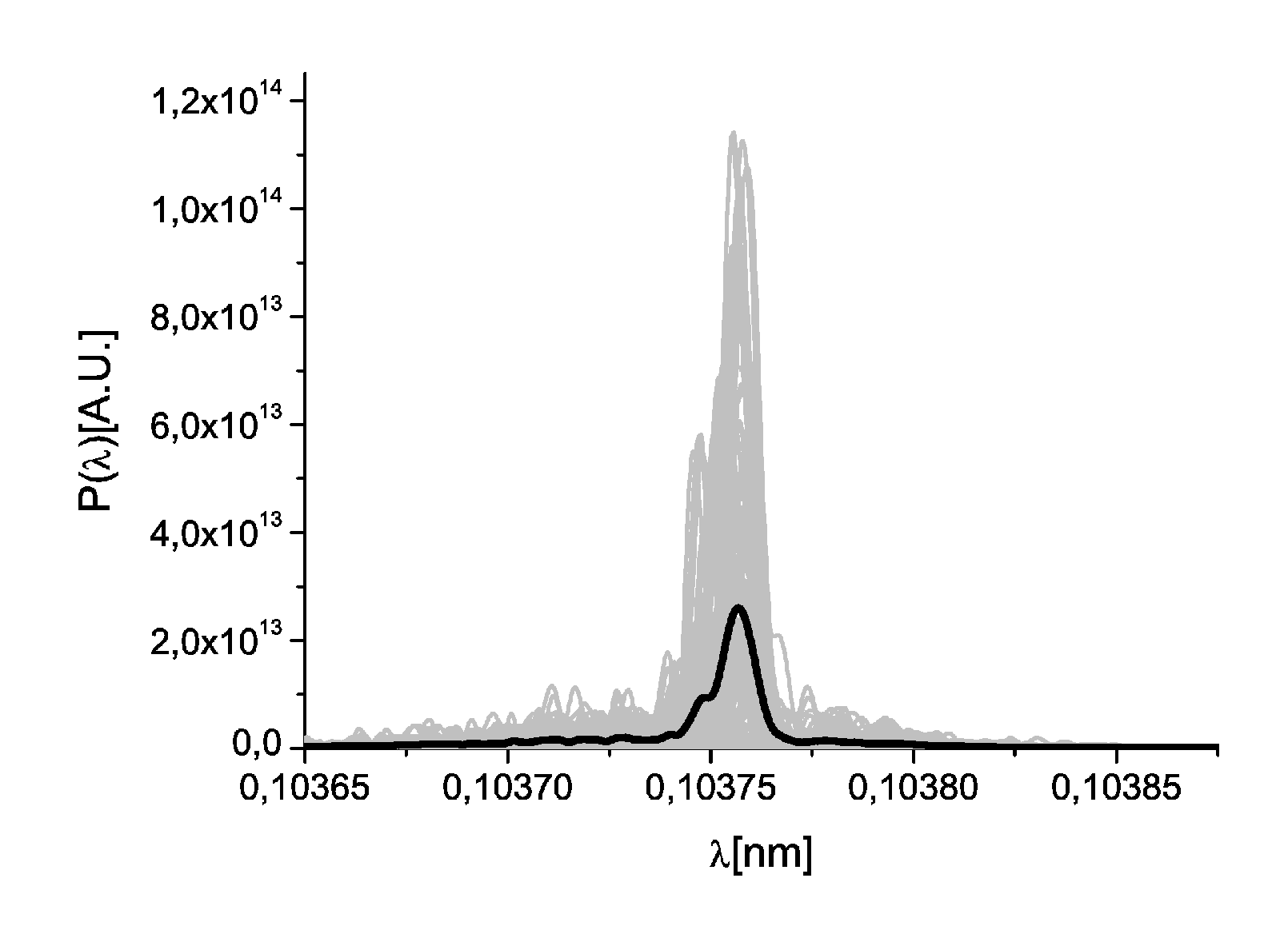}
\caption{Power and spectrum  after the last magnetic chicane. Grey
lines refer to single shot realizations, the black line refers to
the average over a hundred realizations.} \label{12biof213p5}
\end{figure}
Due to the presence of the optical delay, only part of the electron
beam is used to further amplify the radiation pulse in the following
6 undulator cells. The electron beam part which has not lased is
fresh, and can be used for further lasing. In order to do so, after
amplification, the electron beam passes through the final magnetic
chicane, which delays the electron beam. During the previous
amplification, a consistent amount of bunching at the third harmonic
of the fundamental is produced. The power and spectrum of the
radiation pulse at the third harmonic after the last magnetic
chicane are shown in Fig. \ref{12biof213p5}. By delaying the
electron bunch, the magnetic chicane effectively shifts forward the
photon beam with respect to the electron beam. Tunability of such
shift allows the selection of different photon pulse lengths.

\begin{figure}[tb]
\begin{center}
\includegraphics[width=0.5\textwidth]{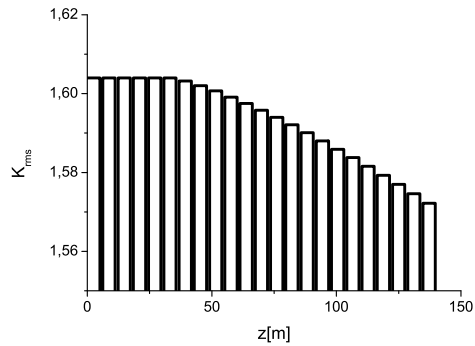}
\end{center}
\caption{Tapering law for the case $\lambda = 0.1$ nm.}
\label{12biof223p5}
\end{figure}

\begin{figure}[tb]
\includegraphics[width=0.5\textwidth]{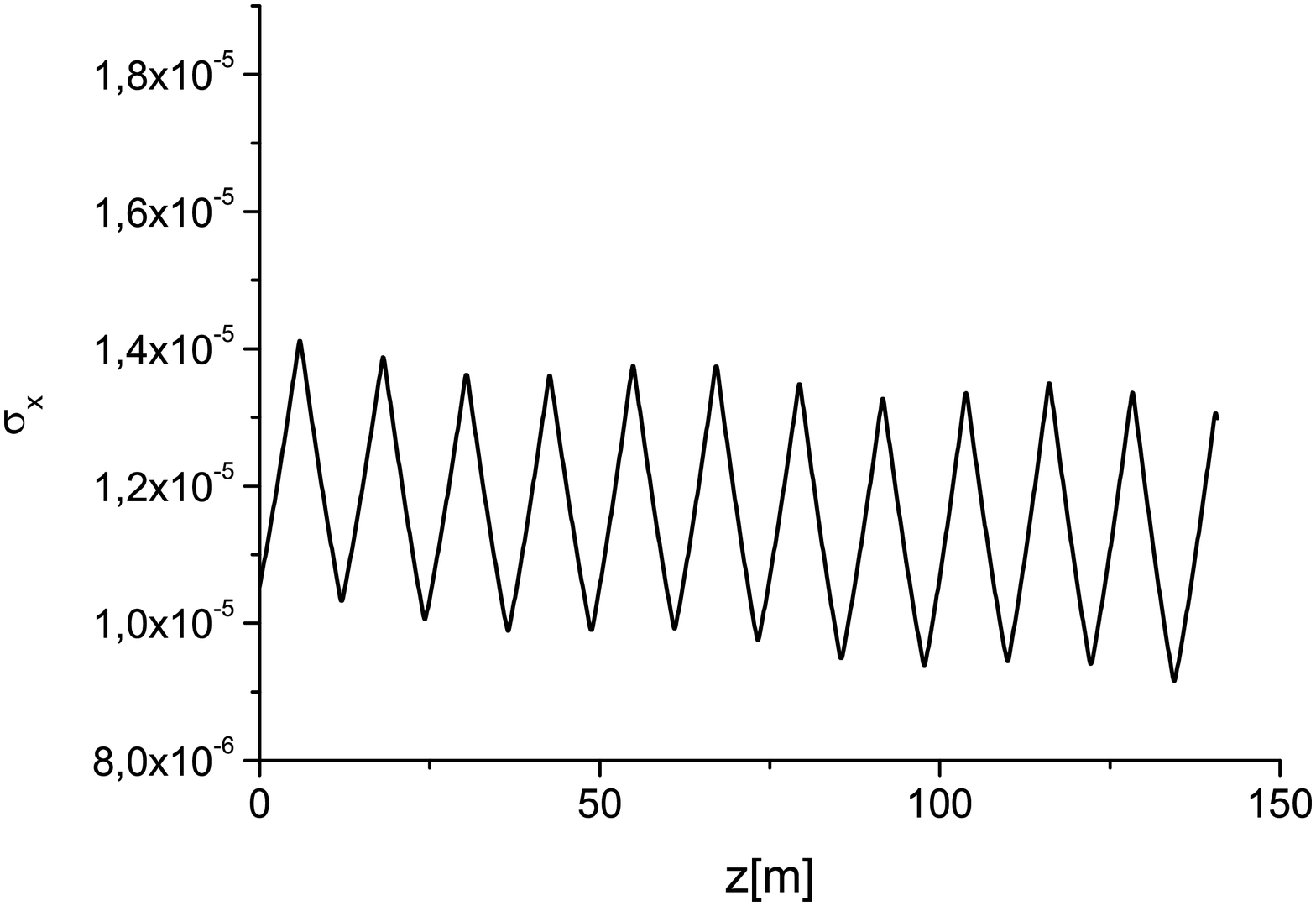}
\includegraphics[width=0.5\textwidth]{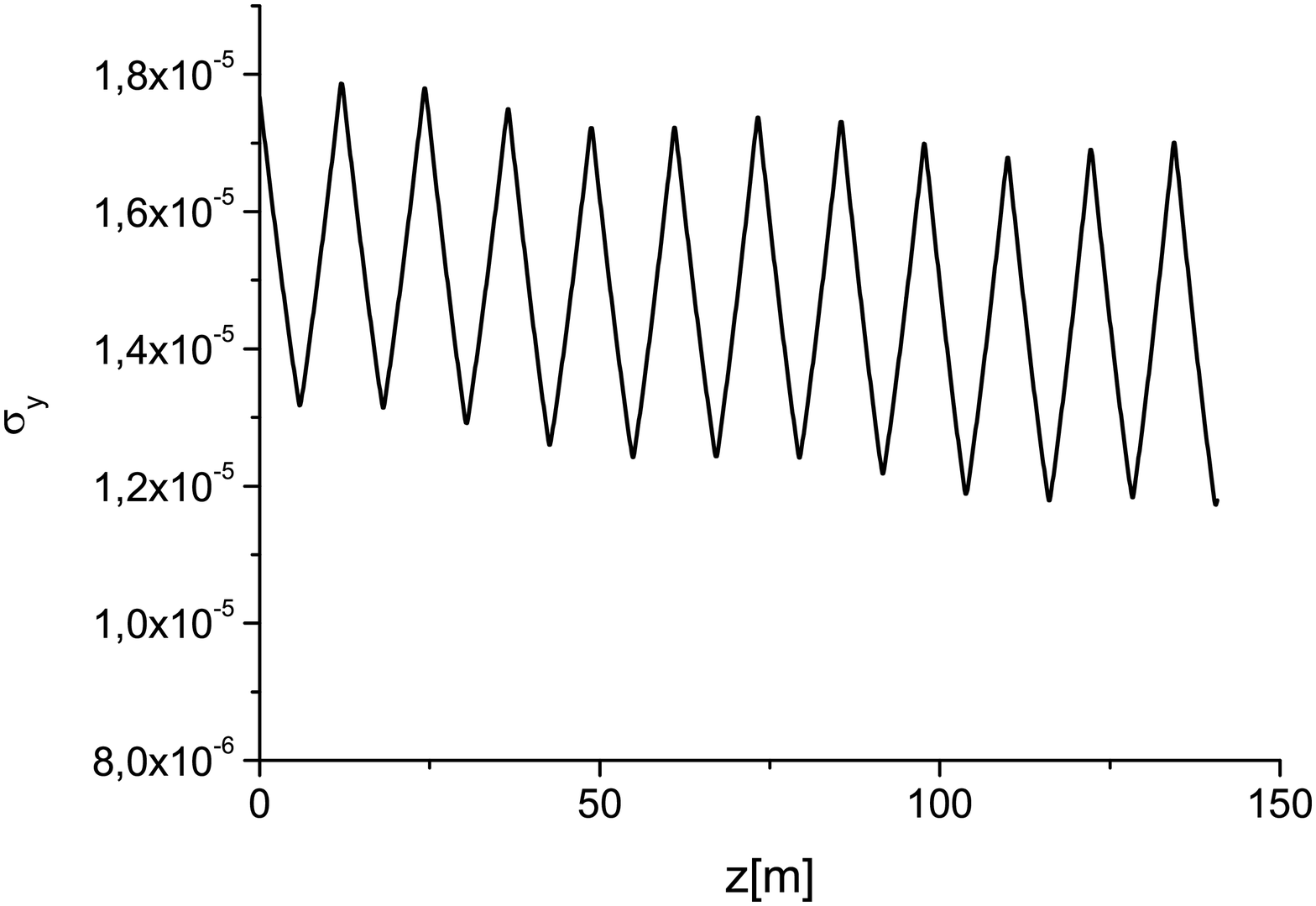}
\caption{Evolution of the horizontal (left plot) and vertical (right
plot) dimensions of the electron bunch as a function of the distance
inside the the tapered part of the undulator at $\lambda = 0.1$ nm.
The plots refer to the longitudinal position inside the bunch
corresponding to the maximum current value. The quadrupole strength
is varied along the undulator axis in order to optimize the output.}
\label{12sigma}
\end{figure}
The last part of the undulator is composed by $23$ cells. It is
tuned at the third harmonic of the fundamental, i.e. around 12 keV
in our case, and partly tapered post-saturation, to increase the
region where electrons and radiation interact properly to the
advantage of the radiation pulse. Tapering is implemented by
changing the $K$ parameter of the undulator segment by segment
according to Fig. \ref{12biof223p5}. The tapering law used in this
work has been implemented on an empirical basis, and the output has
been optimized also by varying the quadrupole strength as shown in
Fig. \ref{12sigma}.

\begin{figure}[tb]
\includegraphics[width=0.5\textwidth]{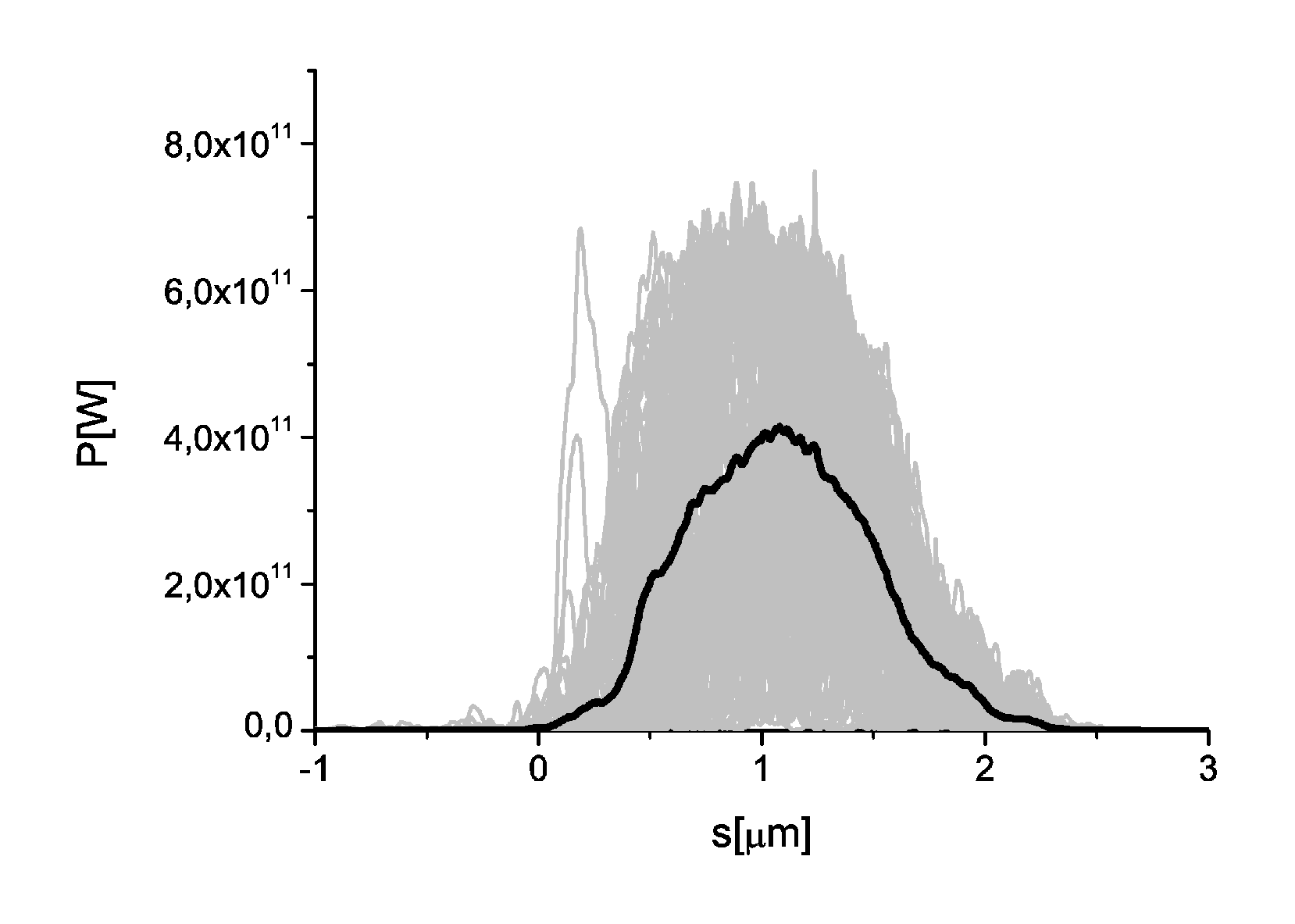}
\includegraphics[width=0.5\textwidth]{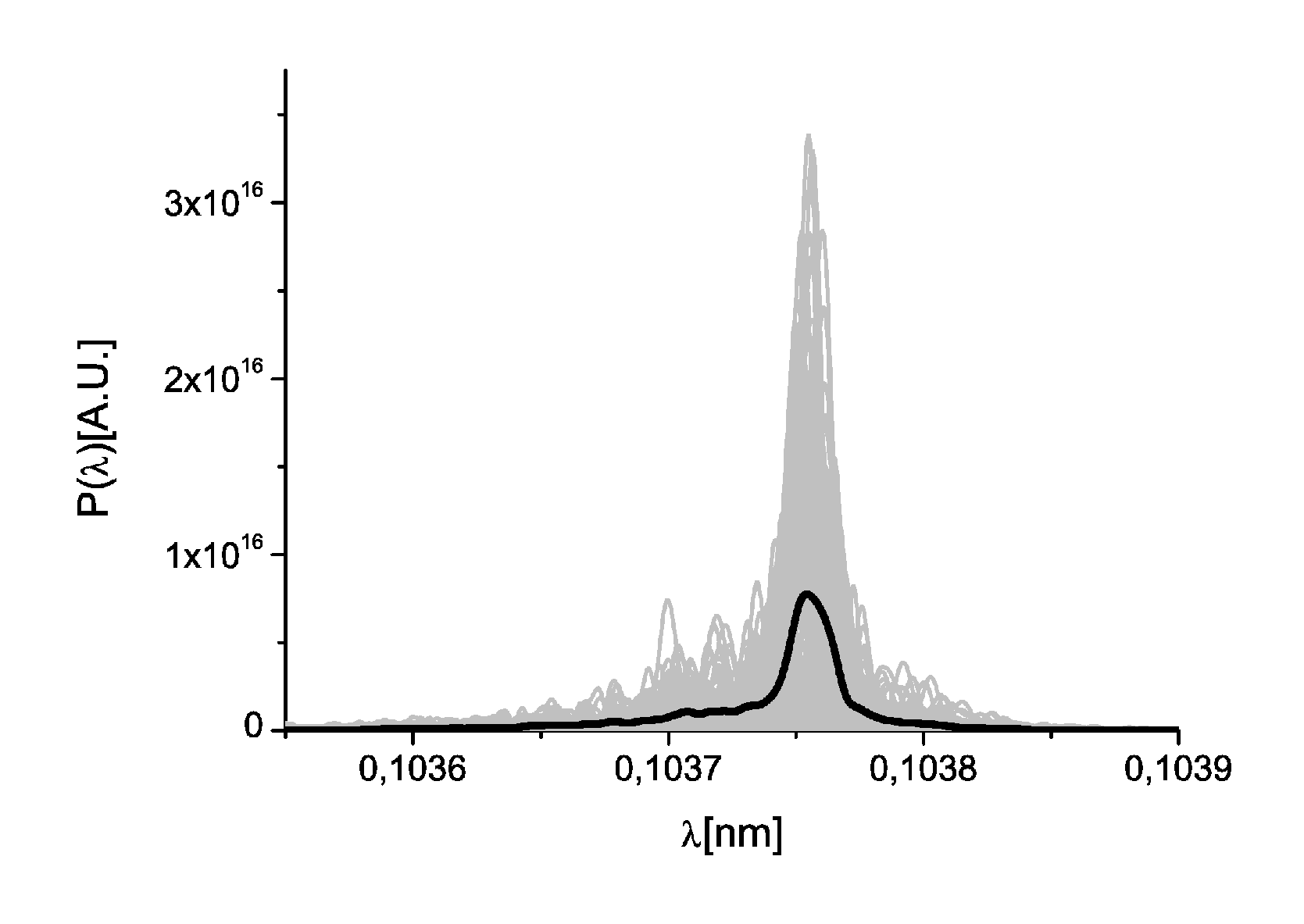}
\caption{Final output. Power and spectrum at the third harmonic
after tapering. Grey lines refer to single shot realizations, the
black line refers to the average over a hundred realizations.}
\label{12biof233p5}
\end{figure}
As usual, combining tapering with monochromatic radiation generation
is particularly effective, since the electron beam does not
experience brisk changes of the ponderomotive potential during the
slippage process. The final output is presented in Fig.
\ref{12biof233p5} in terms of power and spectrum. As one can see,
simulations indicate an output power of about 0.5 TW.

\begin{figure}[tb]
\includegraphics[width=0.5\textwidth]{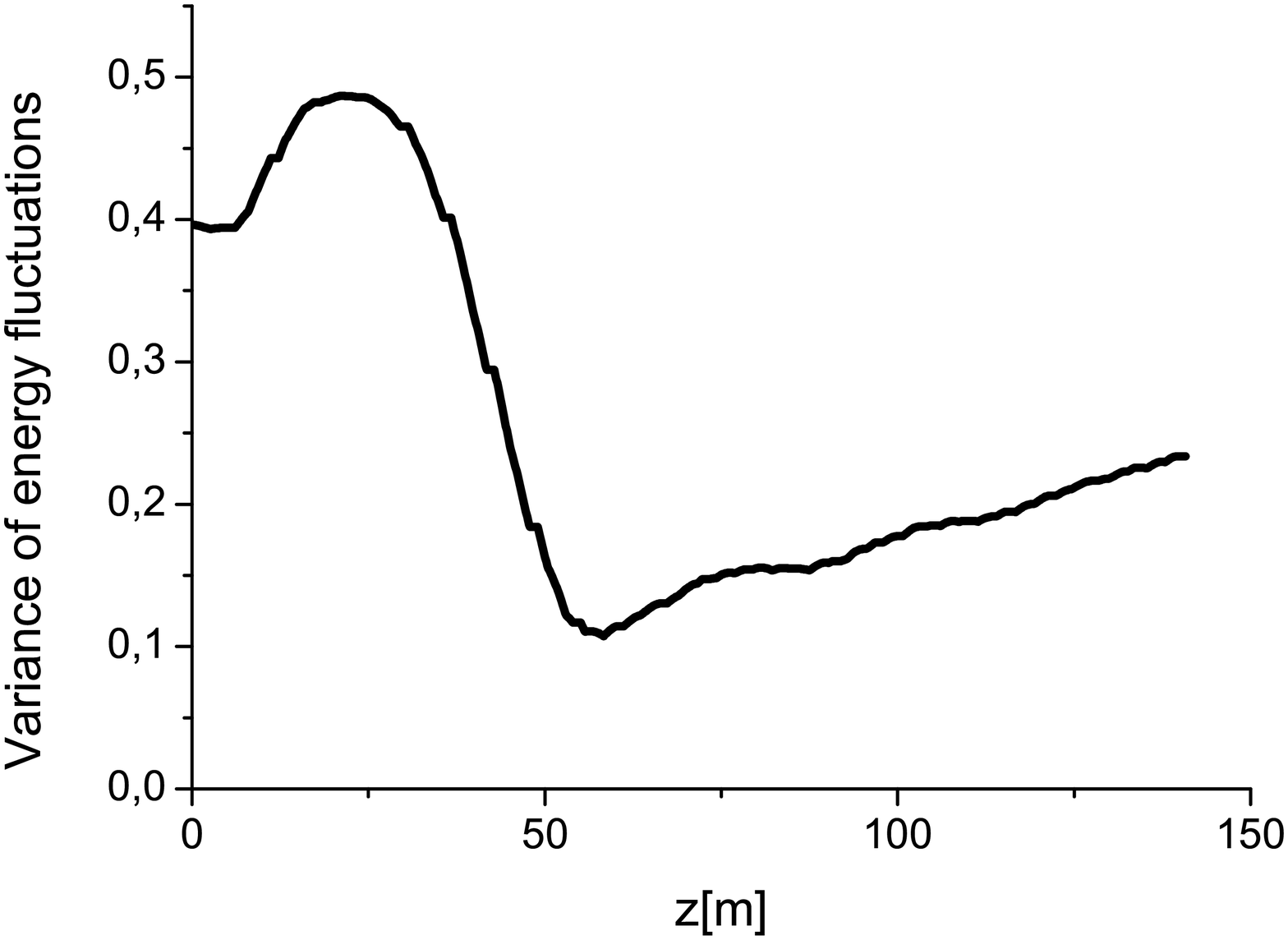}
\includegraphics[width=0.5\textwidth]{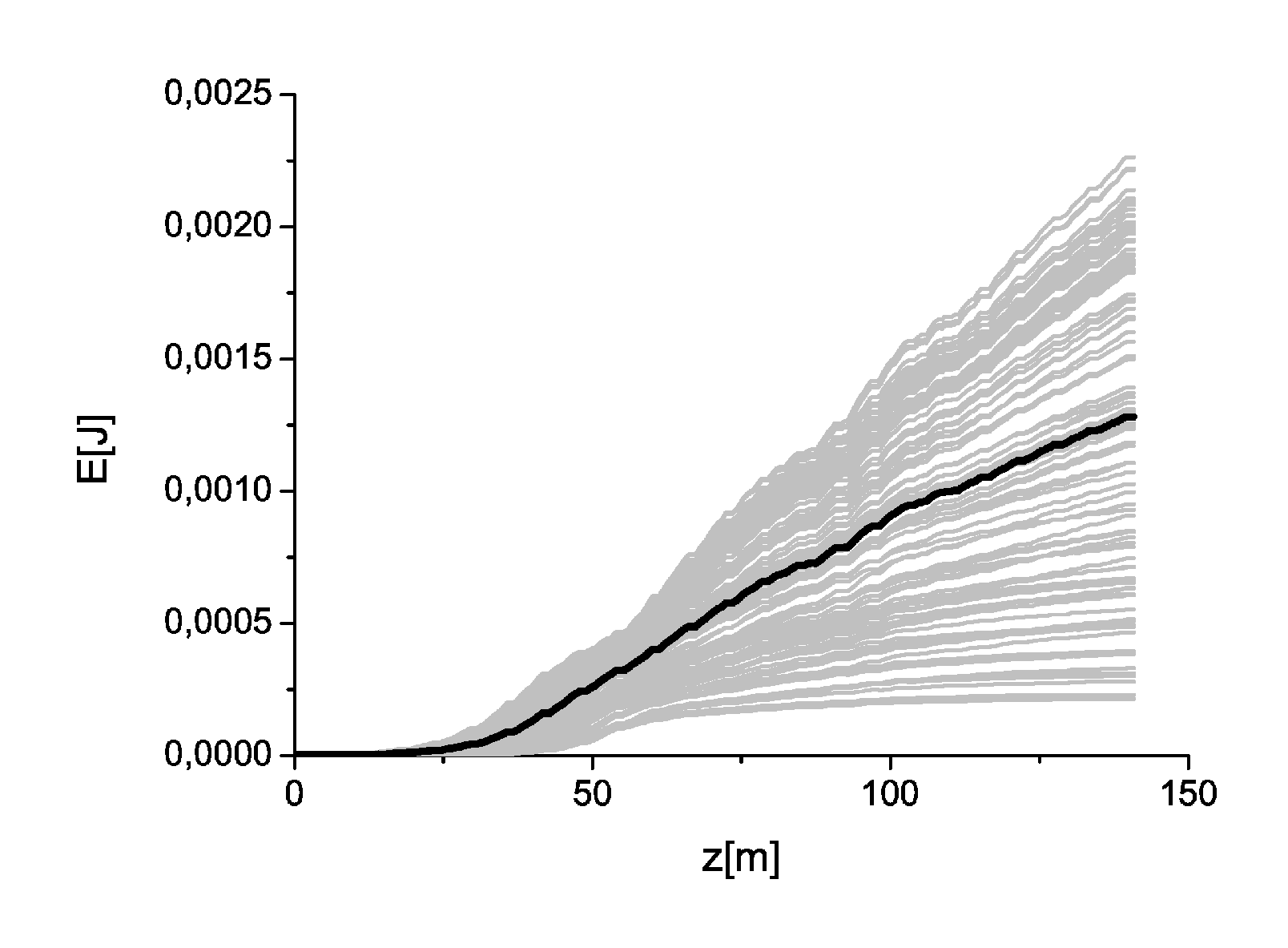}
\caption{Final output. Energy and energy variance of output pulses
for the case $\lambda = 0.1$ nm. In the left plot, grey lines refer
to single shot realizations, the black line refers to the average
over a hundred realizations.} \label{12biof243p5}
\end{figure}

\begin{figure}[tb]
\includegraphics[width=0.5\textwidth]{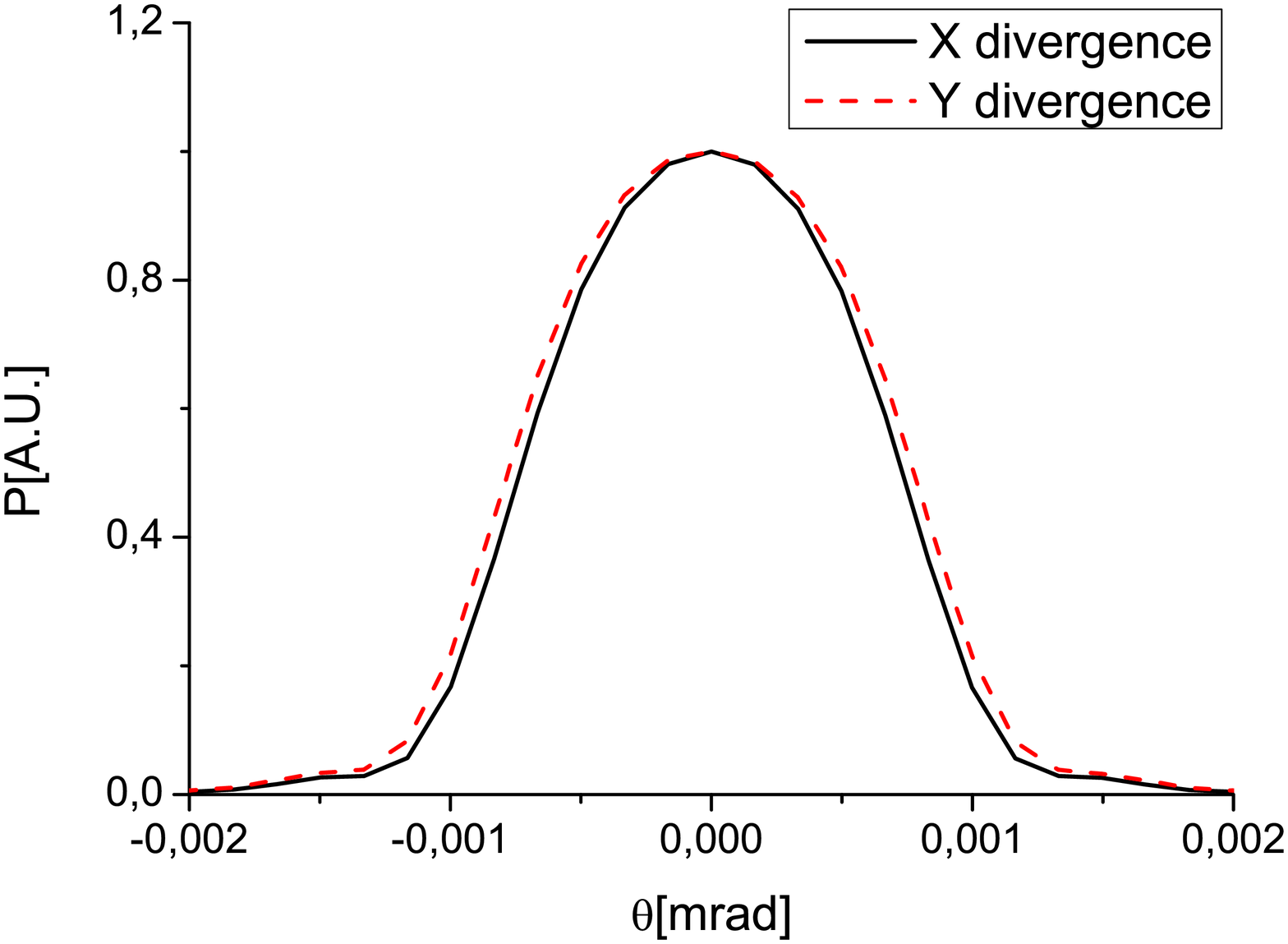}
\includegraphics[width=0.5\textwidth]{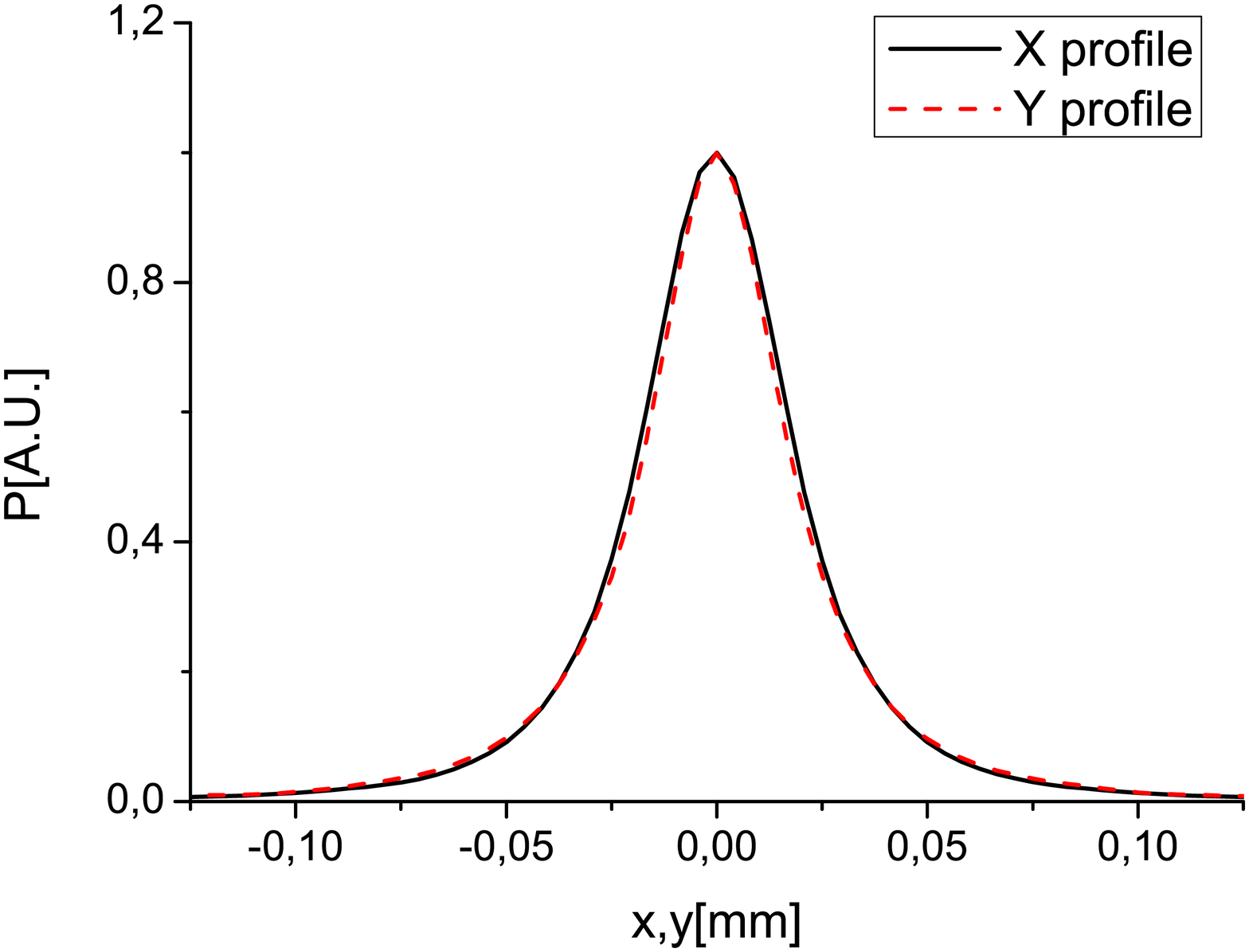}
\caption{Final output. X-ray radiation pulse energy distribution per
unit surface and angular distribution of the X-ray pulse energy at
the exit of output undulator for the case $\lambda = 0.1$ nm.}
\label{12biof2530p5}
\end{figure}
The energy of the radiation pulse and the energy variance are shown
in Fig. \ref{12biof243p5} as a function of the position along the
undulator. The divergence and the size of the radiation pulse at the
exit of the final undulator are shown, instead, in Fig.
\ref{12biof2530p5}. In order to calculate the size, an average of
the transverse intensity profiles is taken. In order to calculate
the divergence, the spatial Fourier transform of the field is
calculated.

\section{Conclusions}

The highest priority for bioimaging experiments at any advanced XFEL
facility is to establish a dedicated beamline for studying
biological objects at the mesoscale, including large macromolecules,
macromolecular complexes, and cell organelles. This requires 2 keV -
6 keV photon energy range and TW peak power pulses. However, higher
photon energies are needed to reach anomalous edges of commonly used
elements (such as Se) for anomalous experimental phasing. Studies at
intermediate resolutions need access to the water window
\cite{BERG}.

A conceptual design of a dedicated bio-imaging beamline based on the
self-seeding scheme developed for European XFEL was suggested in
\cite{OURCC}. The critical attribute of the proposed beamline,
compared with the baseline SASE1 and SASE2 beamlines, is a wider
photon energy range that spans from the water window up to the
K-edge of Selenium (12.6 keV). With the current design of the
European XFEL, the most preferable photon energy range between 3 keV
and 5 keV cannot be used for biological scattering experiments, but
the new proposed beamline could fill this gap operating at those
energies with TW peak power.

The first goal in developing a design for a dedicated bio-imaging
beamline is to make it satisfying all requirements. Once that is
done, the next step is to optimize the design, making it as simple
as possible. In order to improve the original design, here we
propose to extend the photon energy range of the self-seeding setup
with single crystal monochromator to lower photon energies down to 3
keV. An important aspect of this extension is that the self-seeding
scheme with single crystal monochromator is now routinely used in
generating of narrow bandwidth X-ray pulses at the LCLS
\cite{EMNAT}. It combines a potentially wide photon energy range
with a much needed experimental simplicity. Only one X-ray optical
element is needed, and no sensitive alignment is required. The range
of applicability of this novel method is a slightly limited,  at
present, by the availability of a short pulse duration (of about 10
fs or less). However, this range nicely matches that for single
biomolecule imaging.

Optimization of the bio-imaging beamline is performed with extensive
start-to-end simulations, which also take into account effects such
as the spatiotemporal coupling caused by the single crystal
monochromator. One must keep this effect in mind when performing the
design of any self-seeding setup. The spatial shift is proportional
to $\cot(\theta_B)$, and is therefore maximal in the range for small
Bragg angles $\theta_B$. A Bragg geometry close to backscattering
(i.e. $\theta_B$ close to $\pi/2$) would be a more advantageous
option from this viewpoint, albeit with a decrease in the spectral
tunability \cite{SHVID, OURTILT}. It is worth mentioning that this
distortion is easily suppressed by the right choice of crystals
within the photon energy range between 3 keV and 9 keV. Here we
propose to use a set of three diamond crystals. For the C(111), the
C(220) and the C(400) Bragg reflections ($\sigma$-polarization), it
will be possible to respectively cover the photon energy ranges 3
keV - 5 keV, 5 keV - 7 keV, and 7 keV - 9 keV. Finding a solution
suitable for the spectral range between 9 keV and 13 keV is major
challenge due to the large value of $\cot(\theta_B)$ for the C(400)
reflection case. Fortunately, even in this case, this obstacle can
be overcome by using a fresh bunch technique, and exploiting the
self-seeding setup with a C(111) single crystal monochromator, which
is tunable in the photon energy range around 4 keV, in combination
with harmonic generation techniques.

The goal of the present optimized proposal for a dedicated
bio-imaging beamline presented here is to aim for experimental
simplification and performance improvement. The design electron
energy in the most preferable spectral range 3 keV - 5 keV is
increased up to 17.5 GeV. The peak power is shown to reach a maximum
value of 2 TW. The new design takes additional advantage of the fact
that 17.5 GeV is the most preferable operation energy for the SASE1
and the SASE2 beamlines. Because of this, the optimized beamline is
not sensitive to the parallel operation with other European XFEL
beamlines.

\section{Acknowledgements}

We are grateful to Massimo Altarelli, Reinhard Brinkmann, Henry
Chapman, Janos Hajdu, Viktor Lamzin, Serguei Molodtsov and Edgar
Weckert for their support and their interest during the compilation
of this work.

\end{document}